%% file: main.tex
\documentclass[11pt]{article}

\usepackage{enumerate}
\usepackage[bottom]{footmisc}
\usepackage[T1]{fontenc}
\usepackage[utf8]{inputenc}
\usepackage{tikz-feynman}
\usepackage{simpler-wick}
\tikzfeynmanset{compat=1.1.0}
\graphicspath{{img/}}

\usepackage[utf8]{inputenc}
\usepackage[english]{babel}
\usepackage{amsmath, amsfonts, amssymb}
\usepackage{nccmath}
\usepackage{mathrsfs}
\usepackage{dsfont}
\usepackage{bbold}
\usepackage[mathscr]{euscript}
\usepackage{color}
\usepackage{amsthm}
\usepackage{psfrag}
\usepackage[toc,page]{appendix}
\usepackage{graphicx}
\usepackage{url}
\usepackage[hyperfootnotes=false]{hyperref}
\usepackage{multirow}
\usepackage{makecell}
\usepackage{bbm}
\usepackage{comment}
\usepackage{ulem}
\usepackage[backend=bibtex, sorting=none]{biblatex}
\catcode`@=11
\@addtoreset{equation}{section}
\catcode`@=12

\newcommand{\be}{\begin{equation}}
\newcommand{\ee}{\end{equation}}
\newcommand{\bl}{\begin{align}}
\newcommand{\el}{\end{align}}
\newcommand{\ba}{\begin{aligned}}
\newcommand{\ea}{\end{aligned}}
\newcommand{\beqa}{\begin{eqnarray}}
\newcommand{\eeqa}{\end{eqnarray}}

\def\d{\partial}
\def\const{\mbox{const}}

\newcommand{\qu}[1]{``{#1}''} 

\newcommand{\bseq}{\begin{subequations}}
\newcommand{\eseq}{\end{subequations}}

\newcommand{\q}{\mathcal{Q}}

\allowdisplaybreaks[4]
\numberwithin{equation}{section}  

\begin{document}
\begin{titlepage}
\clearpage

\title{{\bf Quantizing non-projectable Ho\v rava gravity \\with Lagrangian path integral} }
\author{D. Blas$^{a,b}$, \, F. Del Porro$^{c,d}$\footnote{francesco.del.porro@nbi.ku.dk}, \, M. Herrero-Valea$^{e}$,  \, J. Radkovski$^{f,g}$\footnote{jradkovski@perimeterinstitute.ca},  \, S. Sibiryakov$^{f,g}$~\\[2mm]
{\small\it $^a$Institut de Fisica d’Altes Energies (IFAE), The Barcelona Institute of Science and Technology,}\\ 
{\small \it Campus UAB, 08193 Bellaterra (Barcelona) Spain}\\
{\small \it  $^b$Instituci\'{o} Catalana de Recerca i Estudis Avan\c{c}ats (ICREA),}\\
{\small \it Passeig Llu\'{i}s Companys 23, 08010 Barcelona, Spain}\\
{\small\it $^c$Center of Gravity, Niels Bohr Institute, Blegdamsvej 17, 2100 Copenhagen, Denmark}\\ 
{\small \it $^d$Niels Bohr International Academy, Niels Bohr Institute,}\\
{\small \it Blegdamsvej 17, 2100 Copenhagen, Denmark}\\ 
{\small\it $^e$Barcelona Supercomputing Center (BSC), Plaça d’Eusebi Güell 1-3, 08034 Barcelona, Spain}\\ 
{\small\it $^f$Department of Physics and Astronomy, McMaster University,}\\
{\small \it 1280 Main St W, Hamilton, ON L8S 4M1, Canada}\\ 
{\small\it $^g$Perimeter Institute for Theoretical Physics, 31 Caroline St N,  Waterloo, ON N2L 2Y5, Canada}
}
\date{}

\maketitle

\begin{abstract}
We formulate the quantum version of \textit{non-projectable} Ho\v rava gravity as a Lagrangian theory with a path integral in the configuration space with an ultra-local in time, but non-local in space, field-dependent measure. Using auxiliary fields, we cast the measure into a local form satisfying several bosonic and fermionic symmetries.
We perform an explicit one-loop computation in the theory in $(2+1)$ dimensions, using for the case study the 
divergent part of the action on a background with non-trivial shift vector; the background spatial metric is taken to be flat and the background lapse function is set to 1. No truncations are assumed at the level of perturbations, for which
we develop a diagrammatic technique and a version of the heat-kernel method. We isolate dangerous linear-in-frequency divergences in the two-point function of the shift, which can lead to spatial non-localities, and explicitly verify 
their cancellation. This leaves a fully local expression for the divergent part of the quadratic effective action, from which we extract the beta functions for the Newton constant and the essential coupling $\lambda$ in the kinetic term of the metric. We formulate the questions that need to be addressed to prove perturbative renormalizability of the non-projectable Ho\v rava gravity.
\end{abstract} 

\thispagestyle{empty}
\end{titlepage}

\newpage

\tableofcontents

\newpage

\input{Sections/Section1} 
\input{Sections/Section2} 
\input{Sections/Section3}

\input{Sections/Section4}
\input{Sections/Section5}

\section*{Acknowledgments}
We thank Andrei Barvinsky, Ted Jacobson, and Simon Lyakhovich for insightful discussions. FDP acknowledges support of the research grant (VIL60819) from VILLUM FONDEN. The Center of Gravity is a Center of Excellence funded by the Danish National Research Foundation under grant No. 184. The work of JR and SS is supported by the 
Natural Sciences and Engineering Research Council (NSERC) of Canada. Research at Perimeter Institute is supported in part by the Government of Canada through the Department of Innovation, Science and Economic Development Canada and by the Province of Ontario through the Ministry of Colleges and Universities.
This publication is part of the R\&D\&i project PID2023-146686NB-C31 funded by MICIU/AEI/10.13039/501100011033/ and by ERDF/EU (DB). 
IFAE is partially funded by the CERCA program of the Generalitat de Catalunya.
This work is supported by ERC grant GravNet (ERC-2024-SyG 101167211, DOI: 10.3030/101167211). Funded by the European Union. Views and opinions expressed are however those of the author(s) only and do not necessarily reflect those of the European Union or the European Research Council Executive Agency. Neither the European Union nor the granting authority can be held responsible for them. 
DB acknowledges the support from the European Research Area (ERA) via the UNDARK project of the Widening participation and spreading excellence programme (project number 101159929).
The work of MHV is supported by an AI4Science fellowship, funded by the Ministerio para la Transformación Digital y de la Función Pública and by the European Union, under the Plan de Recuperación, Transformación y Resiliencia, C005/24-ED CV1.
\appendix

\input{Appendix/AppendixA}

\printbibliography{}

\end{document}

%% file: Sections/Section1.tex
\section{Executive summary}

Ho\v rava gravity was proposed in 2009 as a theory of quantum gravity within perturbative quantum field theory in 3+1 dimensions \cite{Horava2008, Horava2009} (see the reviews \cite{Mukohyama2010, Sotiriou2010, Herrero-Valea2023, Barvinsky2023Review}). The basic assumptions behind this bold possibility are the fundamental breaking of diffeomorphisms to the subgroup of \eqref{FDiffsCoords} and the anisotropic scaling of \eqref{AnisotropicScaling}. Since then, several aspects of the proposal have been scrutinized. There are two main current challenges of the theory:
\begin{itemize}
    \item It would be desirable to endow the proposal with a mechanism to recover Lorentz invariance at scales where it has been tested to high precision. Several mechanisms have been suggested in the past (see below), but the results are inconclusive;
    \item We still lack the proof showing that the scaling arguments put forward in the original proposal survive after quantization for the cases that can reproduce the experimental data, the so-called {\it non-projectable} version. This is highly non-trivial because in addition to the absence of a Lorentz symmetry the theory possesses a complicated constraint structure.
\end{itemize}
In this work, we make a step towards completing the \textit{second} challenge. The root of the problem lies in the appearance of the instantaneous mode brought by the second-class constraints. The mode manifests itself as the  pole with no frequency dependence (cf. \eqref{eq:nn_prop}) present in the propagator of the time-time component of the metric (lapse). In the terminology of \cite{Barvinsky2015}, such propagators are called {\it irregular}. They violate the naive power-counting, producing extra divergences of the frequency integrals in the loop expansion, which may have non-local dependence on the spatial coordinates. 
Yet, the works \cite{Bellorin2022,Bellorin2024} have shown that in fact all the 
{\it irregular frequency divergences} (IFDs) 
brought by the propagation of the instantaneous mode in the loops cancel out, suggesting that the quantum theory is under perturbative control. These works, however, did not take into account the need for the regularization and renormalization, which very well might ruin the cancellation, and correspondingly the locality of the theory. 

To assess this problem we show that starting from the canonical phase space path integral formulation  with second-class constraints in \eqref{eq:pathintegral}, we can, following the reasoning of \cite{HenneauxSlavnov1994},  rewrite the theory as path integral in the \textit{configuration space} \eqref{eq:Z_mu}, with the full weight given by (exponent of) the original action  \eqref{eq:ActionFull} and, in addition, a non-trivial field-dependent measure $\mu$. We uncover a rich symmetry structure of the measure, with several fermionic and bosonic symmetries which strongly restrict the form of possible counter-terms, but still do not fix them completely. Preservation of the measure upon renormalization remains the key open question, whose solution will determine whether the non-projectable Ho\v rava gravity is renormalizable or not.  

We further perform an explicit one-loop computation using two regularization techniques where the IFDs
are manifest. We obtain the divergent part of the one-loop effective action, focusing on the sector with
only the shift field present and we set the computation in a simpler case of $(2+1)$ dimensional theory.\footnote{Ho\v rava gravity in $(2+1)$ dimensions, unlike the general relativity in the same number of dimensions, has propagating degrees of freedom, making its physical properties not so different from those of the $(3+1)$-dimensional theory.} 
Our main results are:
\begin{itemize}
    \item We demonstrate, in our setup, that despite appearing in the intermediate steps of the calculation, the IFDs eventually cancel. We thereby confirm in a simpler Lagrangian setting the previous general results \cite{Bellorin2022} obtained in the Hamiltonian approach.
    \item We develop the diagrammatic technique and the heat-kernel method for computing the remaining local divergences of the one-loop effective action and extract from them the $\beta$-functions for the Newton's constant $G$ and the coupling $\lambda$ (see Eq.~(\ref{eq:ActionFull}) for the definition of the couplings).   
\end{itemize}
We expect no difficulty in extending our computations to 3+1 dimensions, except for the large number of terms involved in the potential of the theory (order $\mathcal{O}(100)$ as compared to $\mathcal{O}(10)$). While we are confident that the cancellation of IFDs at one-loop level persists outside the shift-sector of the theory, 
its extension to higher loops is contingent on the preservation of the form of the measure. 
If this question can be answered in the positive way, non-projectable Ho\v rava gravity would achieve the status of a viable renormalizable field theory of quantum gravity. Studies of the {\it projectable} version, which has been shown to be renormalizable and exhibit asymptotic freedom \cite{Barvinsky2015,Barvinsky2017Renorm,Barvinsky2017AssFreed,Barvinsky2021,Radkovski2023,Barvinsky2023,Barvinsky2024},
provide strong motivation to pursue this effort.

The paper is organized as follows. In Sec.~\ref{sec: Introduction} we review the projectable and non-projectable versions of Ho\v rava gravity and discuss the problems associated with the quantization of the latter. 
In Sec.~\ref{sec: Quantization} we formulate the theory in the phase space, integrate out the canonical momenta, 
and discuss the properties of the resulting path integral measure. We introduce gauge-fixing of spatial diffeomorphisms using the background-field method. Sec.~\ref{sec: Beta Function} is dedicated to a diagrammatic one-loop calculation of the divergence in the effective action for the shift field. We introduce a higher-derivative regularization in time accompanied by a dimensional regularization in space. This combination renders all diagrams, including IFDs, finite and allows us to isolate IFDs as terms linear in the regulator. We verify the cancellation of IFDs and use the remaining local divergences to extract the 
$\beta$-functions of two couplings of the theory.  
In Sec.~\ref{Sec:Heat_Kernel} we develop a heat kernel technique for the non-projectable Ho\v rava gravity. We use it to cross-check the cancellation of IFDs and the
expressions for the $\beta$-functions using yet another regularization, the point-splitting in time. 
We conclude in Sec.~\ref{sec: Conclusion}.

\section{Ho\v rava gravity}
\label{sec: Introduction}

Ho\v rava gravity is a model of quantum gravity formulated in the framework of renormalizable quantum field theory. The proposal emerges from an action principle and a four-dimensional metric as a basic degree of freedom. Two features are key for renormalizability. The first one corresponds to a geometrical structure in the form of a spacetime \textit{foliation} with spacelike leaves.  The foliation \qu{breaks} the diffeomorphism symmetry group of General Relativity (GR) down to its subgroup of the \textit{foliation preserving diffeomorphisms} (FDiffs)
\begin{equation}
\label{FDiffsCoords}
    \mathbf{x} \mapsto \tilde{\mathbf{x}}(t, \mathbf{x}) \, , 
\quad t \mapsto
    \tilde{t}(t) \, ,
\end{equation}
with, crucially, time reparameterizations being \textit{projectable} functions, that is, functions depending on a leaf as a whole. 

The second feature is \textit{anisotropic scaling}, a rather natural possibility given the first one: with time \qu{decoupled} from space by the foliation structure, the theory is allowed to have different scaling laws for the time and space coordinates,
\begin{equation}
\label{AnisotropicScaling}
    {\bf x} \rightarrow b^{-1} \, {\bf x}, \quad t \rightarrow b^{-z} \,
    t \, , 
\end{equation}
with generically dynamical \textit{critical exponent} $z$. Taken together, the two features allow one to formulate the action for the metric with higher spatial derivatives without including extra time derivatives. This improves the convergence of the loop integrals -- and correspondingly the ultraviolet (UV) behavior of the theory -- without introducing Ostrogradsky ghosts and corresponding instabilities as in the, also perturbatively renormalizable, quadratic gravity \cite{Stelle1976, Stelle1977}.

Ho\v rava gravity can be conveniently formulated in the Arnowitt-Deser-Misner (ADM) variables adapted to the foliation \cite{Arnowitt59}, 
\begin{align}
\label{Interval}
    ds^2 = -N^2 dt^2 + \gamma_{ij} (dx^i + N^i dt)(dx^j + N^j dt), \quad i = 1, \dots, d \, \, ,
\end{align}
with \textit{lapse} function $N$, \textit{shift} vector $N^i$, and the \textit{metric} on the leaves $\gamma_{ij}$ transforming (\textit{actively}) under FDiffs \eqref{FDiffsCoords} via
\bseq
\label{FDiffsFields}
\begin{align}
\label{FDiffsFields1}
    &N(t,\mathbf{x}) \rightarrow \tilde{N}(t,\mathbf{x}) = N\big(\tilde{t}(t),\tilde{\mathbf{x}}(\mathbf{x},t)\big) \,  \frac{d \tilde{t}}{dt} \, , \\
    \label{FDiffsFields2}
     &N^i(t,\mathbf{x}) \rightarrow \tilde{N}^i(t,\mathbf{x}) = \Bigg(N^{j}\big(\tilde{t}(t),\tilde{\mathbf{x}}(\mathbf{x},t)\big) \frac{\partial x^{i}}{\partial \tilde{x}^{j}} - \frac{\partial x^i}{\partial \tilde{t}}\Bigg) \, \frac{d \tilde{t}}{d t} \, , \\
     \label{FDiffsFields3}
    &\gamma_{ij}(t, \mathbf{x} ) \rightarrow \tilde{\gamma}_{ij}(t, \mathbf{x} ) = \gamma_{kl}\big(\tilde{t}(t),\tilde{\mathbf{x}}(\mathbf{x},t)\big) \,  \frac{\partial \tilde{x}^k}{\partial x^{i}} \frac{\partial \tilde{x}^l}{\partial x^{j}} \, .
\end{align}
\eseq
The building blocks of the theory are the objects that transform covariantly under the FDiffs above:
\begin{itemize}
    \item The \textit{intrinsic} curvature of the foliation leafs, $R_{ijkl}$,
together with its traces, $R_{kl} \equiv \gamma^{ij} R_{ikjl}$, $R \equiv \gamma^{ij} R_{ij}$.
\item The covariant derivative $\nabla_i$ compatible with metric $\gamma_{ij}$.
\item The \textit{extrinsic} curvature,
\begin{align}
\label{eq: Extrinsic Curvature}
    K_{ij} \equiv \frac{1}{2}D_t \gamma_{ij} \equiv \frac{1}{2N}\left(\dot\gamma_{ij}-\nabla_i N_j-\nabla_j N_i\right) \, , 
\end{align}
with its trace $K \equiv \gamma_{ij} K^{ij}$. Here the dot denotes the time derivative, $\dot{\gamma}_{ij} \equiv \partial_t \gamma_{ij}$.
\item The \textit{acceleration},
\begin{equation}
    a_i = \nabla_i \log N \, , \quad \gamma_{ij} a^i a^j \equiv a^2 \, .
\end{equation}
\end{itemize}
Using these covariant operators, one constructs the most general \cite{Horava2009, Blas2009} action with two time derivatives 
\begin{align}
\label{eq:ActionFull}
    S = \frac{1}{2G}\int dt d^d x\sqrt{\gamma} N \left(K_{ij}K^{ij}-\lambda K^2 - \mathcal{V}[a,R] \right) \, , 
\end{align}
with \textit{potential} $\mathcal{V}$ consisting of all possible FDiff-invariant terms built out of the invariants with $a_i$ and $R_{ijkl}$.
Here $G$ is a familiar Newton constant governing the strength of the interactions and $\lambda$ is a parameter that measures the departure of the kinetic term from the relativistic invariant one. Setting $\lambda=1$ and restricting the potential to $\mathcal{V} = -2 \Lambda + R$ on formally recovers the Einstein--Hilbert action for GR with cosmological constant $\Lambda$, formulated in the ADM variables. In general, $\lambda \neq 1$ and in fact runs with the energy scale. Its $\beta$-function, as well as the one for $G$, will be one of the main results of this paper.

By design \cite{Horava2009}, the theory defined by \eqref{eq:ActionFull} is power-counting renormalizable \textit{if} one assumes that the path integral  at high energies is dominated by a  \textit{free fixed point} with anisotropic scaling \eqref{AnisotropicScaling} with critical exponent
\begin{equation}
    z = d \, ,
\end{equation}
under which the fields scale as
\begin{equation}
\label{eq:Metric_Dimension}
    \gamma_{ij}(t,\mathbf{x}) \rightarrow \gamma_{ij}(b^{-d} \, t, b^{-1} \mathbf{x}) \, , \quad N(t,\mathbf{x}) \rightarrow N(b^{-d} \, t, b^{-1} \mathbf{x}) \, ,
\end{equation}
and, non-trivially,
\begin{equation}
\label{eq:Shift_Dimension}
    N^i(t,\mathbf{x}) \rightarrow b^{d-1} N(b^{-d} \, t, b^{-1} \mathbf{x}) \, .
\end{equation}
That is, the metric and the lapse have scaling dimension zero, whereas the shift has scaling dimension $d-1$.
Since the renormalization properties of the theory are determined by its ultraviolet (UV) behavior, our focus will be on the high-energy limit. We also set $\Lambda=0$ and work in the asymptotically flat spacetime. 

In the simplest anisotropic case of two spatial dimensions, $d=2$, 
the marginal operators with respect to scaling \eqref{AnisotropicScaling} have \textit{four} derivatives. Explicitly, keeping only the marginal terms, we get a potential in \eqref{eq:ActionFull} that can be
parameterized with 7 couplings $\mu_i$  \cite{Herrero-Valea2023,Barvinsky2023Review},
\begin{align}
\label{Potential}
    \mathcal{V}^{\text{high energy}}_{d=2} =  \mu_1 R^2 + \mu_2 a^4 + \mu_3 a^2 R + \mu_4 a^2 \nabla_i a^i + \mu_5 R \nabla_i a^i + \mu_6 \nabla_i a^j \nabla^i a_j + \mu_7 (\nabla_i a^i)^2 \, .
\end{align}
At $d=3$, the potential at high-energies is dominated by the terms with 6 derivatives, such as e.g. $R^3$, of which there are plenty --- $\mathcal{O}(100)$ terms.
At low energies, these higher-order operators are suppressed, the terms with two derivatives start to dominate, and we have relativistic scaling with $z=1$.\footnote{The reader should be familiar with a reverse scenario: QED has $z=1$, but its low-energy Schrödinger EFT we use for atomic physics has $z=2$.}

In \textit{projectable} Ho\v rava gravity, one takes the lapse to be a function of time, consistently with the FDiffs \eqref{FDiffsCoords},
\begin{equation}
\label{eq:Projectability}
    N = N(t) \, .
\end{equation}
This implies $a_i=0$ and only the first term in \eqref{Potential} survives. By quantizing this theory, one can show that it is perturbatively renormalizable \cite{Barvinsky2017Renorm, Barvinsky2015} in any number of dimensions $d$. Moreover, it possesses several asymptotically free UV fixed points (found in \cite{Barvinsky2017AssFreed} for $d=2$ and in \cite{Barvinsky2019, Barvinsky2021} for $d=3$) with a unique point giving rise to a regular Renormalization Group (RG) flow towards the infrared (IR), see \cite{Barvinsky2017AssFreed} for $d=2$ and \cite{Barvinsky2023, Barvinsky2024} for $d=3$.\footnote{See also \cite{DOdorico:2014tyh,DOdorico:2015pil} for non-perturbative studies of the RG.} That is, \textit{projectable} Ho\v rava gravity is perturbatively UV complete. 

However, the \textit{projectable} Ho\v rava gravity is problematic at low energies, where the theory develops a tachyonic instability around the flat background \cite{Koyama2009} or enters the strong coupling regime \cite{Blas2010}, either scenario taking it very far from recovering GR perturbatively. Things improve if one drops the requirement \eqref{eq:Projectability} and allows the lapse to be a proper dynamical field with arbitrary $\mathbf{x}$-dependence, 
\begin{equation}
    N \rightarrow N(t, \mathbf{x}) \, .
\end{equation}
This leads to the \textit{non-projectable} model which
at low energies is just GR plus a stable scalar field that captures the dynamics of the foliation \cite{Blas2010}. The phenomenological constraints on the pure gravity side are rather severe \cite{Blas2014, EmirGumrukcuoglu2017}, but still allow some parameter space where the theory is viable. It is more challenging, however, to limit the percolation of the Lorentz violation into the Standard Model (SM) matter sector \cite{Kostelecky2008, Liberati2012, Liberati2013}. 
A number of mechanisms have been proposed to naturally recover Lorentz invariance at low-energies~\cite{GrootNibbelink2004, Pospelov2010, Pujolas2011, Anber2011, Bednik2013, Kharuk2015,Baggioli:2024vza}, whereas Ref.~\cite{Martynenko2024} has suggested that Lorentz violation in SM at around $10^{16}$~GeV could resolve certain puzzles in cosmic ray physics. 

The question we address in this paper is the UV behavior of the non-projectable model. 
First of all, while the \textit{projectable} model propagates only modes with a regular dispersion relation,
\begin{equation}
\label{DispersionRelationPhysProj}
    \omega^2 \sim q^{2d} \, ,
\end{equation}
the \textit{non-projectable} theory, in addition, has a non-local instantaneous mode corresponding to the fluctuations of the lapse~\cite{Blas2010}. It gives rise to an $\omega$-independent {\it irregular} piece in the lapse propagator~\cite{Barvinsky2015},
\begin{equation}
\label{eq: Irregular Mode}
    \langle N(\omega, q) N (-\omega, -q)\rangle \ni \frac{\const}{q^{2d}} \, ,
\end{equation}
which leads to IFDs and spoils the naive power-counting renormalizability. Technically, the source of the irregular contribution (\ref{eq: Irregular Mode}) is in the $a_i$-dependent terms of the potential $\mathcal{V}$, allowed by the FDiff symmetry. As we discuss in Sec.~\ref{sec: Quantization}, the same terms complicate the algebra of constraints of the theory. In Dirac's terminology, in addition to the \textit{first-class} constraints associated with the gauge symmetries, we get \textit{second-class} constraints. 
One possibility within the canonical quantization would be to solve the second-class constraints explicitly with respect to the lapse and replace 
the Poisson bracket by the Dirac bracket~\cite{Henneaux1994}. In the non-projectable Ho\v rava gravity, however, the solution of the constraints would involve inverse powers of the spatial Laplacian, and the resulting Dirac bracket would be non-local. 

An alternative approach \cite{Senjanovic1976, Henneaux1994}, naturally arising within the functional quantization framework, 
is to work with the phase space path integral and enforce the constraints with $\delta$-functions. This approach was adopted in the previous works on quantum non-projectable theory~\cite{Bellorin2019, Bellorin2021, Bellorin2022, Bellorin2023, Bellorin2024}, where the theory was quantized using the Batalin--Fradkin--Vilkovisky \cite{Fradkin1975,Batalin1983b,Batalin1983c} 
formalism of phase space path integral with the Hamiltonian first-order action. It was further shown how to incorporate the constraining $\delta$-functions into the action with the help of Lagrange multiplier fields. The second-class constraints involve spatial derivatives, which translate into the spatial derivatives of these Lagrange multipliers. 
The latter then have propagators of the form (\ref{eq: Irregular Mode}), giving rise to extra IFDs. Ref.~\cite{Bellorin2022} then demonstrates that the IFDs coming from the lapse propagator are precisely canceled out by the IFDs brought by the Lagrange multipliers in the \textit{all-loop} graphs constructed with the \textit{bare} Hamiltonian action. 

While this result strongly suggests that the model is well-behaved as a quantum theory, there are still some questions to answer in order to demonstrate it fully.  Most notably, it remains to be shown that the structure of the first-order action ensuring the cancellation of IFDs persists after regularization and renormalization. Showing this directly in the phase space appears challenging. This motivates formulating the theory, including the implementation of the second-class constraint, as the Lagrangian path integral typically used in  
the renormalization procedure. Apart from that, the Lagrangian picture has a further technical advantage of requiring fewer fields and, moreover, provides a direct connection with the well-developed purely first-class projectable theory \cite{Barvinsky2015}. 

Motivated by this, we follow \cite{HenneauxSlavnov1994} and demonstrate that, starting from the phase space path integral with second class constraints, one can get to the formulation with the original Lagrangian. The price to pay is the extra field-dependent measure, which is ultra-local in time, but non-local in space. It, loop-by-loop, produces divergent spatially non-local counter-terms with $\delta$-functions in time\footnote{We explicitly show the dimensionality of the $\delta$-function by an index in the round brackets. Since the $\delta$-function of time is one-dimensional, we denote it by $\delta^{(1)}$.} 
$\delta^{(1)}(0)$, which in turn correspond to power law divergences in energy --- the IFDs. These are of the same form as the IFDs generated when the $\omega$-independent propagators \eqref{eq: Irregular Mode} appear in the loops. The results of \cite{Bellorin2022} ensuring the cancellation of IFDs in the Hamiltonian formalism can now be interpreted as the cancellation of divergences between various pieces coming from the Lagrangian action and the path integral measure. 

In relativistic theories with covariant gauges the ultralocal measure produces power-law divergences both in frequency and momentum. One can usually discard divergences like this when employing dimensional regularization \cite{Collins1984}. The situation is less clear already for relativistic theories in non-covariant gauges, which mimic our case. For instance, for the Coulomb gauge, the non-covariant split-dimensional regularization of~\cite{Leibbrandt1996} (see also \cite{Barvinsky1984}) becomes extremely complicated beyond one loop \cite{Leibbrandt1997, Heinrich1999}, and practically it would be challenging to tell whether one can consistently apply it. Moreover, a regularization scheme setting the non-local divergences in the effective action to zero masks the hidden algebraic structure that leads to their cancellation. For these reasons, it is desirable to have an alternative scheme. 

In this work, we propose a hybrid regularization method, dimensional in space, and with higher derivatives in time. It has the advantage of unmasking the cancellation of IFDs, which we demonstrate with the calculation of a particular class of diagrams at one loop. We do not attempt to be general, 
postponing to future the complete one-loop renormalization of the theory and investigation whether it preserves the structure of the measure which ensures cancellation of the IFDs.
Rather, our aim here is to 
make the non-trivial cancellation of IFDs manifest and 
demonstrate
that one \textit{can} do sensible one-loop quantum calculations in the non-projectable Ho\v rava gravity.
To this end, we specialize to the case $d=2$ and derive the shift-dependent divergent part of the one-loop effective action. 
The measure does not contribute to this computation, with IFDs canceling among different diagrams coming from the action alone. 
We then find the simplest set of $\beta$-functions, those for the couplings $G$ and 
$\lambda$.
Similarly to the projectable case \cite{Barvinsky2019}, the coupling $\lambda$ is \textit{essential}, and we check that its $\beta$-function is gauge invariant. Whereas the Newton coupling $G$ is \textit{not essential}, so its $\beta$-function depends on the gauge fixing of FDiffs.


%% file: Sections/Section2.tex
\section{Quantizing the non-projectable model}
\label{sec: Quantization}

In this section, we review the constraint structure of the theory and quantize the model in the background gauge. We start with the analysis of the classical theory \cite{Donnelly2011} (see also \cite{Bellorin2011}) and, following \cite{Bellorin2022, Bellorin2024}, set up the path integral in the phase space restricted by the constraints using the general method of \cite{Senjanovic1976}. Following the spirit of \cite{HenneauxSlavnov1994}, we then integrate out all the canonical momenta to get the partition function in configuration space based on the classical action \eqref{eq:ActionFull} plus the local in time measure. The resulting Lagrangian theory is then quantized in the background field gauge \cite{Abbott1982}, along the lines of \cite{Barvinsky2019}. 

\subsection{Second-class constraints}
\label{Subsec:Constraints}
Taking the Lagrangian \eqref{eq:ActionFull} as a starting point, we can derive the canonical Hamiltonian. After integration by parts, it reads
\begin{equation}
\label{eq: Canonical H}
    H \equiv \int d^d x \, \mathcal{H} \equiv \int d^d x \, \left( \Pi^{ij} \dot{\gamma}_{ij} - \mathcal{L} \right) = \int d^d x \,( {\mathcal{N}} \mathcal{H}_{0} + N^{i} \mathcal{H}_i )\, .
\end{equation}
For reasons that will become clear later, we use a calligraphic notation $\mathcal{N}$ for the lapse.
$\Pi^{ij}$ in \eqref{eq: Canonical H} are the canonical momenta for the metric $\gamma_{ij}$, 
\begin{equation}
    \Pi^{ij} \equiv \frac{\delta S}{\delta \dot{\gamma}_{ij}} = \frac{\sqrt{\gamma}}{2G} ({\cal K}^{ij} - \lambda {\cal K} \gamma^{ij}) \, , 
\end{equation}
where we introduced the calligraphic notation for the extrinsic curvature defined using the lapse ${\cal N}$.  
The functions $\mathcal{H}_0, \mathcal{H}_i$ are given by
\begin{equation}
\label{eq: H0Hi}
    \mathcal{H}_0= \frac{2G}{\sqrt{\gamma}} \left( \Pi_{ij} \Pi^{ij} +  \frac{\lambda}{1-\lambda d} \Pi^2+ \frac{\gamma}{(2 G)^2}\mathcal{V} \right)\, , \quad \mathcal{H}_i = - 2 \gamma_{i k} \nabla_{j} \Pi^{jk} \, .
\end{equation}
Since the action \eqref{eq:ActionFull} contains no time derivatives of lapse and shift, we get two sets of \textit{primary constraints} for their canonical momenta,
\begin{equation}
\label{eq: Primary Constraints}
    \Pi_{\mathcal{N}} = 0, \quad \Pi_{N^i} \equiv \Pi_{i} = 0 \, .
\end{equation}
The existence of primary constraints implies that the Hamiltonian should be defined on the constraint surface~\cite{Henneaux1994}, hence we need to extend it with the Lagrange multipliers enforcing the constraints \eqref{eq: Primary Constraints}:
\begin{equation}
\label{eq: Extended H}
    \mathcal{H} \rightarrow  \mathcal{H} + v \Pi_{\mathcal{N}} + v^{i} \Pi_i \, .
\end{equation}
From the requirement of preserving the constraints in time, we get the \textit{secondary constraints}: 
\begin{align}
\label{eq: Secondary Constraints}
    0 &= \dot{\Pi}_{\mathcal{N}} = \{\Pi_\mathcal{N}, H \} = \frac{\delta H}{\delta \mathcal{N}} = \mathcal{H}_0 - \frac{1}{\mathcal{N}} \nabla_{i} V^{i} \equiv \tilde{\mathcal{H}}_0 \, , \nonumber \\
    0 &= \dot{\Pi}_{i} =  \{\Pi_i, H \} = \frac{\delta H}{\delta N^{i}} = \mathcal{H}_{i}  \, ,
\end{align}
where
\begin{equation} \label{eq:dV}
     V^{i}(x) = \frac{1}{2 G} \frac{\delta}{\delta a_i (x)} \int d^d x \sqrt{\gamma} \mathcal{N} \mathcal{V} \, .
\end{equation}
The constraints $\Pi_i$ and $\mathcal{H}_i$  are \textit{first-class}, meaning that their Poisson brackets with the Hamiltonian and other constraints are proportional to the constraints themselves, i.e., they vanish on the constraint surface. Because of the $\mathcal{N}$-dependence of the potential $\mathcal{V}$ this is not true for $\Pi_\mathcal{N}$ and $\tilde{\mathcal{H}}_0$:
\begin{equation}
\label{eq: Bracket}
    \{\Pi_{\mathcal{N}}, \tilde{\mathcal{H}}_{0}\} \neq 0 \, , \quad \{\tilde{\mathcal{H}}_{0}(x), \tilde{\mathcal{H}}_{0}(y)\} \neq 0 \, \quad  \text{(on-shell)} \, ,
\end{equation}
that is $\Pi_\mathcal{N}$, $\tilde{\mathcal{H}}_0$ are \textit{second-class}.\footnote{They are purely second class only in asymptotically flat spacetimes. One has to be more careful in the general case~\cite{Donnelly2011}.} 

To yield the correct classical dynamics preserving the second-class constraints, the corresponding Lagrange multipliers have to be particular functions of the metric and the lapse  \cite{Donnelly2011}. At the quantum level, we can sidestep this part of the procedure by enforcing the constraints to hold \textit{at all times} with $\delta$-functions.
Let us, for a moment, assume that we gauge-fixed the spatial diffeomorphisms and thus solved the usual issue with the first-class constraints, i.e. gauge symmetries (we will come back to the gauge fixing in Sec.~\ref{subsec: BG quantization}). This way, we can normally define the quantum partition function and apply the standard procedure  \cite{Senjanovic1976} for treating second-class systems (for a textbook treatment, see also \cite{Henneaux1994} and \cite{Rothe2010}). The idea, when applied to the non-projectable model \cite{Bellorin2022, Bellorin2024}, is to start with the phase space Hamiltonian action and define the quantum partition function with the second-class constraints put in by hand:
\begin{equation}
\label{eq:pathintegral}
    Z = \int D \Gamma D \Pi_\mathcal{N} \,  \delta(\Pi_\mathcal{N}) \delta(\tilde{\mathcal{H}}_{0}) \det \{\Pi_{\mathcal{N}}, \tilde{\mathcal{H}}_{0}\} \, \exp\bigg\{i \int dtd^d x   (\Pi^{ij} \dot{\gamma}_{ij} - \mathcal{H})\bigg\} \, .
\end{equation}
Here $\det \{\Pi_{\mathcal{N}}, \tilde{\mathcal{H}}_{0}\}$ is the Jacobian,
\begin{equation}
    \det \{\Pi_{\mathcal{N}}, \tilde{\mathcal{H}}_{0}\} = \det \begin{pmatrix}
    \frac{\delta \tilde{\mathcal{H}}_0}{\delta \mathcal{N}} & \frac{\delta \Pi_\mathcal{N}}{\delta \mathcal{N}} \\
    \frac{\delta \tilde{\mathcal{H}}_0}{\delta \Pi_\mathcal{N}} & \frac{\delta \Pi_\mathcal{N}}{\delta \Pi_\mathcal{N}}
    \end{pmatrix} = \det \bigg(\frac{\delta \tilde{\mathcal{H}}_0}{\delta \mathcal{N}}\bigg)\, ,
\end{equation}
of the measure transformation from the constraint hypersurface into the ambient phase space.

The Hamiltonian density $\mathcal{H}$ is given by \eqref{eq: Canonical H}, 
\begin{equation}
\label{eq: Det M}
    \mathcal{H} = \mathcal{N} \mathcal{H}_{0} + N^{i} \mathcal{H}_i \, ,
\end{equation}
and the phase space measure $D \Gamma$ is defined as follows,\footnote{We do not integrate over 
$\Pi_i$, the momenta conjugate to the shift vector, since they are constrained to vanish.}
\begin{equation}
    D \Gamma \equiv  D \Pi^{ij} D \gamma_{ij} D N^{i} D \mathcal{N}  \, .
\end{equation}
It is convenient to integrate over $\Pi_{\cal N}$ using the $\delta$-function and 
exponentiate the constraint $\tilde{\mathcal{H}}_{0}$ using a Lagrange multiplier which we denote by ${\cal A}$ 
\begin{equation}
\label{eq:qaction}
    Z = \int D \Gamma D \mathcal{A} \, \det \bigg(\frac{\delta \tilde{\mathcal{H}}_0}{\delta \mathcal{N}}\bigg) \, \exp\bigg\{i \int dt d^d x(\Pi^{ij} \dot{\gamma}_{ij} - \mathcal{N}\mathcal{H}_0 - N^{i} \mathcal{H}_i  + \mathcal{A} \tilde{\mathcal{H}}_{0})\bigg\}\, .
\end{equation}
We further introduce a pair of Grassmann variables $(\bar \eta ,\, \eta)$ \cite{Bellorin2022, Bellorin2024} to resolve the determinant \cite{Senjanovic1976}, just like it is done in the Faddeev--Popov method,
\begin{equation} 
      \det \bigg(\frac{\delta \tilde{\mathcal{H}}_0}{\delta \mathcal{N}}\bigg)= \int D \bar \eta D \eta  \, \exp \left\{ i\int dt d^dx \,  \bar \eta \frac{\delta \tilde{ \mathcal H}_0}{\delta \mathcal{N}} \eta\right\}\, .
\end{equation}
This yields,
\begin{equation} \label{eq:Z_A}
    Z = \int D \Gamma    D \mathcal{A}D \bar \eta D \eta\, \exp\bigg\{i \int dt d^d x \left( \Pi^{ij} \dot{\gamma}_{ij} - \mathcal{N} \mathcal{H}_0 - N^{i} \mathcal{H}_i  + \mathcal{A} \tilde{\mathcal{H}}_{0} + \bar \eta \frac{\delta \tilde{ \mathcal H}_0}{\delta \mathcal{N}} \eta \right)\bigg\} \, ,
\end{equation}
which we take as the definition of the partition function for the quantum theory.

\subsection{Back to the Lagrangian: new symmetries} \label{Subsec:susy}

The explicit form of $\mathcal{H}_0, \tilde{\mathcal{H}}_{0}$ and $\mathcal{H}_i$ entering Eq.~(\ref{eq:Z_A}) is given by \eqref{eq: Secondary Constraints}. Importantly, the dependence of the action on the momenta $\Pi_{ij}$ is Gaussian. In fact, all dependence is encoded in the \qu{kinetic part}
\begin{align} \label{eq:action_Pi}
    S_{\rm Kin} = \int dt d^dx \, \Biggl( 2  \mathcal{N} \Pi^{ij} {\cal K}_{ij} - \frac{2G}{\sqrt{\gamma}}&(\mathcal{N} - \mathcal{A}) \left(\Pi_{ij}\Pi^{ij}+\frac{\lambda}{1-\lambda d}\Pi^2 \right)  \Biggr) \, .
\end{align}
The functional integral over $\Pi_{ij}$ can thus be computed explicitly as a saddle point. That procedure leads to the relation
\begin{align}
    \Pi_{ij}= \frac{\sqrt{\gamma}}{2G} \frac{\mathcal{N}}{\mathcal{N}- \mathcal A} ({\cal K}_{ij} - \lambda {\cal K}\gamma_{ij}) \, ,
\end{align}
which can be plugged into \eqref{eq:action_Pi} to get a \qu{modified} kinetic term
\begin{align} \label{eq:S_Kin}
    S_{\rm Kin} = \frac{1}{2 G}\int dt d^d x \sqrt{\gamma}  \, \Bigg( \frac{\mathcal{N}^2}{\mathcal{N}-\mathcal{A}} \left({\cal K}_{ij}^2 - \lambda {\cal K}^2 \right)   \Bigg) \, .
\end{align}
Somewhat surprisingly, this differs from the kinetic term we started with, cf. Eq.~(\ref{eq:ActionFull}). However, we can bring it back to the canonical form of (\ref{eq:ActionFull}) by redefining the lapse,
\begin{align} \label{eq: Field Redef N}
    N = \mathcal{N} - \mathcal{A} \,.
\end{align}
Correspondingly, the extrinsic curvature gets rescaled as,
\begin{align} \label{eq: Field Redef K}
    K_{ij} = \frac{\cal N}{N} \mathcal K_{ij} \,.
\end{align}

To write down the expression for the potential part of the action in a compact way, it is convenient to adopt the \textit{DeWitt notations} \cite{Kiefer:2025udf}. We collect all labels of a field, including its coordinate dependence and possible space-time indices, into a combined index denoted with a capital Latin letter from the middle of the alphabet: $\phi_\alpha({\bf x},t)\mapsto \phi^I$. Given 
two fields $\phi_\alpha ({\mathbf x},t)$ and $\psi_\alpha(\mathbf{x},t)$ with the same tensor structure, we define the following product
\begin{align}
   \phi_I \psi^I \equiv \int dt d^d x \, \sqrt{\gamma} \sum_\alpha \phi_\alpha(\mathbf{x},t) \psi^\alpha(\mathbf{x},t) \, ,
\end{align}
which places the summation over the tensorial and continuous indices on the same footing. 

With these notations, the potential term $S_\mathcal{V}$ of the action $S=S_{\rm Kin} + S_\mathcal{V}$ is expressed as,
\begin{align} \label{eq:S_V}
   S_\mathcal{V}= \frac{1}{2G}\left[-  (\mathcal{N}_I - \mathcal{A}_I)  \, \mathcal{V}^I  +  \mathcal{N}_I \, \frac{\delta \mathcal{V}^I}{\delta \mathcal{N}_J} \mathcal{A}_J + \bar \eta_I \frac{\delta \mathcal{V}^I}{\delta \mathcal{N}_J} \eta_J + \bar\eta_I \frac{\delta \mathcal{V}^J}{\delta \mathcal{N}_I} \eta_J + \mathcal{N}_L \bar \eta_I \frac{\delta^2 \mathcal V^L}{\delta \mathcal{N}_I \delta \mathcal{N}_J} \eta_J \right] \, .
\end{align}
Upon the redefinition (\ref{eq: Field Redef N}), this becomes,
\begin{align} \label{eq:S_V_2}
   S_\mathcal{V}= \frac{1}{2G}\left[-  N_I   \, \tilde{ \mathcal{V}}^I  +  (N_I + \mathcal A_I)\, \frac{\delta \tilde{ \mathcal{V}}^I}{\delta N_J} \mathcal{A}_J + \bar \eta_I \frac{\delta \tilde{ \mathcal{V}}^I}{\delta N_J} \eta_J + \bar\eta_I \frac{\delta \tilde{\mathcal{V}}^J}{\delta N_I} \eta_J + (N_L+ \mathcal A_L) \bar \eta_I \frac{\delta^2 \tilde{ \mathcal{V}}^L}{\delta N_I \delta N_J} \eta_J \right] \, ,
\end{align}
with $\tilde{ \mathcal{V}}\equiv \mathcal V[N+ \mathcal A]$ representing the potential with its appropriate functional dependence after the shift \eqref{eq: Field Redef N}. 

The theory possesses a symmetry under \textit{linear} \textit{nilpotent} transformation $\q$ which maps between the bosonic and fermionic fields and is defined as
\begin{equation} \label{eq:delta_1}
    \q \mathcal A= \eta \,, \qquad \q \mathcal {\bar \eta} =N \,, \qquad \q N= \q \eta =0 \, .
\end{equation}
or, equivalently, as
\begin{equation} \label{eq:delta_2}
    \q = \eta \frac{\delta}{\delta \mathcal A} + N \frac{\delta}{\delta \bar \eta}\,. 
\end{equation}
Let us prove this statement. 
After the field redefinition \eqref{eq: Field Redef N}, the kinetic part of the action does not depend on $\mathcal A$ or $\bar \eta$ and thus is obviously invariant under \eqref{eq:delta_1}. The invariance of $S_{\mathcal V}$, instead, can be shown using the nilpotency of $\q$. To do so, we first notice that, since the potential ${\cal V}$ depends on ${\cal N}$ only through $\nabla_i\log{\cal N}$, it is invariant under a rescaling ${\cal N}\mapsto \alpha{\cal N}$, with a constant $\alpha$. This implies an identity,
\begin{equation}
\label{eq:Vscale}
    {\cal N}_J\frac{\delta{\cal V}^I}{\delta {\cal N}^J}=0\;.
\end{equation}
This, in turn, gives:
\begin{equation} \label{eq:identity}
     \frac{\delta \tilde{\mathcal V}^I}{\delta N_J} N_J= - \frac{\delta \tilde{\mathcal V}^I}{\delta N_J} \mathcal A_J \,.
\end{equation}
Using these relations, the expression (\ref{eq:S_V_2}) can be written as,
\begin{align} 
 S_{\mathcal V}=-\frac{1}{2G} \q \left[ \bar \eta_I \bigg(\tilde{\mathcal V}^I + (N_J+ \mathcal A_J) \frac{\delta \tilde{\mathcal V}^J}{\delta N_I}\bigg) \right]  \,.
\end{align}
Since $\q^2=0$, we obtain $\q S_{\mathcal V}=0$. 

Along the same lines of reasoning, we find that $S$ is \textit{also} symmetric under the action of
\begin{align} 
\label{eq:BRST2}
 \bar \q=- \bar \eta \frac{\delta}{\delta \mathcal A}+N \frac {\delta}{\delta \eta}  \,,
\end{align}
which anticommutes with $\q$
\begin{align} \label{eq:anticummutator_q}
 \{\bar \q, \, \q \}=0  \,.
\end{align}
One can check that $S_{\mathcal V}$, being both $\q$-exact and $\bar \q$-exact, can be written as
\begin{align} 
\label{eq: Sv}
S_{\mathcal V}= \q \bar \q \Xi \,, \qquad \Xi=(N_I+\mathcal A_I) \mathcal V^I[N+\mathcal A]  \,.
\end{align}
The symmetries generated by $\q$, $\bar \q$ resemble the Becchi--Rouet--Stora--Tyutin (BRST) transformations arising in the implementation of the first class constraints through the Faddeev--Popov procedure and playing a key role in the renormalization of gauge theories. An important difference in our case, however, is that these symmetries are spontaneously broken: since $N=1$ in the vacuum, the transformations of the Grassmann fields 
$\bar\eta$ or $\eta$ do not vanish even if these fields themselves are zero. 
Consequently, the implications of these symmetries are more subtle than in the standard BRST procedure. In particular,   
the potential term $S_{\cal V}$ is not trivial, despite being $\q$- and $\bar\q$-exact.

Quite remarkably, on top of the invariance under $\q$ and $\bar \q$, the theory possesses additional fermionic and bosonic symmetries linked to the property (\ref{eq:Vscale}). A straightforward calculation shows that the potential term, and hence the whole action, is invariant under the transformations:
\begin{subequations}
\label{eq:Qhat}
\begin{align}
\label{eq:Qhat1}
\hat{\q}\bar\eta=N+{\cal A}\;,\qquad \hat{\q}{\cal A}=\hat{\q}\eta=\hat{\q}N=0\;,\\
\label{eq:Qhat2}
\bar{\hat{\q}}\eta=N+{\cal A}\;,\qquad \bar{\hat{\q}}{\cal A}=\bar{\hat{\q}}\bar\eta=\bar{\hat{\q}}N=0\;.
\end{align}   
\end{subequations}
Clearly, both these transformations are nilpotent and anti-commute,
\begin{equation}
\label{eq:Qhatcomm}    
\{\hat{\q},\bar{\hat{\q}}\}=0\;.
\end{equation}
They, however, do not anti-commute with $\q$ and $\bar\q$:\footnote{This property implies that the unbroken combinations of the symmetries (\ref{eq:delta_1}), (\ref{eq:BRST2}), (\ref{eq:Qhat}) generated by $(\q-\hat{\q})$ and $(\bar{\q}-\bar{\hat{\q}})$ are {\it not} nilpotent.}
\begin{subequations}
\label{eq:QQhat}
\begin{align}
\label{eq:QQhat1}
&\{\hat{\q},\q \}={\cal C}\;,&&\{\hat{\q},\bar{\q}\}=-\bar{\cal R} \\
\label{eq:QQhat2}
&\{\bar{\hat{\q}},\q \}={\cal R} \;,&& \{ \bar{\hat{\q}},\bar{\q}\}=-\bar{\cal C}\;,
\end{align}    
\end{subequations}
where ${\cal C}$, ${\cal R}$, etc. are new bosonic symmetries
\begin{subequations}
\label{eq:newbos}
\begin{align}
\label{eq:newbos1}
 &{\cal C}\bar\eta=\eta\;,\qquad {\cal C}{\cal A}={\cal C}\eta={\cal C}N=0\;,\\
 \label{eq:newbos2}
 &\bar{\cal C}\eta=\bar\eta\;,\qquad \bar{\cal C}{\cal A}=\bar{\cal C}\bar\eta=\bar{\cal C}N=0\;,\\
 \label{eq:newbos3}
  &{\cal R}{\cal A}=N+{\cal A}\;,\qquad {\cal R}\eta=\eta\;,\qquad{\cal R}\bar\eta={\cal R} N=0\;,\\
  \label{eq:newbos4}
   &\bar{\cal R}{\cal A}=N+{\cal A}\;,\qquad \bar{\cal R}\bar\eta=\bar\eta\;,\qquad\bar {\cal R}\eta=\bar{\cal R} N=0\;.
\end{align}
\end{subequations}
Since the operator acting on $\eta$ and $\bar\eta$ in (\ref{eq:S_V_2}) is symmetric, the invariance of the action with respect to the transformations ${\cal C}$ and $\bar{\cal C}$ is manifest. 

To make the 
invariance with respect to ${\cal R}$ and $\bar{\cal R}$ also manifest, it is
convenient to write the potential part of the action in terms of 
$N$ and ${\cal N}$,   
\begin{align} \label{eq:SV_A_N} 
 S_{\mathcal V}=\frac{1}{2G}\left[-N_I{\cal V}^I-{\cal N}_I\frac{\delta {\cal V}^I}{\delta {\cal N}_J}N_J
 +\bar\eta_I\frac{\delta {\cal V}^I}{\delta {\cal N}_J}\eta_J 
 +\bar\eta_I\frac{\delta {\cal V}^J}{\delta {\cal N}_I}\eta_J 
 +{\cal N}_L\bar\eta_I\frac{\delta^2 {\cal V}^L}{\delta{\cal N}_I\delta {\cal N}_J}\eta_J 
 \right].
\end{align}
Note that in these variables, the dependence of the potential on $N$ is purely linear. 
Since ${\cal V}$ depends on ${\cal N}$ only through $\nabla_i\log {\cal N}$, the potential is invariant with respect to a point-wise transformation $\mathcal R_\epsilon$
\begin{align}
    \mathcal{R}_\epsilon {\cal N} =\epsilon(t) {\cal N} \, , \qquad \mathcal{R}_\epsilon \eta =\epsilon(t) \eta \, ,\qquad \mathcal{R}_\epsilon \bar\eta = \mathcal{R}_\epsilon N=0\,,
\end{align}
with arbitrary constant in space $\epsilon(t)$. The transformation (\ref{eq:newbos3}) is just special case corresponding to a constant transformation function $\epsilon(t)=1$. 
Note that, unlike the FDiff transformation (\ref{FDiffsFields1}), we do not reparameterize the time coordinate, nor do we touch the metric or the ``straight'' lapse function $N$. This implies the invariance of the kinetic term, and hence the whole action.
Similarly, the action is invariant under the \qu{barred} transformation $\bar{\mathcal R}_{\bar\epsilon}$: 
\begin{align}
    \bar{\mathcal R}_{\bar \epsilon} {\cal N} =\bar \epsilon(t) {\cal N} \, , \qquad \bar{\mathcal R}_{\bar\epsilon} \bar\eta =\bar\epsilon(t) \bar\eta \, ,
 \qquad \bar{\mathcal{R}}_{\bar\epsilon} \eta 
   = \bar{\mathcal{R}}_{\bar\epsilon} N=0\,,
\end{align}
for some arbitrary function $\bar \epsilon (t)$, with the special case $\bar\epsilon(t)=1$ giving (\ref{eq:newbos4}). 

While $\mathcal R_\epsilon$ and $\bar{\mathcal R}_{\bar \epsilon} $ are two independent transformations, their difference gives yet one more symmetry involving only the pair of Grassmann variables $\eta$ and $\bar \eta$: 
\begin{equation}
    \big({\cal R}_\epsilon-\bar{\cal R}_\epsilon\big)\eta=\epsilon (t) \eta \,, \qquad \big({\cal R}_\epsilon-\bar{\cal R}_\epsilon\big)\bar{\eta} =-\epsilon(t) \bar{\eta} \,,
\end{equation}
which follows from the invariance of the action under phase rotations of the Grassmann variables and the fact that they enter into the action without any time derivatives. It is straightforward to show that the symmetries ${\cal R}_\epsilon$, $\bar{\cal R}_{\bar \epsilon}$ commute with each other
and form the following algebra with the rest of the transformations: 
\begin{subequations}
\begin{align}
   & [{\cal R}_\epsilon, \q]=[\bar{\cal R}_{\bar\epsilon}, \bar\q]=0\;,&&
    [{\cal R}_\epsilon, \bar\q]=-\epsilon(t) \bar\q\;,&&
    [\bar{\cal R}_{\bar\epsilon}, \q]=-\bar\epsilon(t) \q\;,\\
   & [{\cal R}_\epsilon, \bar{\hat{\q}}]=[\bar{\cal R}_{\bar\epsilon}, {\hat{\q}}]=0\;,&&
    [{\cal R}_\epsilon, \hat\q]=\epsilon(t) \hat{\q}\;,&&
    [\bar{\cal R}_{\bar\epsilon}, \bar{\hat{\q}}]=\bar\epsilon(t) \bar{\hat{\q}}\;,\\
    &[{\cal R}_\epsilon,{\cal C}]=-[\bar{\cal R}_\epsilon,{\cal C}] =\epsilon(t){\cal C}\;,&&
    [{\cal R}_\epsilon,\bar{\cal C}]=-[\bar{\cal R}_\epsilon,\bar{\cal C}] = -\epsilon(t)\bar{\cal C}\;.
\end{align}
\end{subequations}
Finally, the remaining commutators of the transformations ${\cal C}$ and $\bar{\cal C}$ are:
\begin{subequations}
\begin{align}
&[{\cal C},\q]=[{\cal C},\hat\q]=0\;,&& [{\cal C},\bar\q]=-\q\,&&[{\cal C},\bar{\hat\q}]=-\hat\q\;,\\
&[\bar{\cal C},\bar\q]=[\bar{\cal C},\bar{\hat\q}]=0\;,&& [\bar{\cal C},\q]=-\bar\q\,&&[\bar{\cal C},{\hat\q}]=-\bar{\hat\q}\;,\\
&[{\cal C},\bar{\cal C}]={\cal R}-\bar{\cal R}\;.
\end{align}
\end{subequations}

The symmetries discussed above have been
shown at the level of the action $S$, before the gauge fixing of spatial diffeomorphisms. As we will see in Sec.\ref{subsec: BG quantization}, the gauge-fixing and the ghost sectors of the action do not depend on the fields $\mathcal A, \,\eta$ and $\bar \eta$, thus transforming trivially under $\q$, $\hat\q$, ${\cal C}$, ${\cal R}$ and their barred versions. This implies that the symmetries continue to hold upon the gauge fixing.

These symmetries, together with FDiffs, put powerful restrictions on the form of the action, but are not sufficient to completely fix it. Indeed, they are compatible with adding to the potential (\ref{eq:SV_A_N}) terms with an arbitrary dependence on the vector\footnote{Not to be confused with $\nabla_i\log{\cal N}$, which is already present in (\ref{eq:SV_A_N}).} 
$\nabla_i\log N$. 
In other words, we can add to the action another potential term like $N \mathcal{W}[\nabla_i\log N]$, where $\mathcal W$ has the same functional dependence as $\mathcal V$, but with different coupling constants. 
This observation is relevant for renormalization. If such terms arise with divergent coefficients and require counterterms, they will spoil the structure of the action which we derived from path integral implementation of the second-class constraints. In terms of the original variables ${\cal N}$ and ${\cal A}$, they would add non-linear dependence on ${\cal A}$ which then would not be anymore a simple Lagrange multiplier. This is potentially dangerous since a non-linear dependence on ${\cal A}$ is likely to 
spoil the delicate cancellation of non-local divergencies shown in \cite{Bellorin2022}.
Whether such divergences actually arise or not, is beyond the scope of this paper. We content ourselves with a remark that proving their absence remains the main unsolved obstacle on the way to a complete proof of renormalizability of the non-projectable Ho\v rava gravity.

\subsection{Second-class constraint as an ultra-local in time measure} \label{s:non_local_measure}

While we have brought the kinetic part of the action to the standard Lagrangian form (\ref{eq:ActionFull}) by the field redefinition (\ref{eq: Field Redef N}), (\ref{eq: Field Redef K}), the potential part (\ref{eq:S_V_2}) still looks very different from that in Eq.~(\ref{eq:ActionFull}). We now argue that, in accord with the general result of \cite{HenneauxSlavnov1994}, the fields ${\cal A}$, $\eta$ and $\bar\eta$ can be integrated out leaving behind the original Lagrangian action and an ultra-local in time measure depending on the metric $\gamma_{ij}$ and the lapse $N$. 

To this end, we Taylor expand all terms in (\ref{eq:S_V_2}) in powers of the field ${\cal A}$. The zeroth-order term, which does not contain any ${\cal A}$ or Grassmann fields simply coincides with the original potential, as in (\ref{eq:ActionFull}). Importantly, the terms linear in ${\cal A}$ cancel out. The remaining contributions are quadratic or higher order in ${\cal A}$, $\eta$ and $\bar\eta$. Thus, integrating over these fields at fixed $\gamma_{ij}$ and $N$ produces a sum of multi-loop diagrams made of the $\langle {\cal A}{\cal A}\rangle$ and $\langle \bar\eta\eta\rangle$ propagators and various vertices involving ${\cal A}$, $\eta$, $\bar\eta$ in external metric and lapse background. Since the potential part (\ref{eq:S_V_2}) does not contain any time derivatives, the $\langle {\cal A}{\cal A}\rangle$ and $\langle \bar\eta\eta\rangle$ propagators are proportional to the $\delta$-function of time, and hence the loop diagrams made of them depend on the metric and lapse on a single time slice. Overall, these diagrams are proportional to $\delta^{(1)}(0)$, i.e. they are ultra-local in time. On the other hand, their spatial dependence is non-local, since the $\langle {\cal A}{\cal A}\rangle$ and $\langle \bar\eta\eta\rangle$ propagators have non-trivial momentum dependence (see Eq.~(\ref{Aetaprops}) from Appendix~\ref{app:A} for the case $d=2$). 
Thus, upon integrating out ${\cal A}$ and the Grassmann variables, the partition function takes the form,  
\begin{equation} \label{eq:Z_mu}
    Z = \int D\gamma_{ij} D N^{i} D N \, \mu[N;\gamma_{ij}] \, e^{i S} \, , 
\end{equation}
where $S$ is the action \eqref{eq:ActionFull} we started with, before implementing the constraints, and $\mu[N;\gamma_{ij}]$ is an ultra-local in time, but non-local in space measure. 

Appearance of an ultra-local in time measure is a common property in Lagrangian path integral quantization of theories with second-class constraints \cite{HenneauxSlavnov1994}. In relativistic theories, this measure is often discarded, since Lorentz invariance implies that it must be also ultra-local in space, i.e. depends on fields only at a single point and is proportional to $\delta^{(d+1)}(0)$. In diagrammatic expansion, this leads to power-law divergences in the $(d+1)$-dimensional momentum, which are consistently eliminated by the usual dimensional regularization. 

The situation in Ho\v rava gravity is more subtle. It is tempting to suggest that the measure can still be neglected if we adopt dimensional regularization, now specifically in the time direction. This, however, needs to be combined with the dimensional regularization in the space directions needed to regulate the usual UV divergences, so we would be forced to consider an analog of the split dimensional regularization \cite{Leibbrandt1996} whose consistency beyond one loop requires careful analysis of the potential mixed divergences \cite{Barvinsky1984, Leibbrandt1997, Heinrich1999}. We thus prefer to keep the measure explicitly and adopt an alternative regularization in the time direction which makes the power-law divergences prominent. In practice, we work with the localized form of the measure, before integrating out ${\cal A}$, $\eta$ and $\bar\eta$ fields, i.e. we work with the potential term (\ref{eq:S_V_2}).

The non-local in space, ultra-local in time divergences from the measure $\mu$ are expected to cancel similar divergences arsing from the action $S$ due to the irregular pieces in the lapse propagator. Note, however, that the measure $\mu$ does not depend on the shift vector $N^i$, since the latter does not appear in the potential part of the action (\ref{eq:S_V_2}). Thus, at one loop, the correlators of $N^i$ will not receive any contributions from the measure, and the non-local divergences in them arising from the action must cancel among themselves. This is indeed the case, as we verify explicitly in Secs.~\ref{sec: Beta Function}, \ref{Sec:Heat_Kernel}.

\subsection{First-class constraints}
\label{subsec: BG quantization}

Having dealt with the second-class constraint, let us go back to the first-class constraints $\mathcal{H}_i$, which generate the gauge symmetry of spatial diffeomorphisms. Here we implement the background field method \cite{Abbott1982} (see also \cite{Weinberg1996}) following closely the discussion in \cite{Barvinsky2015}. The major differences compared to \cite{Barvinsky2015} will be the presence of the auxiliary fields ${\cal A}$, $\eta$, $\bar\eta$ and introduction of the background for the lapse, as well as the time-reparametrization symmetry that we choose to keep \textit{intact}. To make the presentation more transparent, we focus on the terms that contribute to the one-loop effective action and specify to the case $d=2$, which will be explicitly studied in the subsequent sections. 

We expand
\begin{equation}
        \gamma_{ij} = \bar{\gamma}_{ij} + h_{ij} \,, \qquad N_i =  \bar{N}_i + n_i \, , \qquad N = \bar N + n \, ,
\end{equation}
whereas the auxiliary fields ${\cal A}$, $\eta$, $\bar\eta$ are kept background-free.
Note that we perturb the shift with \textit{lower} indices. This is done simply for computational efficiency. We raise and lower indices with the background metric $\bar{\gamma}_{ij}$. This implies for the perturbed shift with the upper index: 
\begin{equation}
\label{eq:Shift Pert Upper}
     \gamma^{ij} N_{j} \equiv N^{i} = \bar N^{i} + \delta N^{i} \,,~~ ~~~~~~~
    \delta N^{i} =n^i - h^{i j} \bar N_{j} + h^{ik}h_{kj}\bar N^j-h^{ij}n_j+\ldots\, ,
\end{equation}
where dots stand for higher-order terms that do not contribute at one loop.

In the background field approach \cite{Abbott1982}, we construct a gauge fixing term invariant under the background field gauge transformation ${\cal D}_{\text{bg}}$ identical to the spatial diffeomorphism part of \eqref{FDiffsFields},
\begin{equation}
    \bar \gamma_{ij} \xrightarrow{{\cal D}_{\text{bg}}} \bar \gamma_{kl} \frac{\partial \tilde{x}^k}{\partial x^{i}} \frac{\partial \tilde{x}^l}{\partial x^{j}} \, ,~~~~~~~
    \bar N^i \xrightarrow{{\cal D}_{\text{bg}}} \Big(\bar N^{j} \frac{\partial x^{i}}{\partial \tilde{x}^{j}} - \frac{\partial x^i}{\partial \tilde{t}}\Big) \, .
\end{equation}
Since we want to keep the time-reparameterization symmetry ${\cal T}$ unfixed we also require covariance,
\begin{equation}
     \bar N^i \xrightarrow{\cal T} \bar N^{i} \frac{d \tilde{t}}{d t} \, ,~~~~~~~ \bar{N} \xrightarrow{\cal T} \bar N\frac{d\tilde{t}}{dt} \, .
\end{equation}
This implies that $\delta N^i$ from \eqref{eq:Shift Pert Upper} transforms non-trivially \textit{both} under ${\cal D}_{\text{bg}}$ and under time reparametrizations:
\begin{equation}
    \delta N^i \xrightarrow{{\cal D}_{\text{bg}}} \frac{\partial \tilde x^{i}} {\partial x^{j}} \delta N^i \, ,~~~~~~~
     \delta N^i  \xrightarrow{\cal T} \frac{d \tilde{t}}{d t} \delta N^i  \, .
\end{equation}
To construct an invariant gauge fixing term, we will need a time-reparameterization scalar 
    \begin{equation}
\label{ShiftPertScalar}
    \delta \hat{N}^i \equiv\frac{\delta N^i}{\bar{N}} \, ,
\end{equation}
and its covariant time derivative
\begin{equation}
\label{CovDerbg}
    \bar D_t \delta \hat{N}^i \equiv \frac{\delta \dot{\hat{N}}^i - \bar N^{k} \bar\nabla_k  \delta \hat{N}^i +  \delta \hat{N}^k \bar\nabla_k \bar N^i}{\bar{N}} \, .
\end{equation}
Here bar on the covariant derivatives means that they are defined using the background metric $\bar\gamma_{ij}$.

Next we introduce a pair of Faddeev--Popov ghosts $c^i$, $\bar c_i$ and the Nakanishi--Lautrup field $b_i$, which we assume to be scalars under the time-reparameterizations. We also introduce the Slavnov operator $\mathbf{s}$ defining the BRST transformations of all field fluctuations, 
\begin{subequations}
\label{BRSTFields}
\begin{gather}
\label{BRSThN}
    \mathbf{s} h_{ij} = \bar{\gamma}_{jk}\bar{\nabla}_{i} c^{k} + \bar{\gamma}_{ik}\bar{\nabla}_{j} c^{k} 
+ \bar{\nabla}_{i}c^{k} h_{j k} + \bar{\nabla}_{j}c^{k} h_{i k} +
c^{k} \bar{\nabla}_{k} h_{ij} \, , \\
 \mathbf{s} \delta N^{i}  = \bar{N}\bar{D}_t c^{i} - \delta N^{j} \bar{\nabla}_{j}c^{i} +
    c^{j}\bar{\nabla}_{j}\delta N^i \, ,\\
    \mathbf{s} n = c^i \bar{\nabla}_i n \,, ~~~~
    \mathbf{s} {\cal A} = c^i \bar{\nabla}_i {\cal A} \,, ~~~~
    \mathbf{s} \eta = c^i \bar{\nabla}_i \eta \,, ~~~~
    \mathbf{s} \bar\eta = c^i \bar{\nabla}_i \bar\eta \,,
    \end{gather}
\end{subequations}
as well as of the ghosts and the Nakanishi--Lautrup field,
\begin{equation}
\label{BRSTghosts}
 \mathbf{s}c^{i} = c^{j} \bar{\nabla}_{j} c^{i} \, , \quad  
\mathbf{s} \bar{c}_{i} = b_{i} \, , \quad \mathbf{s}  b_{i} =0 \, .
\end{equation}
Assuming that $\mathbf{s}$ acts trivially on the background fields, it is straightforward to show that 
$\mathbf{s}$ is nilpotent, $\mathbf{s}^2 = 0$. The transformations (\ref{BRSTFields}) are nothing but an infinitesimal version of FDiffs, with the coordinate increment replaced by the ghost field $c^i$. Thus they leave the action $S$ invariant. 

The gauge-fixed quantum action in $d=2$ can now be written as,
\begin{equation}
    S_{q} = S + \frac{1}{2 G} \int d td^2 x  \, \mathbf{s} \Psi \, ,
\end{equation}
where the \textit{gauge-fixing fermion} $\Psi$ is a covariant (with respect to time reparameterizations) version of the one used in \cite{Barvinsky2017AssFreed},
\begin{equation}
\label{eq: Gauge Fixing Fermion}
    \Psi = \sqrt{\bar \gamma} \bar N \left( 2 \bar{c}_i F^{i} - \frac{1}{\sigma} \bar{c}_{i} \mathcal{O}^{ij} b_{j} \right) \, .
\end{equation}
Here\footnote{Note an opposite sign of the operator ${\cal O}_{ij}$ compared to \cite{Barvinsky2015} (where it was denoted by ${\cal O}_{ij}^{-1}$). This difference arises because we work in real time, whereas Ref.~\cite{Barvinsky2015} performed a Wick rotation to the ``Euclidean'' time.}
\begin{equation}
\label{GaugeFixingFunction}
       F^i=  \bar{D}_t \delta \hat{N}^i + \frac{1}{2 \sigma} \mathcal{O}^{ij} ( \bar{\nabla}_k h^{k}_{j} - \lambda \bar{\nabla}_j h) \,, \qquad \mathcal{O}^{ij}= \bar{\gamma}^{ij} \bar{\Delta} + \xi \bar{\nabla}^i \bar{\nabla}^j \, ,
\end{equation}
and $\sigma, \xi$ are constant gauge-fixing parameters.
The choice of $\Psi$ is motivated by computational convenience: it leads to cancellation of terms mixing the metric and the shift perturbations in the quadratic action.

Integrating out the Nakanishi--Lautrup field $b_i$, we get the \textit{quantum} Lagrangian extended by the gauge-fixing piece plus the ghost contribution:
\begin{equation}
\label{eq: Quantum Lagrangian}
    \mathcal{L}_q =\mathcal{L} + \mathcal{L}_{\text{gf}} + \mathcal{L}_{\text{gh}} \,,
    ~~~~~~\mathcal{L}_{\text{gf}}=  \frac{\sigma}{2 G}   \,  F^i (\mathcal{O}^{-1})_{ij} F^j \,, ~~~~~~
    {\cal L}_{\rm gh}=-\frac{1}{G} \bar c_i \,\mathbf{s}F^i\,.
\end{equation}
In the next section, we illustrate the above formalism in an explicit computation of the divergences of the one-loop effective action for $\bar{N}_i$. 
Explicit form of the gauge-fixing and ghost Lagrangians in the background relevant for this calculation is given in Appendix~\ref{s:gauge fixing}.

%% file: Sections/Section3.tex
\section{One-loop effective action from Feynman diagrams}
\label{sec: Beta Function}
We take $\bar{N}_i$ to be the function of $\mathbf{x}$ only \cite{Barvinsky2019},
\begin{align} \label{eq:background_shift}
    \bar{N}_i=\bar{N}_i(\mathbf{x}) \, ,
\end{align}
and set the background for the metric and the lapse to the simplest possible one, namely
\begin{equation} \label{eq:background_gamma_N}
    \bar{\gamma}_{ij} = \delta_{ij} \, , \quad \bar{N} = 1 \, .
\end{equation}
In what follows, we will not distinguish the lower and upper indices and will use only the lower ones. The summation with the flat background metric is assumed.

The background (\ref{eq:background_shift}), (\ref{eq:background_gamma_N}) is sufficiently general to capture the local $\bar N_i$-dependent divergences in the effective action, which one expects by power counting to reproduce the form of the bare action,
\begin{equation} \label{eq:S_Ni}
    S_{\bar N_i} = \frac{1}{2 G}\int dt d^2x \left( \frac{1}{2} \partial_i \bar{N}_j \partial_i \bar{N}_j + \left( \frac{1}{2}- \lambda \right) \partial_i \bar{N}_i \partial_j \bar{N}_j  \right) \, .
\end{equation}
As discussed above, there can also be IFDs due to irregular terms in the lapse propagator. We will indeed see such divergences arise at the intermediate steps of the calculation, but eventually they will cancel. The general proof of the cancellation of IFDs at one loop has been provided in \cite{Bellorin2022} within the Hamiltonian path integral approach. Here we focus on explicitly verifying the cancellation within the subset of terms quadratic in the shift field $\bar N_i$.

Below, we derive the Feynman rules and compute the one-loop correction to the two terms in (\ref{eq:S_Ni}) by calculating the two-point function of $\bar{N}_i$ in $d=2$. 
Our computation follows closely the one performed in \cite{Barvinsky2019}. Since the measure $\mu[N;\gamma_{ij}]$ in (\ref{eq:Z_mu}) does not depend on $\bar N_i$, it does not contribute into the effective action (\ref{eq:S_Ni}). Equivalently, the perturbations of the fields   
${\cal A}$ and $\eta$, $\bar\eta$ decouple from $\bar N_i$, so we omit them in most of the calculations. They are reinstated only in Sec.~\ref{subsec: Irreg Diag} where a coupling between them and the shift is introduced by the regulator needed to isolate IFDs.
We use symbolic computer algebra provided by Mathematica \cite{Mathematica} and various xAct packages \cite{xAct, xPerm, xPert, xTras, Fieldsx} to manipulate the formulas. 
The notebook with calculations is available at~\cite{github}.

\subsection{Feynman rules} \label{s:diag_approach}

\subsubsection{Propagators}
To compute the propagators, we expand the Lagrangian around the background up to second order in the quantum fluctuations and look at the $\bar{N}_i$-independent part: 
\begin{equation}\label{eq:Lqflat}
\begin{split}
    \mathcal{L}^{(2)}_{\rm flat}=& \frac{1}{8G}( \dot{h}_{ij} \dot{h}_{ij} - \lambda \dot{h}^2) + \dot{n}_i \pi_i  +\frac{1}{G}\dot{\bar{c}}_i \dot{c}_i  -  \frac{1}{2G} \left( \mu_1+\frac{\lambda^2(1+\xi)}{4 \sigma} \right) h \Delta^2 h \\
    & + \frac{1}{2G} \left(2 \mu_1+\frac{\lambda(1+\xi)}{2 \sigma} \right) h_{ij} \partial_{ij} \Delta h  - \frac{1}{8G \sigma}h_{ij} \partial_{jk} \Delta h_{ik} - \frac{1}{2 G} \left( \mu_1 + \frac{\xi}{4 \sigma} \right) h_{ij} \partial_{ijkl}h_{kl}\\
    &- \frac{\mu_{67}}{2G} n \Delta^2 n + \frac{\mu_5}{2G} n( \Delta^2 h - \Delta \partial_{ij} h_{ij} ) - \frac{1}{4G} n_i \Delta n_i  - \frac{1-2 \lambda}{4 G} n_i \partial_{ij} n_j\\
    &- \frac{G}{2 \sigma} \pi_i \Delta \pi_i - \frac{G\xi}{2 \sigma} \pi_i \partial_{ij} \pi_j - \frac{1}{2 G \sigma } \bar{c}_i \Delta^2 c_{i} - \frac{1}{2 G \sigma} \Big(1 - 2 \lambda + 2 \xi (1-\lambda)\Big) \bar{c}_i  \Delta \partial_{ij} c_j  \,.
\end{split}
\end{equation}
Here $\Delta=\delta_{ij}\partial_i\partial_j$ is the Laplace operator in flat space, the object $\partial_{i_1 i_2  \cdots i_n}$ stands for $\partial_{i_1} \partial_{i_2} \cdots \partial_{i_n}$, and
\be
\mu_{67} \equiv \mu_6 + \mu_7 \,. \label{eq:mu67}
\ee
Due to our choice of the gauge-fixing function, the terms mixing $n_i$ and $h_{ij}$ have canceled out between (\ref{eq:Lgf}) and the original Lagrangian (\ref{eq:ActionFull}), leading to  
three kinematically decoupled field sectors: $\{h_{ij}, n\}$, $\{n_i, \pi_j \}$ and $\{\bar c_i, c_j \}$. We find the propagators by going to the Fourier space $(\partial_t, \partial_i) \to (-i \omega, i q_i)$ and inverting the differential operator matrix in the quadratic Lagrangian \eqref{eq:Lqflat} sector by sector. 

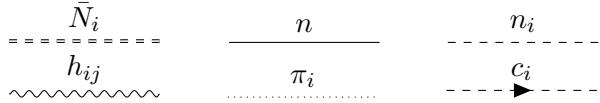
\begin{figure}
\begin{center}
\begin{tikzpicture}
  \begin{feynman}
    \vertex (i) at (-1, 0) ;
    \vertex (f) at (1, 0) ;
    \diagram* {
      (i) -- [double, dashed, edge label=$\bar N_i$] (f),
    };
  \end{feynman}
\end{tikzpicture} \qquad
\begin{tikzpicture}
  \begin{feynman}
    \vertex (i) at (-1, 0) ;
    \vertex (f) at (1, 0) ;
    \diagram* {
      (i) -- [ edge label=$ n$] (f),
    };
  \end{feynman}
\end{tikzpicture} \qquad
\begin{tikzpicture}
  \begin{feynman}
    \vertex (i) at (-1, 0) ;
    \vertex (f) at (1, 0) ;
    \diagram* {
      (i) -- [ dashed, edge label=$ n_i$] (f),
    };
  \end{feynman}
\end{tikzpicture} \nonumber \\
\begin{tikzpicture}
  \begin{feynman}
    \vertex (i) at (-1, 0) ;
    \vertex (f) at (1, 0) ;
    \diagram* {
      (i) -- [photon, edge label=$ h_{ij}$] (f),
    };
  \end{feynman}
\end{tikzpicture}  \qquad
\begin{tikzpicture}
  \begin{feynman}
    \vertex (i) at (-1, 0) ;
    \vertex (f) at (1, 0) ;
    \diagram* {
      (i) -- [dotted, edge label=$ \pi_i$] (f),
    };
  \end{feynman}
\end{tikzpicture} \qquad
\begin{tikzpicture}
  \begin{feynman}
    \vertex (i) at (-1, 0) ;
    \vertex (f) at (1, 0) ;
    \diagram* {
      (i) -- [fermion, dashed, edge label=$ c_i$] (f),
    };
  \end{feynman}
\end{tikzpicture} 
\end{center}\caption{Graphic notation for the fields.}
 \label{f:legs}
\end{figure}

Before writing down the result, let us introduce a few notations. There are three pole structures arising in the propagators. One of them corresponds to the physical scalar graviton mode propagated by the theory \cite{Blas2010},   
\begin{align}
\label{eq:physical mode}
    \mathcal{P}_s (\omega, q) = \big[-\omega^2 + \mu_s q^4\big]^{-1} \, ,
\end{align}
where 
\begin{equation}
\mu_s \equiv\frac{1-\lambda}{1-2\lambda} \left(4 \mu_1 - \frac{\mu_5^2}{\mu_{67}}\right) \,. \label{eq:mus} 
\end{equation}
Note that this pole does not depend on the gauge parameters $\sigma$ and $\xi$. By contrast, the two other pole structures are associated with gauge modes and are gauge-dependent,
\begin{align}
\label{eq:gauge_modes}
   \mathcal{P}_1(\omega, q)= \bigg[-\omega^2 + \frac{1}{2 \sigma }q^4\bigg]^{-1}  \,, \quad 
   \mathcal{P}_2(\omega, q) = \bigg[-\omega^2 +  \frac{\rho}{\sigma} q^4\bigg]^{-1} \, ,
\end{align}
with 
\begin{align}
\rho \equiv (1-\lambda)(1+\xi) \label{eq:rho} \,.
\end{align}
In terms of these structures, we have for the propagators: 
\begin{itemize}
    \item Sector  $\{h_{ij}, n\}$:
\begin{subequations}
\label{eq:hnsec_prop}
\begin{align}
   &\langle nn \rangle = - 2i G \, \left[ \frac{1- \lambda}{1- 2 \lambda} \frac{\mu_5^2}{2\mu^2_{67}} \mathcal{P}_{s} + \frac{1}{2\mu_{67} q^4} \right] \, ,\label{eq:nn_prop} \\
   &\langle n h_{ij} \rangle = -2iG \,  \frac{\mu_5}{\mu_{67}} \, \mathcal{P}_s  \, \left(\frac{1-\lambda}{1-2 \lambda} \delta_{ij} - \hat{q}_i \hat{q}_j \right) \,, \label{eq:nh_prop}\\
    &\langle  h_{ij} h_{kl} \rangle =  -2 i G  \biggl[  2 \mathcal{P}_s \biggl(\frac{1-\lambda}{1-2 \lambda}\delta_{ij}\delta_{kl} - \delta_{ij} \hat{q}_k \hat{q}_l -  \delta_{kl} \hat{q}_i \hat{q}_j \biggr) + \mathcal{P}_1 Q_{ijkl}  \nonumber \\
    &\qquad\qquad\qquad\qquad
    +\biggl( \frac{2 \mathcal{P}_2}{1-\lambda} - 4 \mathcal{P}_1 + \frac{2 (1-2 \lambda)}{1- \lambda} \mathcal{P}_s \biggr) \hat{q}_i \hat{q}_j \hat{q}_k \hat{q}_l  \biggr] \, ,
    \label{eq:hh_prop}
\end{align}
\end{subequations}
where we have introduced the tensors,\footnote{Eq.~(\ref{eq:hh_prop}) has been simplified using an identity that holds for an arbitrary unit vector $\hat{q}$ in $d=2$ \cite{Barvinsky2015}:
\begin{equation}
\label{UnitMatrix}
     \delta_{ik} \delta_{jl} + \delta_{jk} \delta_{il} =  2 \delta_{ij}\delta_{kl}-2 \delta_{ij} \hat{q}_k \hat{q}_l - 2 \delta_{kl} \hat{q}_i \hat{q}_j + \delta_{ik} \hat{q}_l \hat{q}_j + \delta_{il} \hat{q}_k \hat{q}_j + \delta_{jk} \hat{q}_l \hat{q}_i + \delta_{jl} \hat{q}_k \hat{q}_i \, .
\end{equation}
This identity will not hold if we use the dimensional regularization with $d\neq 2$. The difference is insignificant at one loop as long as we are interested in regular local divergences. The situation is more subtle in the presence of IFDs and we will use exact propagators when dealing with them in Sec.~\ref{subsec: Irreg Diag}.}
\begin{equation}
\label{eq: Q_Tensor}
    Q_{ijkl}=\hat{q}_i\hat{q}_k \delta_{jl} +\hat{q}_i\hat{q}_l \delta_{jk}+\hat{q}_j\hat{q}_k \delta_{il}+\hat{q}_j\hat{q}_l \delta_{ik} \, , \qquad \hat{q}_i= \frac{q_i}{q} \, .
\end{equation}
\item The $\{n_i, \pi_j\}$ sector:
\begin{subequations}
\begin{align}
    &\langle n_i n_j \rangle = \frac{i G}{\sigma} q^2 \biggl( \mathcal{P}_1 (\delta_{ij} - \hat{q}_i \hat{q}_j) + \frac{\rho}{1-\lambda} \mathcal{P}_2 \hat{q}_i \hat{q}_j \biggr) \,, \label{eq:nini_prop} \\
   & \langle \pi_i \pi_j \rangle = \frac{i}{2 G}  q^2 \biggl( \mathcal{P}_1 (\delta_{ij} - \hat{q}_i \hat{q}_j) + 2 (1 - \lambda) \mathcal{P}_2 \hat{q}_i \hat{q}_j \biggr) \,, \label{eq:pipi_prop} \\
   & \langle n_i(\omega,q) \pi_j(-\omega,-q) \rangle = - \langle \pi_i(\omega,q) n_j(-\omega,-q) \rangle= - \omega \biggl( \mathcal{P}_1 (\delta_{ij} - \hat{q}_i \hat{q}_j) + \mathcal{P}_2 \hat{q}_i \hat{q}_j \biggr)\,.\label{eq:nipi_prop}
\end{align}
\end{subequations}
\item The ghosts:
\begin{align}\label{eq:ghost_prop}
   \langle \Bar{c}_i c_j \rangle =- \langle  c_i \Bar{c}_j \rangle = - 
i G  \biggl( \mathcal{P}_1 (\delta_{ij} - \hat{q}_i \hat{q}_j) + \mathcal{P}_2 \hat{q}_i \hat{q}_j \biggr) \, .
\end{align}
\end{itemize}
For completeness, we also give in Appendix~\ref{app:A} the propagators of the auxiliary fields $\langle{\cal A}{\cal A}\rangle$ and $\langle\bar \eta\eta\rangle$.  
All other propagators vanish identically. In the above expressions, we have suppressed the arguments of the fields, except in the case of the mixed propagator $\langle n_i \pi_j\rangle$, where one has to be careful with the overall sign. We further have dropped the overall momentum-conserving delta-function and have multiplied by $-i$, so that these expressions exactly correspond to internal lines in momentum space Feynman rules. We also introduce graphic notations for different field lines shown in Fig.~\ref{f:legs}. 
The mixed propagators will be naturally depicted with \qu{half and half} lines. 

The propagators (\ref{eq:hnsec_prop})--(\ref{eq:ghost_prop}) drastically simplify with the choice
\begin{equation}
\label{eq:physical_gauge}
\sigma= \frac{1}{2 \mu_s} \,, \quad \rho= \frac{1}{2} \,,
\end{equation}
which gives $\mathcal P_1=\mathcal P_2 = \mathcal P_s$. We call this particular gauge choice \textit{uniform gauge} in what follows. 

As mentioned earlier, there is an instantaneous mode contained in the 
lapse propagator $\langle n n \rangle$ with a pole at $q=0$.\footnote{Note that our basis of fields differs from the one employed in \cite{Bellorin2022}, where the non-locality is removed from the lapse propagator at the expense of mixing lapse with the auxiliary field ${\cal A}$, see Appendix~\ref{app:A} for details.}
This kind of pole structure still ensures the right high-momentum scaling of the propagator, but 
does not suppress frequency integrals. 
Following \cite{Anselmi:2008bq,Barvinsky2015} we call such propagators {\it irregular}. The divergence coming from the integration over frequency --- the IFD --- can have
arbitrary momentum dependence and thus be non-local in space \cite{Barvinsky2015}.
However, as shown within the Hamiltonian formulation in \cite{Bellorin2022,Bellorin2024}, and as we will demonstrate on a particular example below, a full sum of all the diagrams will lead to the cancellation of IFDs.

\subsubsection{Vertices and diagrams} \label{s:vertices}

The second ingredient is the interaction vertices. These are 
obtained from the Lagrangian \eqref{eq: Quantum Lagrangian} expanded up to the quadratic order in $\bar N_i(\bold x)$ and the field perturbations. This expansion is lengthy and is relegated to Appendix~\ref{app: InteractionLagrangian}. It gives rise to 
a set of three-legged vertices with a single external field  $\bar N_i$, shown in Fig. \ref{f:3pt}, and a set of four-legged ones with two external background fields in Fig.~\ref{f:4pt}.
The momentum-space expressions associated with these vertices are cumbersome and we do not give them explicitly. They can be found in the Mathematica notebook at \cite{github}.
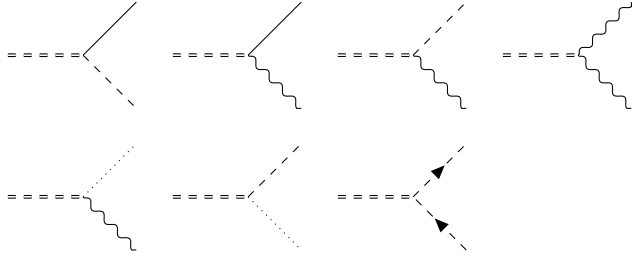
\begin{figure}[ht]
\begin{center}
 \begin{tikzpicture}
    \matrix[column sep=0.2cm, row sep=0.2cm] {
      \node {
        \begin{tikzpicture}[baseline=(a)]
  \begin{feynman}[inline=(d.base)]
  \vertex (a) ;
    \vertex [right=of a] (b) ;
    \vertex [above right=of b] (f2) ;
    \vertex [below right=of b] (f3) ;
    \diagram* {
      (a) -- [double, dashed] (b),
      (b) -- [] (f2),
      (b) -- [dashed] (f3),
    }; 
  \end{feynman}
\end{tikzpicture}
      }; &
      \node {
        \begin{tikzpicture}[baseline=(a)]
  \begin{feynman}[inline=(d.base)]
  \vertex (a) ;
    \vertex [right=of a] (b) ;
    \vertex [above right=of b] (f2) ;
    \vertex [below right=of b] (f3) ;
    \diagram* {
      (a) -- [double, dashed] (b),
      (b) -- [] (f2),
      (b) -- [photon] (f3),
    }; 
  \end{feynman}
\end{tikzpicture}
      }; &
      \node {
      \begin{tikzpicture}[baseline=(a)]
  \begin{feynman}[inline=(d.base)]
  \vertex (a) ;
    \vertex [right=of a] (b) ;
    \vertex [above right=of b] (f2) ;
    \vertex [below right=of b] (f3) ;
    \diagram* {
      (a) -- [double, dashed] (b),
      (b) -- [dashed] (f2),
      (b) -- [photon] (f3),
    }; 
  \end{feynman}
\end{tikzpicture}
      }; &
      \node {
        \begin{tikzpicture}[baseline=(a)]
  \begin{feynman}[inline=(d.base)]
  \vertex (a) ;
    \vertex [right=of a] (b) ;
    \vertex [above right=of b] (f2);
    \vertex [below right=of b] (f3) ;
    \diagram* {
      (a) -- [double, dashed] (b),
      (b) -- [photon] (f2),
      (b) -- [photon] (f3),
    }; 
  \end{feynman}
\end{tikzpicture} 
      }; \\
      \node {
        \begin{tikzpicture}[baseline=(a)]
  \begin{feynman}[inline=(d.base)]
  \vertex (a);
    \vertex [right=of a] (b) ;
    \vertex [above right=of b] (f2) ;
    \vertex [below right=of b] (f3) ;
    \diagram* {
      (a) -- [double, dashed] (b),
      (b) -- [dotted] (f2),
      (b) -- [photon] (f3),
    }; 
  \end{feynman}
\end{tikzpicture}
      }; &
      \node {
        \begin{tikzpicture}[baseline=(a)]
  \begin{feynman}[inline=(d.base)]
  \vertex (a) ;
    \vertex [right=of a] (b) ;
    \vertex [above right=of b] (f2);
    \vertex [below right=of b] (f3) ;
    \diagram* {
      (a) -- [double, dashed] (b),
      (b) -- [dashed] (f2),
      (b) -- [dotted] (f3),
    }; 
  \end{feynman}
\end{tikzpicture}
      }; &
      \node {
       \begin{tikzpicture}[baseline=(a)]
  \begin{feynman}[inline=(d.base)]
  \vertex (a) ;
    \vertex [right=of a] (b) ;
    \vertex [above right=of b] (f2) ;
    \vertex [below right=of b] (f3) ;
    \diagram* {
      (a) -- [double, dashed] (b),
      (b) -- [dashed, fermion] (f2),
      (f3) -- [dashed, fermion] (b),
    }; 
  \end{feynman}
\end{tikzpicture}
      };  \\
    };
  \end{tikzpicture}
\end{center}\caption{Interaction vertices with one background shift vector $\bar N_i$.}
 \label{f:3pt}
\end{figure}
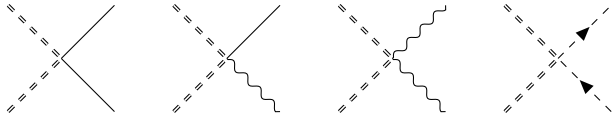
\begin{figure}[ht]
\begin{center}
  \begin{tikzpicture}
    \matrix[column sep=0.5cm] {
      \node{
        \begin{tikzpicture}
  \begin{feynman}[inline=(d.base)]
  \vertex (a);
    \vertex [above left=of a](i1) ;
    \vertex [below left=of a] (i2) ;
    \vertex [above right=of a] (f2) ;
    \vertex [below right=of a] (f3);
    \diagram* {
      (i1) -- [double,dashed] (a),
      (i2) -- [double,dashed] (a),
      (a) -- [] (f2),
      (a) -- [] (f3),
    }; 
  \end{feynman}
\end{tikzpicture}
      }; &
      \node{
        \begin{tikzpicture}
  \begin{feynman}[inline=(d.base)]
  \vertex (a);
    \vertex [above left=of a](i1) ;
    \vertex [below left=of a] (i2);
    \vertex [above right=of a] (f2) ;
    \vertex [below right=of a] (f3) ;
    \diagram* {
      (i1) -- [double,dashed] (a),
      (i2) -- [double,dashed] (a),
      (a) -- [] (f2),
      (a) -- [photon] (f3),
    }; 
  \end{feynman}
\end{tikzpicture}
      }; &
      \node{
       \begin{tikzpicture}
  \begin{feynman}[inline=(d.base)]
  \vertex (a);
    \vertex [above left=of a](i1);
    \vertex [below left=of a] (i2);
    \vertex [above right=of a] (f2);
    \vertex [below right=of a] (f3) ;
    \diagram* {
      (i1) -- [double,dashed] (a),
      (i2) -- [double,dashed] (a),
      (a) -- [photon] (f2),
      (a) -- [photon] (f3),
    }; 
  \end{feynman}
\end{tikzpicture}
      }; &
      \node{
        \begin{tikzpicture}
 \begin{feynman}[inline=(d.base)]
  \vertex (a);
    \vertex [above left=of a](i1) ;
    \vertex [below left=of a] (i2);
    \vertex [above right=of a] (f2) ;
    \vertex [below right=of a] (f3) ;
    \diagram* {
      (i1) -- [double,dashed] (a),
      (i2) -- [double,dashed] (a),
      (a) -- [dashed, fermion] (f2),
      (f3) -- [dashed,fermion] (a),
    }; 
  \end{feynman}
\end{tikzpicture} 
      }; \\
    };
  \end{tikzpicture}
\end{center}\caption{Interaction vertices with two background shift vectors $\bar N_i$.}
 \label{f:4pt}
\end{figure}

Combining vertices with propagators, we obtain  
17 divergent diagrams contributing to the two-point function of $\bar N_i$. Among them, there are 4 ``bubble'' diagrams obtained from the four-legged vertices by closing the legs for the fluctuations in a loop, see Fig.~\ref{f:bubbles}. The remaining 11 diagrams are of the ``fish'' type, being constructed with a pair of three-legged vertices, see Fig.~\ref{f:fish}. 
Only 3 diagrams involve the irregular propagator $\langle n n \rangle$. Note that there are no graphs with two irregular internal lines. This is simply because there is no vertex with one $\bar{N}_i$ and two $n$'s (see Appendix~\ref{app: InteractionLagrangian}). We will analyse the irregular diagrams separately in Sec.~\ref{subsec: Irreg Diag}. Before doing so, we discuss the regularization of local divergences.
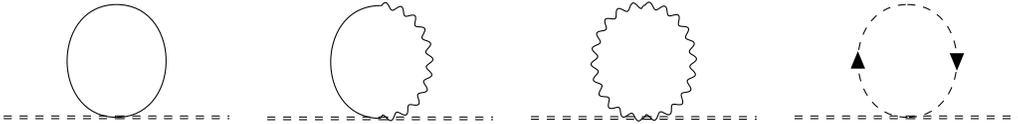
\begin{figure}[htb]
\begin{center}
  \begin{tikzpicture}
    \matrix[column sep=0.2cm] {
      \node{
        \begin{tikzpicture}[scale=0.75]
  \begin{feynman}
    \vertex (i) at (-1, 0) ;
    \vertex (a) at (0, 0);
    \vertex (m) at (1, 0);
    \vertex (u) at (1, 2);
    \vertex (b) at (2, 0);
    \vertex (f) at (3, 0) ;
    \diagram* {
      (i) -- [double, dashed] (m),
      (m) -- [double, dashed] (f),
      (m) -- [solid, half left] (u),
      (u) -- [solid, half left] (m),
    };
  \end{feynman}
\end{tikzpicture}
      }; &
      \node{
        \begin{tikzpicture}[scale=0.75]
  \begin{feynman}
    \vertex (i) at (-1, 0) ;
    \vertex (a) at (0, 0);
    \vertex (m) at (1, 0);
    \vertex (u) at (1, 2);
    \vertex (b) at (2, 0);
    \vertex (f) at (3, 0) ;
    \diagram* {
      (i) -- [double, dashed] (m),
      (m) -- [double, dashed] (f),
      (m) -- [solid, half left] (u),
      (u) -- [photon, half left] (m),
    };
  \end{feynman}
\end{tikzpicture}
      }; &
      \node{
       \begin{tikzpicture}[scale=0.75]
  \begin{feynman}
    \vertex (i) at (-1, 0) ;
    \vertex (a) at (0, 0);
    \vertex (m) at (1, 0);
    \vertex (u) at (1, 2);
    \vertex (b) at (2, 0);
    \vertex (f) at (3, 0) ;
    \diagram* {
      (i) -- [double, dashed] (m),
      (m) -- [double, dashed] (f),
      (m) -- [photon, half left] (u),
      (u) -- [photon, half left] (m),
    };
  \end{feynman}
\end{tikzpicture}
      }; &
      \node{
        \begin{tikzpicture}[scale=0.75]
  \begin{feynman}
    \vertex (i) at (-1, 0) ;
    \vertex (a) at (0, 0);
    \vertex (m) at (1, 0);
    \vertex (u) at (1, 2);
    \vertex (b) at (2, 0);
    \vertex (f) at (3, 0) ;
    \diagram* {
      (i) -- [double, dashed] (m),
      (m) -- [double, dashed] (f),
      (m) -- [dashed, fermion, half left] (u),
      (u) -- [dashed, fermion, half left] (m),
    };
  \end{feynman}
\end{tikzpicture} 
      }; \\
    };
  \end{tikzpicture}
\end{center}\caption{Bubble diagrams contributing into the two-point function of $\bar N_i$.}
 \label{f:bubbles}
\end{figure}
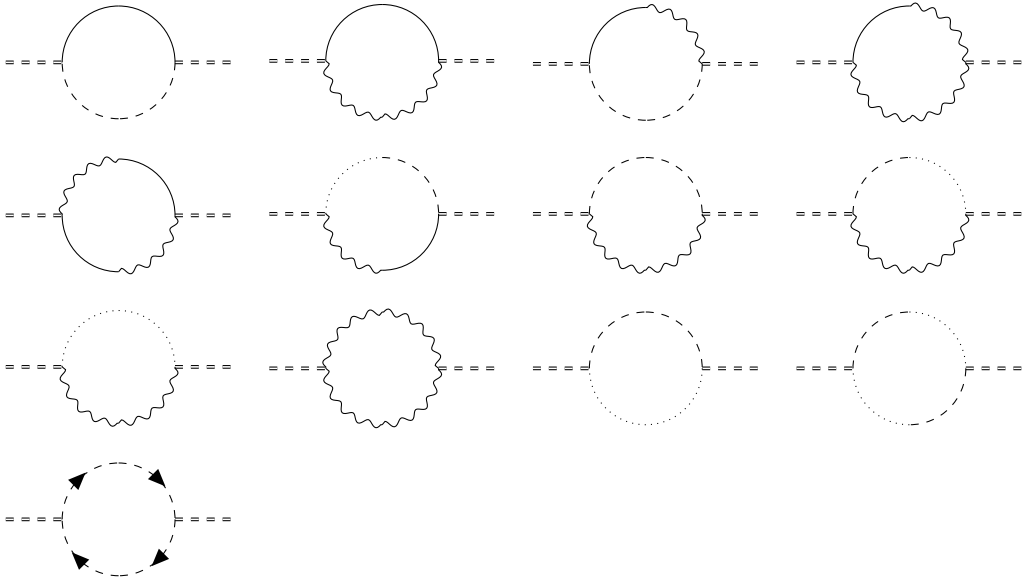
\begin{figure}[htb]
\begin{center}
 \begin{tikzpicture}
    \matrix[column sep=0.2cm, row sep=0.2cm] {
      \node {
        \begin{tikzpicture}[scale=0.75]
  \begin{feynman}
    \vertex (i) at (-1,0) ;
    \vertex (a) at (0, 0);
    \vertex (u) at (1, 1);
    \vertex (d) at (1, -1);
    \vertex (b) at (2, 0);
    \vertex (f) at (3,0) ;
    \diagram* {
      (i) -- [double, dashed] (a),
      (b) -- [double, dashed] (f),
      (a) -- [quarter left] (u),
      (u) -- [quarter left] (b),
      (b) -- [dashed, quarter left] (d),
      (d) -- [dashed, quarter left] (a),
    };
  \end{feynman}
\end{tikzpicture}
      }; &
      \node {
        \begin{tikzpicture}[scale=0.75]
  \begin{feynman}
    \vertex (i) at (-1,0) ;
    \vertex (a) at (0, 0);
    \vertex (u) at (1, 1);
    \vertex (d) at (1, -1);
    \vertex (b) at (2, 0);
    \vertex (f) at (3,0) ;
    \diagram* {
      (i) -- [double, dashed] (a),
      (b) -- [double, dashed] (f),
      (a) -- [quarter left] (u),
      (u) -- [quarter left] (b),
      (b) -- [photon, quarter left] (d),
      (d) -- [photon, quarter left] (a),
    };
  \end{feynman}
\end{tikzpicture}
      }; &
      \node {
       \begin{tikzpicture}[scale=0.75]
  \begin{feynman}
    \vertex (i) at (-1,0) ;
    \vertex (a) at (0, 0);
    \vertex (u) at (1, 1);
    \vertex (d) at (1, -1);
    \vertex (b) at (2, 0);
    \vertex (f) at (3,0) ;
    \diagram* {
      (i) -- [double, dashed] (a),
      (b) -- [double, dashed] (f),
      (a) -- [quarter left] (u),
      (u) -- [photon, quarter left] (b),
      (b) -- [dashed, quarter left] (d),
      (d) -- [dashed, quarter left] (a),
    };
  \end{feynman}
\end{tikzpicture}
      }; &
      \node {
       \begin{tikzpicture}[scale=0.75]
  \begin{feynman}
    \vertex (i) at (-1,0) ;
    \vertex (a) at (0, 0);
    \vertex (u) at (1, 1);
    \vertex (d) at (1, -1);
    \vertex (b) at (2, 0);
    \vertex (f) at (3,0) ;
    \diagram* {
      (i) -- [double, dashed] (a),
      (b) -- [double, dashed] (f),
      (a) -- [quarter left] (u),
      (u) -- [photon, quarter left] (b),
      (b) -- [photon, quarter left] (d),
      (d) -- [photon, quarter left] (a),
    };
  \end{feynman}
\end{tikzpicture} 
      }; \\
      \node {
        \begin{tikzpicture}[scale=0.75]
   \begin{feynman}
    \vertex (i) at (-1,0) ;
    \vertex (a) at (0, 0);
    \vertex (u) at (1, 1);
    \vertex (d) at (1, -1);
    \vertex (b) at (2, 0);
    \vertex (f) at (3,0) ;
    \diagram* {
      (i) -- [double, dashed] (a),
      (b) -- [double, dashed] (f),
      (a) -- [photon, quarter left] (u),
      (u) -- [solid, quarter left] (b),
      (b) -- [photon, quarter left] (d),
      (d) -- [solid, quarter left] (a),
    };
  \end{feynman}
\end{tikzpicture}
      }; &
      \node {
        \begin{tikzpicture}[scale=0.75]
  \begin{feynman}
    \vertex (i) at (-1,0) ;
    \vertex (a) at (0, 0);
    \vertex (u) at (1, 1);
    \vertex (d) at (1, -1);
    \vertex (b) at (2, 0);
    \vertex (f) at (3,0) ;
    \diagram* {
      (i) -- [double, dashed] (a),
      (b) -- [double, dashed] (f),
      (a) -- [dotted, quarter left] (u),
      (u) -- [dashed, quarter left] (b),
      (b) -- [solid, quarter left] (d),
      (d) -- [photon, quarter left] (a),
    };
  \end{feynman}
\end{tikzpicture}
      }; &
      \node {
        \begin{tikzpicture}[scale=0.75]
   \begin{feynman}
    \vertex (i) at (-1,0) ;
    \vertex (a) at (0, 0);
    \vertex (u) at (1, 1);
    \vertex (d) at (1, -1);
    \vertex (b) at (2, 0);
    \vertex (f) at (3,0) ;
    \diagram* {
      (i) -- [double, dashed] (a),
      (b) -- [double, dashed] (f),
      (a) -- [dashed, quarter left] (u),
      (u) -- [dashed, quarter left] (b),
      (b) -- [photon, quarter left] (d),
      (d) -- [photon, quarter left] (a),
    };
  \end{feynman}
\end{tikzpicture}
      }; &
      \node {
       \begin{tikzpicture}[scale=0.75]
  \begin{feynman}
    \vertex (i) at (-1,0) ;
    \vertex (a) at (0, 0);
    \vertex (u) at (1, 1);
    \vertex (d) at (1, -1);
    \vertex (b) at (2, 0);
    \vertex (f) at (3,0) ;
    \diagram* {
      (i) -- [double, dashed] (a),
      (b) -- [double, dashed] (f),
      (a) -- [dashed, quarter left] (u),
      (u) -- [dotted, quarter left] (b),
      (b) -- [photon, quarter left] (d),
      (d) -- [photon, quarter left] (a),
    };
  \end{feynman}
\end{tikzpicture}
      }; \\
 \node {
        \begin{tikzpicture}[scale=0.75]
   \begin{feynman}
    \vertex (i) at (-1,0) ;
    \vertex (a) at (0, 0);
    \vertex (u) at (1, 1);
    \vertex (d) at (1, -1);
    \vertex (b) at (2, 0);
    \vertex (f) at (3,0) ;
    \diagram* {
      (i) -- [double, dashed] (a),
      (b) -- [double, dashed] (f),
      (a) -- [dotted, quarter left] (u),
      (u) -- [dotted, quarter left] (b),
      (b) -- [photon, quarter left] (d),
      (d) -- [photon, quarter left] (a),
    };
  \end{feynman}
\end{tikzpicture}
      }; &
      \node {
        \begin{tikzpicture}[scale=0.75]
 \begin{feynman}
    \vertex (i) at (-1,0) ;
    \vertex (a) at (0, 0);
    \vertex (u) at (1, 1);
    \vertex (d) at (1, -1);
    \vertex (b) at (2, 0);
    \vertex (f) at (3,0) ;
    \diagram* {
      (i) -- [double, dashed] (a),
      (b) -- [double, dashed] (f),
      (a) -- [photon, quarter left] (u),
      (u) -- [photon, quarter left] (b),
      (b) -- [photon, quarter left] (d),
      (d) -- [photon, quarter left] (a),
    };
  \end{feynman}
\end{tikzpicture}
      }; &
      \node {
        \begin{tikzpicture}[scale=0.75]
   \begin{feynman}
    \vertex (i) at (-1,0) ;
    \vertex (a) at (0, 0);
    \vertex (u) at (1, 1);
    \vertex (d) at (1, -1);
    \vertex (b) at (2, 0);
    \vertex (f) at (3,0) ;
    \diagram* {
      (i) -- [double, dashed] (a),
      (b) -- [double, dashed] (f),
      (a) -- [dashed, quarter left] (u),
      (u) -- [dashed, quarter left] (b),
      (b) -- [dotted, quarter left] (d),
      (d) -- [dotted, quarter left] (a),
    };
  \end{feynman}
\end{tikzpicture}
      }; &
      \node {
       \begin{tikzpicture}[scale=0.75]
   \begin{feynman}
    \vertex (i) at (-1,0) ;
    \vertex (a) at (0, 0);
    \vertex (u) at (1, 1);
    \vertex (d) at (1, -1);
    \vertex (b) at (2, 0);
    \vertex (f) at (3,0) ;
    \diagram* {
      (i) -- [double, dashed] (a),
      (b) -- [double, dashed] (f),
      (a) -- [dashed, quarter left] (u),
      (u) -- [dotted, quarter left] (b),
      (b) -- [dashed, quarter left] (d),
      (d) -- [dotted, quarter left] (a),
    };
  \end{feynman}
\end{tikzpicture}
     }; \\
\node {
        \begin{tikzpicture}[scale=0.75]
  \begin{feynman}
    \vertex (i) at (-1,0) ;
    \vertex (a) at (0, 0);
    \vertex (u) at (1, 1);
    \vertex (d) at (1, -1);
    \vertex (b) at (2, 0);
    \vertex (f) at (3,0) ;
    \diagram* {
      (i) -- [double, dashed] (a),
      (b) -- [double, dashed] (f),
      (a) -- [dashed, fermion, quarter left] (u),
      (u) -- [dashed, fermion, quarter left] (b),
      (b) -- [dashed, fermion, quarter left] (d),
      (d) -- [dashed, fermion, quarter left] (a),
    };
  \end{feynman}
\end{tikzpicture}
       }; \\
    };
  \end{tikzpicture}
\end{center}\caption{Fish diagrams contributing into the two-point function of $\bar N_i$.}
 \label{f:fish}
\end{figure}

\subsection{Regular local divergences}
\label{subsec: regularization}
We work in Fourier space and refer to the frequency and momentum running in the loop as $(\omega, \, q_i)$. The external momentum carried by the background field $\bar N_i$ is denoted by $Q_i$. Since the background $\bar N_i(\bold x)$ is time-independent, the external frequency vanishes. As we saw above, the diagrams and, consequently, the loop integrals can be classified into two groups: bubbles in Fig. \ref{f:bubbles} and fishes in Fig. \ref{f:fish}. Schematically, a bubble diagram will lead to a loop integral of the form 
\begin{align} \label{eq:bubble_integral}
    \int \frac{d\omega d^2 q}{(2\pi)^3} \Pi(\omega,q)V^{(4)}(\omega, q, Q),
\end{align}
where $\Pi$ is a particular propagator forming the loop and $V^{(4)}$ is a four-legged vertex. Fish diagrams have two propagators and two vertices, leading to a more involved expression for the integrand,
\begin{align}  \label{eq:fish_integral}
    \int \frac{d\omega d^2 q}{(2\pi)^3} \Pi_1(\omega,q)\Pi_2(-\omega,Q-q)V_1^{(3)}(\omega, q, Q)V_2^{(3)}(\omega, q-Q) \, ,
\end{align}
with $\Pi_{1,2}$ standing for the propagators and $V^{(3)}_{1,2}$ for the three-legged vertices. Let us first discuss the diagrams with only regular propagators, which we call \textit{regular diagrams}. 
Diagrams with at least one $1/q^4$-propagator which we call \textit{irregular diagrams} 
will be dealt with in the next subsection.

We extract the one-loop counter-terms to the effective action in \eqref{eq:S_Ni}  by expanding the integrals \eqref{eq:bubble_integral} and \eqref{eq:fish_integral} in the external momentum $Q_i$ and keeping only the quadratic terms. The loop integrals take the form,\footnote{The integration over the directions of $q_i$ amounts to averaging over angles. For our purposes, we only need
\begin{align}
\label{eq: AveragingAngles}
    \overline{q_{i}q_{j}} =  \frac{1}{d}q^2 \delta_{ij} \,, \qquad \overline{q_{i}q_{j}q_kq_l}=\frac{1}{d(d+2)}q^4 ( \delta_{ij} \delta_{kl}+\delta_{ik} \delta_{jl}+\delta_{il} \delta_{kj}) \,,
\end{align}
which we write in general number of dimensions for later use.}
\begin{align} \label{eq:typical_term}
    \int \frac{d\omega \, d^2 q}{(2\pi)^3} \omega^{2a} q^{4b} \big[{\cal P}_I(\omega,q)\big]^A \big[{\cal P}_J(\omega,q)\big]^B\;,
\end{align}
with ${\cal P}_I$, ${\cal P}_J$,  $(I,J)=s,1,2$, being the pole structures (\ref{eq:physical mode}), (\ref{eq:gauge_modes}), and non-negative powers $a,b,A,B\ge0$. 
The overall scaling dimension of the diagrams contributing into the renormalization of the shift-dependent action (\ref{eq:S_Ni}) must be zero. 
Due to the scaling of the poles $\mathcal P_I$, this implies that the powers must satisfy the relation,
\be
A+B=a+b+1 \,.
\ee
The regularization of the UV-divergences is then performed through the Schwinger time pa\-ra\-me\-te\-ri\-za\-tion of the pole structures. Using the standard formula,
\begin{align}
    \big[{\cal P}_I(\omega,q)\big]^A = \int_0^\infty ds \frac{s^{A-1}}{\Gamma(A)}e^{-s [{\cal P}_I(\omega,q)]^{-1} } \,,
\end{align}
and, after the \textit{Wick rotation}, 
\begin{equation}
    \omega \rightarrow i \omega \, ,
\end{equation}
we compute the Gaussian integrals over $\omega$ and $q$, integrate over one of the Schwinger parameters $s$ and pull out the overall integral over the Schwinger time. The final expression for the logarithmically divergent contribution to the effective action is then
\begin{equation}
\label{eq:Action Div}
\Delta  \Gamma_{\rm div}= \int d t \int \frac{d^2Q}{(2\pi)^2}\bar N_i(Q_k) \bar N_j(-Q_k)\left( C_1 Q^2 \delta_{ij} + C_2 Q_i Q_j \right) \times \int \frac{ds}{s} \,,
\end{equation}
with $\bar{N}_i(Q_i)$ to be understood as the Fourier transforms of $\bar{N}_i$. Below, we derive the coefficients $C_{1,2}$ and use them to obtain the $\beta$-functions for the couplings $\lambda$ and $G$. Before computing the latter, we, however, have to deal with the irregular diagrams.

\subsection{Irregular frequency divergences}
\label{subsec: Irreg Diag}
The irregular contribution contained in the $\langle nn \rangle$ propagator brings about a power-law divergence in frequency loop integrals. As mentioned in Sec. \ref{sec: Introduction}, one in principle could use split-dimensional regularization (see e.g. \cite{Collins1984, Leibbrandt1996, Heinrich1999,Anselmi:2008bq, Lambert2022}) to set the one-loop integrals with this mode to zero \textit{individually} --- similarly to what one usually does in a relativistic theory. As a result, no irregular divergence appears at all. This is relatively straightforward only at one-loop level. Already at two loops, the procedure becomes much more involved \cite{Barvinsky1984, Leibbrandt1997, Heinrich1999}. 
Another drawback of the split dimensional regularization is that it hides 
the remarkable algebraic cancellation between the IFDs coming from different graphs. 

In order to flesh this algebra out, we therefore employ a different regularization procedure.\footnote{
A procedure similar in spirit was used in \cite{Baulieu1998, Niegawa2006, Baulieu2007} for the quantization of the Yang--Mills theory in Coulomb gauge, where the irregular frequency divergences were controlled by considering a family of gauges interpolating between the Coulomb and Lorentz gauge.} A convenient option is to use the dimensional regularization only in space and combine it with regularization by higher derivatives in time. This is implemented by adding to the bare Lagrangian a term invariant under FDiffs, as well as all the symmetries of Sec. \ref{sec: Quantization},
\begin{equation}
    \mathcal{L} \rightarrow  \mathcal{L} + \mathcal{L}_{\text{reg}} \, , 
\end{equation}
with
\begin{equation}
\label{eq: HigherDerTerm}
    \mathcal{L}_{\text{reg}}=  -\sqrt{\gamma} \frac{\mu_{67}}{2G} \mathcal{Q} \bar{\mathcal{Q}} \bigg[\, \mathcal{N}\, \frac{\big( D_t \mathfrak{a}_i\big)^2}{\Lambda^2} \bigg]\, , \quad \mathfrak{a}_i = \nabla_i \log \mathcal{N}  \, ,
\end{equation}
where the factor $\mu_{67}$ has been extracted for convenience and ${\cal N}=N+{\cal A}$, 
cf.~(\ref{eq: Field Redef N}). The covariant derivative of $\mathfrak{a}_i$ is given by
\begin{equation}
\label{eq: covdiva}
    D_t \mathfrak{a}_i = \frac{1}{N} \left(\dot{\mathfrak{a}}_i - N^k \nabla_k \mathfrak{a}_i + \mathfrak{a}^k \nabla_k N_i \right) \, .
\end{equation}
The term \eqref{eq: HigherDerTerm} has the same form as the potential \eqref{eq: Sv}, and so is invariant under all the symmetries $\mathcal{Q}$, $\hat{\mathcal{Q}}$, ${\cal C}$, $\mathcal{R}$ and their conjugates. One can also check the invariance explicitly using the formulae of Sec.~\ref{sec: Quantization}. 

As we are going to see shortly, addition of this regulator term modifies the propagators in such a way that the frequency integrals become convergent. At the same time, it also introduces new vertices which must be included in the calculation. We reserve a detailed study of this regularization, including its applicability in higher loops, for future work. Here, we employ it to isolate the one-loop IFDs and demonstrate their cancellation. 

We can conveniently derive the new contributions to the vertices and to the propagators by first specifying to our background $N_i = \bar{N}_i$, then explicitly applying the $\mathcal{Q}, \bar{\mathcal{Q}}$ and finally keeping only the terms quadratic in perturbations. The result is
\begin{equation}
\label{eq:Lregquad}
\mathcal{L}^{(2)}_{\text{reg}} =  \frac{\mu_{67}}{2 G \Lambda^2}\Big( \bar{D}_t \partial_i n \bar{D}_t \partial_i n -  \bar{D}_t \partial_i \mathcal{A} \bar{D}_t \partial_i \mathcal{A} - 2 \bar{D}_t \partial_i \bar{\eta} \bar{D}_t \partial_i \eta \Big)\, .
\end{equation}
As in \eqref{eq:S_V_2}, the linear in $\mathcal{A}$ term cancels.
The regulator \eqref{eq: HigherDerTerm} together with the spatial dimensional regularization lead to the modification of all the propagators involving $h_{ij}$ and $n$, as well as $\mathcal{A}$ and $\bar{\eta}, \eta$. These are modified in three ways. First, we should replace the constant coupling $\mu_{67}$ with a frequency and momentum dependent form-factor, 
\begin{equation}
\label{eq:mu67sub}
    \mu_{67} \rightarrow \tilde{\mu}_{67}(\omega,q)= \mu_{67} \, \Big(1- \omega^2/\big(\Lambda^2 q^2\big) \Big) \, .
\end{equation}
Secondly, the pole  $\tilde{\mathcal{P}}_s$ corresponding to the physical mode gets modified via
\begin{equation}
    \tilde{\mathcal{P}}_s^{-1} \rightarrow \big(\tilde{\mathcal{P}}_s^d\big)^{-1} = - \omega^2 + (1+\Delta_d) \tilde{\mu}_s q^4 \, , \quad \Delta_d = (d-2) \frac{1-\lambda}{1-\lambda d} \, .
\end{equation}
where in $\tilde{\mu}_s$ we also have to make the substitution \eqref{eq:mu67sub}.
Finally, we get new terms proportional to $d-2$.
In particular, for the lapse-lapse propagator we have
\begin{equation}
\label{eq:nn_prop_reg}
    \langle n n  \rangle  = -iG \, \frac{1}{ \tilde{\mu}_{67}q^4} - i G \, (1+\Delta_d)\frac{1-\lambda}{1-2\lambda } \frac{\mu_5^2}{\tilde{\mu}^2_{67}} \tilde{\mathcal{P}}^d_{s} \equiv \langle n n  \rangle_{\text{irr}} + \langle n n  \rangle_{\text{reg}}\, .
\end{equation}
Combining \eqref{eq:Lregquad} with the linearized part of the potential term \eqref{eq:S_V_2} we obtain the regularized propagators for the fields $\mathcal{A}$ and $\eta, \bar{\eta}$, see Eqs.~(\ref{Aetaprops}) from Appendix \ref{app:A}.
They coincide with the first, irregular, part of the $\langle n n \rangle$ propagator, taken with the opposite sign.
For metric-metric propagator a lengthy but straightforward calculation yields,
\begin{align}
    \langle h_{ij} h_{kl} \rangle=  \, -2i G\Bigg[& 2 \tilde{\mathcal{P}}^d_s \biggl(\frac{1\!-\!\lambda}{1\!-\!2 \lambda}\delta_{mn}\delta_{kl}\! -\! \delta_{mn} \hat{q}_k \hat{q}_l\! -\!  \delta_{kl} \hat{q}_m \hat{q}_n \biggr) + \mathcal{P}_1 Q_{mnkl} \nonumber \\
      &+ \biggl( \frac{2 \mathcal{P}_2}{1-\lambda} - 4 \mathcal{P}_1 + \frac{2 (1-2 \lambda)}{1- \lambda} \tilde{\mathcal{P}}^{d}_s \biggr)\hat{q}_m \hat{q}_n \hat{q}_k \hat{q}_l  + (d-2)\tilde{\mathcal{P}}^d_s A_{ijkl} + \frac{1}{\omega^2
    }Z_{ijkl} \Bigg]\, .
\end{align}
The tensors $\hat{q}_i, Q_{ijkl}$ are given in Eqs.~\eqref{eq: Q_Tensor} and the new tensor structure $A_{ijkl}$ is 
\begin{align}
    A_{ijkl} = \frac{2\lambda^2}{(1-\lambda d)(1-2\lambda)}\Bigg( 1 - &\frac{(1-2\lambda)^2}{\lambda^2} \frac{\tilde{\mu}_s q^4}{\omega^2} \Bigg)  \delta_{ij} \delta_{kl} + 2 \frac{1-2\lambda}{1-\lambda d} \frac{\tilde{\mu}_s q^4}{\omega^2} \left( \delta_{ij} \hat{q}_k \hat{q}_l + \delta_{kl} \hat{q}_i \hat{q}_j \right) \, .
\end{align}
The tensor  
\begin{equation}
    Z_{ijkl} =  2 \delta_{ij}\delta_{kl} -\delta_{ik} \delta_{jl} - \delta_{jk} \delta_{il} - 2 \delta_{ij} \hat{q}_k \hat{q}_l - 2 \delta_{kl} \hat{q}_i \hat{q}_j + \delta_{ik} \hat{q}_l \hat{q}_j + \delta_{il} \hat{q}_k \hat{q}_j + \delta_{jk} \hat{q}_l \hat{q}_i + \delta_{jl} \hat{q}_k \hat{q}_i \, .
\end{equation}
vanishes in $d=2$ (c.f. \eqref{UnitMatrix}), but is non-zero in general.
The mixed metric-lapse propagator $\langle h_{ij} n \rangle$ is also modified, but we do not present it here since 
it does not appear in the irregular diagrams.

As for the vertices, the first term in \eqref{eq:Lregquad} modifies the quartic vertex with two $\bar N_i$ and two lapse fluctuations $n$, and brings in a new vertex with one $\bar{N}_i$ and two lapses $n$ (see Eqs.~(\ref{eq:newHigherDerV}) from Appendix~\ref{app: InteractionLagrangian}). From the rest of \eqref{eq:Lregquad} we get two new quartic vertices with two $\bar N_i$: one with two $\mathcal{A}$'s and one with $\bar{\eta} \eta$. From the signs in \eqref{eq:Lregquad} it is easy to see that these are equal to each other and coincide with the negative of the new contribution to the vertex with two $\bar N_i$ and two $n$ \eqref{eq:newHigherDerV}. Finally we get two new cubic vertices with one $\bar{N}_i$: one with two $\mathcal{A}$'s and one with $\eta \bar{\eta}$. These are equal to each other and equal to minus the vertex with one $\bar{N}_i$ and two $n$ from \eqref{eq:newHigherDerV}. For completeness, all the vertices with $\mathcal{A}$ and $\bar{\eta},\eta$ are provided in Eqs.~\eqref{eq:newHigherDerVAeta},\eqref{eq:newHigherDerVAeta1}. We will see shortly that these signs lead to certain cancellations between diagrams.

There are now 8 Feynman diagrams containing irregular contributions, which we show in Figs.~\ref{f:irreg_diagrams} and \ref{f:irreg_diagrams_Aeta}.. 
The first three diagrams of Fig.~\ref{f:irreg_diagrams} contain a finite part in the limit $\Lambda \rightarrow \infty$ coming from the second term $\langle n n  \rangle_{\text{reg}}$ in the propagator \eqref{eq:nn_prop_reg}. These contributions are fully regular and we do not keep track of them in this subsection. 
\begin{figure}[ht]
\centering
\renewcommand{\arraystretch}{1.5}
\begin{tabular}{cc}
\(\displaystyle
\Gamma^1=\quad\begin{tikzpicture}[scale=0.75]
  \begin{feynman}
    \vertex (i) at (-1, 0) ;
    \vertex (a) at (0, 0);
    \vertex (m) at (1, 0);
    \vertex (u) at (1, 2);
    \vertex (b) at (2, 0);
    \vertex (f) at (3, 0) ;
    \diagram* {
      (i) -- [double, dashed] (m),
      (m) -- [double, dashed] (f),
      (m) -- [solid, half left] (u),
      (u) -- [solid, half left] (m),
    };
  \end{feynman}
\end{tikzpicture}
\) \hspace{2cm}
\(\displaystyle
\Gamma^2=\quad
\raisebox{-1.8em}{
\begin{tikzpicture}[scale=0.75]
 \begin{feynman}
    \vertex (i) at (-1,0) ;
    \vertex (a) at (0, 0);
    \vertex (u) at (1, 1);
    \vertex (d) at (1, -1);
    \vertex (b) at (2, 0);
    \vertex (f) at (3,0) ;
    \diagram* {
      (i) -- [double, dashed] (a),
      (b) -- [double, dashed] (f),
      (a) -- [quarter left] (u),
      (u) -- [quarter left] (b),
      (b) -- [photon, quarter left] (d),
      (d) -- [photon, quarter left] (a),
    };
  \end{feynman}
\end{tikzpicture}}
\) \\[2.0em]   

\(\displaystyle
\Gamma^3=\quad
\raisebox{-1.8em}{
\begin{tikzpicture}[scale=0.75]
 \begin{feynman}
    \vertex (i) at (-1,0) ;
    \vertex (a) at (0, 0);
    \vertex (u) at (1, 1);
    \vertex (d) at (1, -1);
    \vertex (b) at (2, 0);
    \vertex (f) at (3,0) ;
    \diagram* {
      (i) -- [double, dashed] (a),
      (b) -- [double, dashed] (f),
      (a) -- [quarter left] (u),
      (u) -- [quarter left] (b),
      (b) -- [dashed, quarter left] (d),
      (d) -- [dashed, quarter left] (a),
    };
  \end{feynman}
\end{tikzpicture}}
\) \hspace{2cm}
\(\displaystyle
\Gamma^4=\quad
\raisebox{-1.8em}{
\begin{tikzpicture}[scale=0.75]
 \begin{feynman}
    \vertex (i) at (-1,0) ;
    \vertex (a) at (0, 0);
    \vertex (u) at (1, 1);
    \vertex (d) at (1, -1);
    \vertex (b) at (2, 0);
    \vertex (f) at (3,0) ;
    \diagram* {
      (i) -- [double, dashed] (a),
      (b) -- [double, dashed] (f),
      (a) -- [quarter left] (u),
      (u) -- [quarter left] (b),
      (b) -- [quarter left] (d),
      (d) -- [quarter left] (a),
    };
  \end{feynman}
\end{tikzpicture}}
\) \\
\end{tabular}
\caption{Irregular diagrams with the lapse $n$.}
\label{f:irreg_diagrams}
\end{figure}
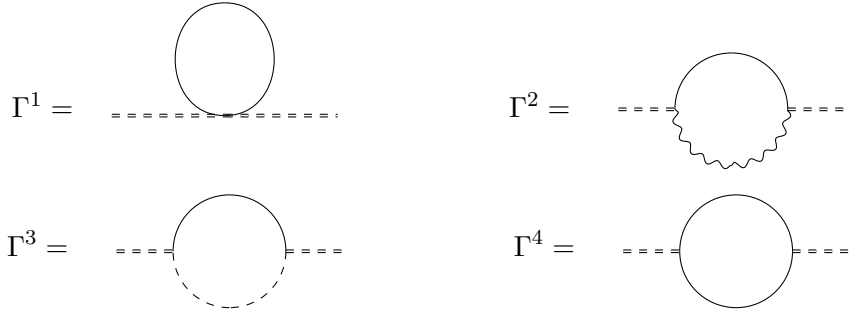
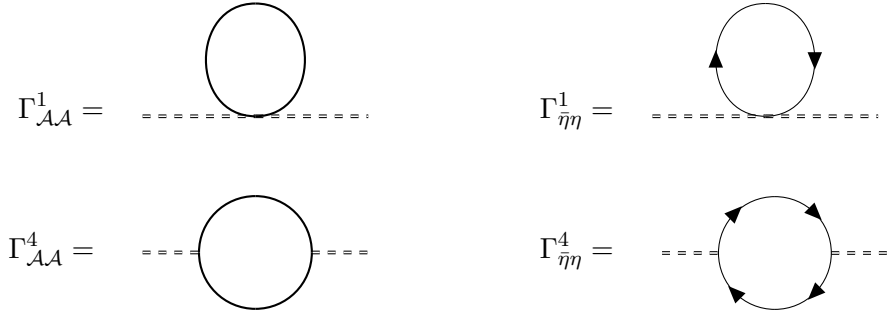
\begin{figure}[ht]
\centering
\renewcommand{\arraystretch}{1.5}
\begin{tabular}{cc}
\(\displaystyle
\Gamma^{1}_{\mathcal{A}\mathcal{A}}=\quad\begin{tikzpicture}[scale=0.75]
  \begin{feynman}
    \vertex (i) at (-1, 0) ;
    \vertex (a) at (0, 0);
    \vertex (m) at (1, 0);
    \vertex (u) at (1, 2);
    \vertex (b) at (2, 0);
    \vertex (f) at (3, 0) ;
    \diagram* {
      (i) -- [double, dashed] (m),
      (m) -- [double, dashed] (f),
      (m) -- [thick, half left] (u),
      (u) -- [thick, half left] (m),
    };
  \end{feynman}
\end{tikzpicture}
\) \hspace{2cm}
\(\displaystyle
\Gamma^{1}_{\bar{\eta}\eta}=\quad\begin{tikzpicture}[scale=0.75]
  \begin{feynman}
    \vertex (i) at (-1, 0) ;
    \vertex (a) at (0, 0);
    \vertex (m) at (1, 0);
    \vertex (u) at (1, 2);
    \vertex (b) at (2, 0);
    \vertex (f) at (3, 0) ;
    \diagram* {
      (i) -- [double, dashed] (m),
      (m) -- [double, dashed] (f),
      (m) -- [fermion, half left] (u),
      (u) -- [fermion, half left] (m),
    };
  \end{feynman}
\end{tikzpicture}
\) \\[2.0em]   

\(\displaystyle
\Gamma^4_{\mathcal{A} \mathcal{A}}=\quad
\raisebox{-1.8em}{
\begin{tikzpicture}[scale=0.75]
 \begin{feynman}
    \vertex (i) at (-1,0) ;
    \vertex (a) at (0, 0);
    \vertex (u) at (1, 1);
    \vertex (d) at (1, -1);
    \vertex (b) at (2, 0);
    \vertex (f) at (3,0) ;
    \diagram* {
      (i) -- [double, dashed] (a),
      (b) -- [double, dashed] (f),
      (a) -- [thick, quarter left] (u),
      (u) -- [thick, quarter left] (b),
      (b) -- [thick, quarter left] (d),
      (d) -- [thick, quarter left] (a),
    };
  \end{feynman}
\end{tikzpicture}}
\) \hspace{2cm}
\(\displaystyle
\Gamma^4_{\bar{\eta}\eta}=\quad
\raisebox{-1.8em}{
\begin{tikzpicture}[scale=0.75]
 \begin{feynman}
    \vertex (i) at (-1,0) ;
    \vertex (a) at (0, 0);
    \vertex (u) at (1, 1);
    \vertex (d) at (1, -1);
    \vertex (b) at (2, 0);
    \vertex (f) at (3,0) ;
    \diagram* {
      (i) -- [double, dashed] (a),
      (b) -- [double, dashed] (f),
      (a) -- [fermion, quarter left] (u),
      (u) -- [fermion, quarter left] (b),
      (b) -- [fermion, quarter left] (d),
      (d) -- [fermion, quarter left] (a),
    };
  \end{feynman}
\end{tikzpicture}}
\) \\
\end{tabular}
\caption{Irregular diagrams with auxiliary fields $\mathcal{A}$ and $\eta, \bar{\eta}$. The thick solid line denotes the field $\mathcal{A}$, whereas the lines with arrows stand for the fermions $\eta$, 
$\bar{\eta}$.}
\label{f:irreg_diagrams_Aeta}
\end{figure}

Using the relevant pieces of the interaction Lagrangian \eqref{eq:Ni_n}, \eqref{eq:NiNj_n}, \eqref{eq:newHigherDerV} and Wick-rotating $\omega \rightarrow i \omega$, we find the expression for the the first diagram in Fig.~\ref{f:irreg_diagrams},\footnote{Recall that $\Gamma$'s carry the indices of the background fields $\bar{N}_i$.} 
\begin{align}
\label{eq:Bubble Irreg}
    \Gamma^1_{ij} =& \frac{i}{2\mu_{67}} \left[ Q^2 \delta_{ij} + (1-2\lambda) Q_i Q_j \right] \int \frac{ d^d q}{(2 \pi)^d} \int \frac{d \omega}{2 \pi} \frac{1}{q^2 (q^2 + \omega^2 / \Lambda^2)}    \nonumber \\ 
 &+i \frac{2}{d} \, Q^2 \delta_{ij}  \int \frac{ d^d q}{(2 \pi)^d} \int \frac{d \omega}{2 \pi} \frac{1}{\Lambda^2 (q^2+\omega^2/\Lambda^2)}  + \dots  \,,
\end{align}
where dots stand for regular terms.
Here we have averaged over the spatial angles $\Omega_d$ with a dimensionally regularized measure, 
\begin{equation}
    d^d q =  q^{d-1} dq \, d \Omega_d\, ,
\end{equation}
using formulae \eqref{eq: AveragingAngles}. 
The last terms in this expression cancels by the diagrams in the upper row of Fig.~\ref{f:irreg_diagrams_Aeta},
\begin{equation}
\big(\Gamma^{1}_{\mathcal{A}\mathcal{A}}\big)_{ij} + \big(\Gamma^{1}_{\bar{\eta}\eta}\big)_{ij} = -i \frac{2}{d} \, Q^2 \delta_{ij}  \int \frac{ d^d q}{(2 \pi)^d} \int \frac{d \omega}{2 \pi} \frac{1}{\Lambda^2 (q^2+\omega^2/\Lambda^2)} = - i \frac{1}{d} Q^2 \delta_{ij} \int \frac{ d^d q}{(2 \pi)^d} \frac{1}{\Lambda q} \, .
\end{equation}
Integrating the remaining contribution over frequency we obtain,
\begin{align}
\label{eq:Gamma1_IntOmega}
\Gamma^1_{ij}+\big(\Gamma^{1}_{\mathcal{A}\mathcal{A}}\big)_{ij} + \big(\Gamma^{1}_{\bar{\eta}\eta}\big)_{ij} = \frac{i}{2\mu_{67}} \left[ Q^2 \delta_{ij} + (1-2\lambda) Q_i Q_j \right] \int \frac{ d^d q}{(2 \pi)^d} \frac{\Lambda}{2 q^3 }  + \dots \, .
\end{align}
This expression converges at $q\to\infty$. Instead, it diverges at low momenta and needs an IR regulator. Imposing a momentum cutoff $\Lambda_{\text{IR}}$ we obtain,
\begin{align}
\label{eq: Bubble_Irreg_Reg}
\Gamma^1_{ij} =  \frac{c_d}{3-d}\frac{i}{2\mu_{67}} \left[ Q^2 \delta_{ij} + (1-2\lambda) Q_i Q_j \right] \times \frac{\Lambda}{2\Lambda_{\text{IR}}^{3-d}} + \dots \,,
\end{align}
with
\begin{equation}
    c_d = \int \frac{d \Omega_d}{(2 \pi)^d} \, .
\end{equation}
Note that this expression is polynomial in the external momentum $Q_i$ and hence, strictly speaking, local. Still, it is problematic due to its dependence on the IR cutoff which signals a mixing between UV and IR divergences. Fortunately, we are going to see that it cancels by a contribution from the fish diagrams.

We now turn to the diagram $\Gamma^2$. 
The vertices are still given by \eqref{eq: NbarLapseMetricVertex}. We have 
\begin{align}
\label{eq:Gamma2_init}
 &\Gamma^2_{ij} =\!
     i\!\!\int\!\!\frac{ d^d q}{(2 \pi)^d} \!\int \frac{d \omega}{2 \pi} \Big( \frac{1}{2}\delta_{ki} Q_l + \frac{1}{2}\delta_{li} Q_k - \lambda\delta_{kl} Q_i \Big) \frac{ \omega^2 \Big( \frac{1}{2}\delta_{mj} Q_n + \frac{1}{2}\delta_{nj} Q_m - \lambda\delta_{mn} Q_j\Big) }{2 \mu_{67} (\mathbf{q}+\mathbf{Q})^2\big((\mathbf{q}+\mathbf{Q})^2 + \omega^2/\Lambda^2\big)} \langle h_{kl} h_{mn} \rangle  +\dots
\end{align}
Dots again stand for 
$\Lambda$-finite regular terms coming from the second part of the propagator \eqref{eq:nn_prop_reg}. 
We perform the Wick rotation, rescale the frequency variable, $\omega \rightarrow \Lambda \, \tilde{\omega}$ and expand the brackets carefully, using the $d$-dimensional contractions. Using simple algebraic manipulations, we can extract the piece explicitly divergent at $\Lambda \rightarrow \infty$. The integrand of \eqref{eq:Gamma2_init} has a complicated dependence on the external momentum and thus the resulting divergent term itself a priori need not be a polynomial of $Q_i$. And indeed the individual contributions will contain non-localities, e.g. of the form
\begin{equation}
     \frac{\Lambda}{Q^2} \int \frac{ d^d q}{(2 \pi)^d} \int \frac{d \tilde{\omega}}{2 \pi}   \frac{ Q^2 q^2 -q_i q_j Q_i Q_j}{q^2(\mathbf{q}+\mathbf{Q})^2\big((\mathbf{q}+\mathbf{Q})^2 + \tilde{\omega}^2\big)} \propto \frac{\Lambda}{\sqrt{Q^2}} \, .
    \end{equation}
Yet the full expression is perfectly local. In total we get
\begin{align}  
\label{eq: Fish Irreg metric}
      \Gamma^2_{ij}=-\frac{c_d}{3-d}\frac{i}{2\mu_{67}} \left[ Q^2 \delta_{ij} + (1-2\lambda) Q_i Q_j \right] \times \frac{\Lambda}{2\Lambda_{\text{IR}}^{3-d}}  +  \hat{\Gamma}^2_{ij} + \dots \, .
\end{align}
The $\Lambda$-divergent piece in the first line precisely cancels with the bubble \eqref{eq: Bubble_Irreg_Reg}. The term $\hat{\Gamma}^2_{ij}$ stands for the new finite at $\Lambda\to \infty$ piece coming from the loop integral displayed in \eqref{eq:Gamma2_init}. In Appendix~\ref{app:irreg_beta} we show that this new piece contains a \textit{local} logarithmic divergence in momentum which contributes to the effective action in the limit $\Lambda \rightarrow \infty$

The vertices entering the diagram $\Gamma^{3}_{ij}$ do not contain any time derivatives, see 
Eq.~(\ref{eq: NbarLapseShiftVertex}). As a consequence, the integral over $\omega$ converges even without any additional regularization. It does, however, contribute to the overall logarithmic divergence. This contribution is computed in Appendix~\ref{app:irreg_beta}.

Finally, using vertices from \eqref{eq:newHigherDerV1}, (\ref{eq:newHigherDerVAeta}), (\ref{eq:newHigherDerVAeta1}), and the propagators \eqref{eq:nn_prop_reg}, (\ref{Aetaprops}) we obtain the last diagram $\Gamma^4_{ij}$ and its counterparts with the auxiliary fields, $\big(\Gamma^4_{\mathcal{A}\mathcal{A}}\big)_{ij},\big(\Gamma^4_{\bar{\eta}\eta}\big)_{ij}$. Similarly to the case of bubbles, the latter diagrams cancel the term with two $\langle nn \rangle_{\text{irr}}$ in the full sum $\Gamma^4_{ij}+\big(\Gamma^4_{\mathcal{A}\mathcal{A}}\big)_{ij}+\big(\Gamma^4_{\bar{\eta}\eta}\big)_{ij}$. The remaining expression is suppressed by at least two powers of $\Lambda$ and vanishes in the limit $\Lambda\to\infty$. Thus, it does not  
contribute to the effective action \eqref{eq:Action Div}.

All in all, we have shown that, though divergences linear in $\Lambda$ arise in individual diagrams, they cancel in the final result. It is worth emphasizing that the parts of the diagrams shown in Fig.~\ref{f:irreg_diagrams} involving the irregular part of the lapse propagator $\langle nn\rangle_{\rm irr}$ do not cancel completely: their parts finite at $\Lambda\to \infty$ (but still divergent at $d\to 2$) survive and provide non-trivial contribution into the effective action.

\subsection{Renormalization group running of $\lambda$ and $G$} \label{subsec:RG}
With no IFDs remaining, the derivation of the $\beta$-functions proceeds as usual. Given the one-loop divergent correction to the effective action (\ref{eq:Action Div}), 
one compares with \eqref{eq:S_Ni} and finds the expressions for the renormalized couplings:
\be
\label{eq: Renorm Couplings}
G_R=G-4 C_1 G^2 \int \frac{ds}{s} \,, \quad \lambda_R=\lambda- 2 G \left[ C_2 + (2 \lambda-1) C_1 \right] \int \frac{ds}{s} \,.
\ee
Since the scaling dimension of the Schwinger parameter $s$ is $(-4)$, we follow
\cite{Barvinsky2017AssFreed,Barvinsky2019} (see also \cite{Liao1994}) and identify 
\begin{equation}
\label{eq:S_To_Log}
\int \frac{ds}{s} \to \log \left( \frac{\Lambda_{\rm UV}^4}{k_*^4} \right) \,.
\end{equation}
Here $\Lambda_{\rm UV}$ is an unspecified UV cutoff of the theory and $k_*$ is the RG scale \cite{Peskin:1995ev}. Physically, in the Wilsonian picture, we interpret this logarithm as a result of integrating out the virtual high-frequency modes between $k_*$ and $\Lambda_{\text{UV}}$. The energy-dependence of the QFT observables is captured by the running of the couplings with the RG scale $k_*$ \cite{Weinberg1996}. The running is quantified by the derivatives, i.e., the $\beta$-functions:
\be \label{eq:betas}
 \beta_G= k_* \frac{d G_R}{dk_*}= 16 C_1 G^2 \,, \qquad \beta_\lambda= k_* \frac{d \lambda_R}{dk_*}= 8 G \left[ C_2 + (2 \lambda-1) C_1 \right] \,.
\ee

The expressions for the counter-terms $C_{1,2}$ are not illuminating and 
we refer the reader to Eqs.~\eqref{eq:C1}, \eqref{eq:C2}) in 
Appendix~\ref{app:C} for the full expressions in a generic gauge. 
The expression for the $\beta$-function of the coupling $G$ is straightforwardly related to the counter-term $C_1$. It is gauge dependent since the coupling $G$ is {\it not essential}: it can be changed by adding to the action a term that vanishes on the equations of motion, cf.~\cite{Barvinsky2017AssFreed,Barvinsky2019}. For general values of the gauge parameters $\sigma$, $\xi$, it can be obtained from Eq.~(\ref{eq:C1}). 
Here we quote the result in the \textit{uniform gauge} (\ref{eq:physical_gauge}):
\begin{equation}
\label{eq:beta_G}
\beta_G= -\frac{G^2 }{16 \pi (1-2 \lambda)\sqrt{\mu_s}} \left[(23 - 30 \lambda)  - 2\frac{\mu_5}{\mu_{67}}(5-6 \lambda) - 4\frac{\mu_1}{\mu_{67}}(1-2 \lambda) \right]\,.
\end{equation}  

On the other hand, the coupling $\lambda$ is \textit{essential} \cite{Barvinsky2017AssFreed,Barvinsky2019} and its $\beta$-function must be gauge invariant. This is indeed the case: the dependence of the counter-terms $C_{1,2}$ on the gauge parameters cancel out in the combination entering $\beta_\lambda$, which provides a strong cross-check of our calculation. We have, 
\be \label{eq:beta_lambda}
\beta_\lambda= \frac{G}{32 \pi \sqrt{\mu_s}} \left[  \big( 15 - 14 \lambda \big)  - 2 \frac{\mu_5}{\mu_{67}}(7-6\lambda) + 4 \frac{\mu_1}{\mu_{67}}(1+2\lambda)\right]\,.
\ee
Let us dwell on this result. The first term comes from the same diagrams that exist in the projectable theory, that is, the diagrams without lapse. Sending $\mu_{6}$ and/or $\mu_7$ to $\infty$ at the level of the path integral enforces the conditions (see \eqref{Potential}),
\be
(\nabla_i a_j)(\nabla^i a^j)=0~~~~\text{and/or}~~~~ (\nabla_i a^i)^2 =0 \,.
\ee
Since we take all field fluctuations to vanish at spatial infinity, and $a_i$ is twist-free, these conditions imply $a_i=0$, and thus the projectable limit. Taking this limit in \eqref{eq:beta_lambda} we indeed reproduce the result of the projectable model \cite{Barvinsky2017AssFreed},
\be
\lim_{\mu_{67} \to \infty} \beta_\lambda= \frac{(15 - 14 \lambda)}{64 \pi }\sqrt{\frac{1-2\lambda}{1-\lambda}} \,\frac{G}{\sqrt{\mu_1}}\,.
\ee

The second and third term in the square bracket \eqref{eq:beta_lambda} come from the new class of diagrams with the propagating lapse. Note that from the form of the $\langle n n \rangle$ propagator \eqref{eq:nn_prop} one could expect the terms proportional to $\mu_5^2$ in the expression for the beta-functions. Curiously, in the final result for the whole logarithmically-divergent part of the effective action (in an arbitrary gauge) such terms cancel out.\footnote{Note that there is still $\mu_5^2$ dependence inside the $\mu_{s}$ in the common denominator.} Further, as explained in Appendix~\ref{app:irreg_beta}, the $\beta$-functions receive non-vanishing contributions from the $\Lambda$-finite \qu{leftover} of the irregular parts of the diagrams $\Gamma^2$ and $\Gamma^3$ from Fig.~\ref{f:irreg_diagrams}.\footnote{Recall that the diagram $\Gamma^4$ does not contribute.} This demonstrates that the irregular part of the lapse propagator $\langle nn\rangle_{\rm irr}$ is physical and cannot be simply discarded. 

\bigskip

The explicit check of the cancellation of non-local divergences and the $\beta$-functions \eqref{eq:beta_G}, \eqref{eq:beta_lambda} are the main results of this section. 

%% file: Sections/Section4.tex
\section{Effective action from the heat kernel}
\label{Sec:Heat_Kernel}
In this section we introduce an alternative method for one-loop calculations in non-projectable Ho\v rava gravity based on the heat kernel technique, see \cite{Vassilevich:2003xt,Barvinsky2021,Barvinsky:2017mal} and references therein. Its advantage is resummation in a compact form of multiple Feynman diagrams contributing into the effective action. This approach is most naturally formulated in ``Euclidean'' time, so we perform the Wick rotation $t\to-i\tau$, $N_i\to iN_i$. Another difference from the previous section is in the regularization. As we will see below, the heat kernel approach naturally leads to the point-splitting regularization in Euclidean time. The cancellation of IFDs then will be seen as the cancellation between the singular terms proportional to $\delta(\Delta \tau) \Bigr|_{\Delta \tau \rightarrow 0}$. To simplify the calculation, in this section, we will adopt the \textit{uniform gauge} \eqref{eq:physical_gauge}. The method can be generalized to an arbitrary gauge with an appropriate modification~\cite{Barvinsky2025}. The Mathematica notebook with the computation can be found at \cite{github}.

\subsection{Setting up the calculation}

We again consider the background where only the shift has non-trivial value $\bar N_i(\mathbf{x})$ and 
write the effective Euclidean action at one loop as the Gaussian integral over the fluctuations,
\begin{equation}
\label{eq: Gammafluct}
    e^{-\Delta \Gamma} = \int D \Psi D c  D \bar{c} \exp{\Bigg\{-\int d\tau  d^2 x\left(\frac{1}{2}\Psi^{A} \mathbb{D}_{A B}[\partial_{\tau}, \partial_i ; \bar{N}_{i}] \Psi^{B} + \bar{c}^{i} \mathbb{B}_{ij}[\partial_{\tau}, \partial_i ; \bar{N}_{i}] c^{j}  \right)\Bigg\}} \, .
\end{equation}
Here we have combined all the field fluctuations, except ghosts, into a single multi-vector,\footnote{
Since in this section we do not introduce a higher-derivative regulator, the auxiliary fields ${\cal A}$, $\eta$, $\bar\eta$ decouple to the background shift and thus produce a field-independent contribution into the effective action. We do not consider them explicitly in what follows.} 
\begin{equation}
\label{eq:fieldorder}
    \Psi^{A} = (n_i, \pi_i, h_{ij},n) \, ,
\end{equation}
and redefined,
\begin{equation}
    (n_i, h_{ij}, n) \rightarrow \sqrt{2 G} (n_i, h_{ij}, n)\, , \quad\pi_i \rightarrow \frac{1}{\sqrt{2G}} \pi_i\,,
    \quad (\bar{c}^i, c^j) \rightarrow \sqrt{G} \, (\bar{c}^i, c^j) \, , 
\end{equation}
to absorb the factors of $G$. Integrating over the fluctuations we obtain,
\begin{equation}
\label{eq: DeltaGammaFieldsGhosts}
    \Delta \Gamma = \frac{1}{2}\log \det \mathbb{D} - \log \det \mathbb{B}  \equiv \Delta \Gamma_{\text{fields}} + \Delta \Gamma_{\text{ghosts}}\, .
\end{equation}
We postpone the discussion of the ghost contributions to Sec.~\ref{ssec:hkghosts} and consider here the 
fields determinant. 

The quadratic form $\mathbb{D}$ consists of three terms,
\begin{equation}
    \mathbb{D} = \mathbb{D}^{(0)} + \mathbb{D}^{(1)} + \mathbb{D}^{(2)} \, ,
\end{equation}
with zero, one and two background fields $\bar{N}^{i}$ respectively. All the $\mathbb{D}$'s are differential operators built out of $\partial_{i}$ and $\partial_{\tau}$ with coefficients depending on the background field $\bar{N}_i$ and its derivatives,
\begin{equation}
    \mathbb{D} = \mathbb{D}[\partial_{\tau}, \partial_{t} \, ; \bar{N}_i, \partial_{i} \bar{N}_j] \, .
\end{equation}
In what follows, we mostly omit the arguments of $\mathbb{D}$ to avoid cluttered expressions.

The $\mathbb{D}^{(0)}$-part is read off from the quadratic Lagrangian \eqref{eq:Lqflat}, whereas the terms $\mathbb{D}^{(1,2)}$ represent the interaction Lagrangian pieces that can be found in Appendix~\ref{app: InteractionLagrangian}. In these pieces one has to bring all derivatives to act on the fluctuations to the right using integration by parts.
Thanks to the gauge fixing \eqref{GaugeFixingFunction}, which at quadratic level decouples $h_{ij}$ from $n_i$ (and correspondingly $\pi_i$), the quadratic form $\mathbb{D}^{(0)} $ is block-diagonal.  Explicitly, in the uniform gauge, and with the ordering of fields (\ref{eq:fieldorder}), it reads 
\begin{equation}
\label{eq:Dnot}
    \mathbb{D}^{(0)} = \begin{pmatrix}
         (2\lambda\! - \!1)\partial^i \partial^j \!-\!\delta^{ij}\Delta  &  i \delta^{ij} \partial_{\tau} & 0 & 0  \\
        -i \delta^{ij}  \partial_{\tau}& {\mu_s} \left(\frac{1-2\lambda}{2(1-\lambda)}\partial^i \partial^j\! - \!\delta^{ij} \Delta \right) & 0 & 0 \\
        0 & 0 &  \frac{1}{2} \left(G^{ij kl}( \mu_s \Delta^2\!-\!\partial_{\tau}^2)\! +\! \frac{\mu_5^2}{\mu_{67}} \mathcal{D}^{ij} \mathcal{D}^{kl} \right) & \mu_5 \mathcal{D}^{ij} \Delta  \\
        0 & 0 & \mu_5 \mathcal{D}^{kl}\Delta  & 2 \mu_{67} \Delta^2
    \end{pmatrix} 
\end{equation}
where we have used the identity (\ref{UnitMatrix}) to simplify the quadratic form in the metric sector.
Here $G^{ijkl}$ is the DeWitt metric in the space of metric fluctuations $h_{ij}$,
\begin{equation}
    G^{ijkl} = \frac{1}{2} \delta^{ik}\delta^{jl} + \frac{1}{2} \delta^{jk}\delta^{il} - \lambda \, \delta^{ij} \delta^{kl} \, , \quad (G^{-1})_{ijkl} = \frac{1}{2} \delta_{ik}\delta_{jl} + \frac{1}{2} \delta_{jk}\delta_{il} + \frac{\lambda}{1-2\lambda} \delta_{ij} \delta_{kl} \, ,
\end{equation}
and the differential operator $\mathcal{D}^{ij}$ is defined as
\begin{equation}
    \mathcal{D}^{ij} \equiv -\delta^{ij}\Delta  + \partial^{i} \partial^{j} \,.
\end{equation}
The quadratic forms $\mathbb{D}^{(1)}$ and $\mathbb{D}^{(2)}$, on the other hand, will have non-zero entries in the upper right and lower left blocks that mix the $\{n^i,\, \pi^i\}$ and $\{h_{ij}, n\}$ sectors.

Next, we would like to write the 
functional determinant as the functional trace,
\begin{equation}
\label{eq:LogDetTrLog}
    \log \det \mathbb{D} = \text{Tr} \log \mathbb{D} \, ,
\end{equation}
and use the standard formula
\begin{equation}
    \text{Tr} \log \mathbb{D} = -\int \frac{ds}{s} \text{Tr} \, e^{-s \, \mathbb{D}} \, .
\end{equation}
However, a complication arises. The quadratic form 
$\mathbb{D}_{AB}$ has entries of different dimensions.
This is a consequence of the fact that the shift, and correspondingly its fluctuation $n_{i}$, has scaling dimension $1$, whereas $h_{ij}$ and $n$ are dimensionless, see Eqs.~\eqref{eq:Metric_Dimension}, \eqref{eq:Shift_Dimension}.
Hence, the formula (\ref{eq:LogDetTrLog}) cannot be directly applied: the expression $\log \mathbb{D}$ is understood as a Taylor series in $\mathbb{D}$ and therefore involves matrix multiplication, which is ill-defined for the matrices with entries of different dimension. 
As we show in the next subsection, this issue is resolved by redefining 
$\mathbb{D}$, multiplying it on the left and on the right by constant matrices,  
\begin{equation}
\label{eq:D_to_LDR}
\mathbb{D} \to \mathbb{L}\mathbb{D}\mathbb{R} \, \, .
\end{equation}
This redefinition promotes $\mathbb{D}$ to a proper operator, changing its determinant only by a field-independent, and thus irrelevant, factor. The following discussion of the heat kernel
procedure refers to the redefined $\mathbb{D}$.

By definition, the functional trace is, 
\begin{equation}
     \text{Tr} \, e^{-s \, \mathbb{D}} = \int d \mathbf{x} d\tau \,   \text{tr } \Big[e^{- s \, \mathbb{D}\left[\partial_\tau, \partial_i \right]} \delta(\mathbf{x} - \mathbf{x}' ) \delta(\tau - \tau')\Big] \Bigr|_{\mathbf{x}' \rightarrow \mathbf{x}, \, \tau' \rightarrow \tau} \, .
\end{equation}
Here we first compute the matrix trace, as well as the trace over the tensor indices $i,j, \dots$. We then take the coincident limit, and finally integrate over space and time. 
Since we consider time-independent background, we can  
represent the $\delta$-function in time with its Fourier integral and write,
\begin{equation}
\label{eq: TrLogFourier}
    \text{Tr} \log \mathbb{D} = -\int \frac{ds}{s} \int d\mathbf{x} d\tau    \int \frac{d\omega}{2 \pi} \, \text{tr } \Big[ K(s| \,  \omega; \mathbf{x},\mathbf{x}') \Big]\Bigr|_{\mathbf{x}' \rightarrow \mathbf{x}} e^{i \omega (\tau - \tau'))} \Bigr|_{\tau' \rightarrow \tau} \, ,
\end{equation}
with 
\begin{equation}
    K(s| \,  \omega; \mathbf{x},\mathbf{x}') = e^{- s \, \mathbb{D}\left[i \omega, \partial_i \right]} \delta(\mathbf{x} - \mathbf{x}' ) \, .
\end{equation}
The task of computing the effective action then reduces to finding the trace of the \textit{heat kernel} $K$. 
In particular, the logarithmic divergence is given by the zeroth-order in the expansion of $K$ in powers of $s$. 

We can find $K$ perturbatively in powers of the background using the fact that it satisfies the equation 
\begin{equation}
\label{HeatEquation}
    - \partial_s K = \mathbb{D} K  \, ,
\end{equation}
with initial condition
\begin{equation}
\label{InitCond}
    K \Bigr|_{s\rightarrow 0} =   \, \delta(\mathbf{x} - \mathbf{x}') \mathbbm{1}\, ,
\end{equation}
where $\mathbbm{1}$ is the unit matrix acting on the multi-vector $\Psi^A$, Eq.~(\ref{eq:fieldorder}).
The solution is given by the Duhamel series organized as an expansion in powers of the background $\bar{N}_i$,
\begin{align}
\label{eq:Duhamel}
     K \Bigr|_{\mathbf{x}' \rightarrow \mathbf{x}}= &e^{-s\mathbb{D}^{(0)}}\delta(\mathbf{x}-\mathbf{x}') \Bigr|_{\mathbf{x}' \rightarrow \mathbf{x}}
     - \int_0^s ds' \,  e^{-(s-s') \mathbb{D}^{(0)}} \mathbb{D}^{(1)} e^{-s' \mathbb{D}^{(0)}} \delta(\mathbf{x}-\mathbf{x}') \Bigr|_{\mathbf{x}' \rightarrow \mathbf{x}} \nonumber \\
     &- \int_0^s ds' \,  e^{-(s-s') \mathbb{D}^{(0)}} \mathbb{D}^{(2)} e^{-s' \mathbb{D}^{(0)}} \delta(\mathbf{x}-\mathbf{x}') \Bigr|_{\mathbf{x}' \rightarrow \mathbf{x}} \nonumber \\
    &+ \int^s_{0} ds' \int_0^{s'} ds''  \, e^{-(s-s') \mathbb{D}^{(0)}} \mathbb{D}^{(1)} e^{-(s'-s'') \mathbb{D}^{(0)}} \mathbb{D}^{(1)}e^{-s'' \mathbb{D}^{(0)}}\delta(\mathbf{x}-\mathbf{x}')\Bigr|_{\mathbf{x}' \rightarrow \mathbf{x}} + \dots 
\end{align}
where we have denoted with dots the higher order terms $\mathcal{O}\big( (\bar{N}_i)^3 \big)$. The latter do not contribute to the quadratic in $\bar N_i$ part of the effective action \eqref{eq:Action Div}, which in $\mathbf{x}$-space is given by,
\begin{equation}
\label{eq:eff_action_heat_kernel}
    \Delta \Gamma_{\text{log div}} = \int d \mathbf{x} d \tau \Big( C_1 \partial_i \bar N_j \partial_{i} \bar N_j + C_2 \partial_i \bar N_i \partial_{j} \bar N_j \Big) \times \int \frac{ds}{s} \, .
\end{equation}
By covariance, the linear in $\bar{N}_i$ term in the first line of \eqref{eq:Duhamel} must be proportional to $\partial_i \bar{N}_i$ and therefore vanishes upon taking the functional trace \eqref{eq: TrLogFourier}.
The two remaining non-trivial terms when plugged in \eqref{eq: TrLogFourier} comprise the fields part of the effective action,\footnote{One can show that the $\Delta \Gamma^{(2)} $ term is equal to the sum of the \qu{bubble} diagrams, whereas the second term --- to the sum of the \qu{fish}.} 
\begin{equation}
\label{eq:delta_gamma_fields}
    \Delta \Gamma_{\text{fields}}  \equiv  \Delta \Gamma^{(2)}_{\text{fields}} + \Delta \Gamma^{(1,1)}_{\text{fields}} \, .
\end{equation}
From \eqref{eq:eff_action_heat_kernel} we also see that in both terms we must only keep the pieces with \textit{two} derivatives of the background. 
We are going to see that both of these terms contain \textit{irregular} pieces that contain extra divergences, on top of the overall logarithmic divergence produced by the integral $\int\frac{ds}{s}$. 
Similarly to what happens in the diagrammatic approach of Sec. \ref{sec: Beta Function}, they cancel out. Before computing the effective action we, however, must address the redefinition (\ref{eq:D_to_LDR}).

\subsection{Diagonalizing $\mathbb{D}^{(0)}$}
In addition to having entries of different dimension, the quadratic form (\ref{eq:Dnot}) is not \textit{diagonal}. This complicates the exponentiation and, therefore, presents a nuisance when doing the actual computation. We fix both problems simultaneously by multiplying $\mathbb{D}$ on the left with $\mathbb{L}$ and on the right with $\mathbb{R}$, where $\mathbb{L}$ and $\mathbb{R}$ are background-independent matrices with constant non-zero determinant, see Eq.~(\ref{eq:D_to_LDR}). 
The choice,
\begin{equation}
    \mathbb{L}= \begin{pmatrix}
        \delta^{j}_{i} & 0 & 0 & 0 \\
       0 & \delta^{j}_{i} & 0 & 0 \\
       0 & 0 & - \frac{1}{2}\left(\delta_{i}^{k} \delta_{j}^{l} + \delta_{j}^{k} \delta_{i}^{l} \right) \Delta &  \frac{\mu_5}{2\mu_{67}} \mathcal{D}^{ij} \\
       0 & 0 & 0 & - \Delta
    \end{pmatrix} \, ,
\end{equation}
and 
\begin{equation}
    \mathbb{R} = \begin{pmatrix}
        \mu_s\left(\frac{1-2\lambda}{2(1-\lambda)} \Delta^2 \partial_i \partial_j -\delta_{ij} \Delta^3\right) &  -i \delta_{ij} \Delta^2\partial_{\tau} & 0 & 0 \\
        i \delta_{ij}  \Delta^2\partial_{\tau}&    (2\lambda-1)\Delta^2\partial_i \partial_j -\delta_{ij}\Delta^3 & 0 & 0 \\
        0 & 0 & -2 (G^{-1})_{ijkl} \Delta & 0 \\
        0 & 0 & \frac{\mu_5}{\mu_{67}} \left(\partial_k \partial_l-\frac{1-\lambda}{1-2\lambda} \delta_{kl} \Delta \right) & - \Delta
    \end{pmatrix} \, ,
\end{equation}
makes the dimensions right and conveniently brings $\mathbb{D}^{(0)}$ to a diagonal form. Of course, 
the contributions $\mathbb{D}^{(1,2)}$ must also be transformed in the same way. We keep the same notation for the transformed $\mathbb{D}$'s to avoid cluttered expressions. 

Explicitly, the transformed operator $\mathbb{D}^{(0)}$ is given by, 
\begin{equation}
\label{eq: New D}
    \mathbb{D}^{(0)} = \, \begin{pmatrix}
    \delta_i^j \mathbb{d}_{s}& 0 & 0 & 0 \\
    0 & \delta_i^j\mathbb{d}_{s} & 0 & 0 \\
    0 & 0 &  I^{kl}_{ij}\mathbb{d}_{s}  & 0 \\
    0 & 0 & 0 &   \mathbb{d}_{67} 
    \end{pmatrix} \, ,
\end{equation}
where we have defined
\begin{equation}
\label{eq:dispersion_relations_operators}
    \mathbb{d}_{s}[\partial_{\tau}, \partial_i] \equiv \Delta^2 (- \partial_{\tau}^2 + \mu_s \Delta^2) \, , \quad \mathbb{d}_{67}[\partial_i] \equiv 2 \mu_{67}\Delta^4 \, .
\end{equation}
It is diagonal, with first three entries being proportional to the 
dispersion relations. Note that these relations are the same for all the modes in the uniform gauge. The instantaneous mode is contained in the lower right corner. Note also the overall extra factor of $\Delta^2$ in \eqref{eq:dispersion_relations_operators}. This is a price to pay for diagonalizing the $\{h_{ij},n\}$ sector. 
An important consequence is that the \qu{proper time} $s$ corresponding to this $\mathbb{D}^{(0)}$ has  scaling dimension $(-8)$. This will have implications for the correct identification of the logarithmic divergences.

\subsection{Computing the fields determinant}
\label{sec:detComp}

Let us compute the two terms in \eqref{eq:delta_gamma_fields}. We start with the $\Delta \Gamma^{(2)}$-term.
\subsubsection{Effective action: Contribution $\Delta \Gamma^{(2)}_{\text{fields}}$}
Using the cyclicity of the trace, we bring the leftmost exponent to the right,
\begin{equation}
\label{eq: Bghosts}
    \Delta \Gamma^{(2)}_{\text{fields}}  = - \int \frac{ds}{s}\int d \tau d \mathbf{x} \int \frac{d \omega }{2 \pi}  \int_0^s ds' \, \text{tr}\, \mathbb{D}^{(2)}\left[i\omega, \partial_i \right] e^{-s \mathbb{D}^{(0)}\left[i\omega, \partial_i \right]} \delta(\mathbf{x}-\mathbf{x}')\Bigr|_{\mathbf{x}' \rightarrow \mathbf{x}} e^{i \omega (\tau - \tau'))} \Bigr|_{\tau' \rightarrow \tau}\, .
\end{equation}
Representing the spatial $\delta$-function with the Fourier integral, we write
\begin{equation}
\label{eq: BtoFourier}
     \text{tr}\, \mathbb{D}^{(2)}\left[i \omega, \partial_i \right] e^{-s \mathbb{D}^{(0)}} \delta(\mathbf{x}-\mathbf{x}')\Bigr|_{\mathbf{x}' \rightarrow \mathbf{x}} =  \int \frac{d^2q}{(2 \pi)^2}  \, \text{tr}\,\mathbb{D}^{(2)} \big[i \omega, i q_i \big] e^{-s \mathbb{D}^{(0)}[i \omega, i q_i]} \, ,
\end{equation}
Because of the particular structure \eqref{eq: New D}, after taking the matrix trace, there will be \textit{regular} terms with
\begin{equation}
    e^{-s q^4 (\omega^2 + \mu_s q^4)} \, ,
\end{equation}
as well as the \textit{irregular} ones coming with the exponent
\begin{equation}
    e^{-s \mu_{67} q^8} \, .
\end{equation}
For a fixed $s$, the $\omega$-integral for the regular terms is convergent and we can take the limit $\tau' \rightarrow\tau$. We then evaluate one by one the integrals over $\omega$, $q_i$, and $s'$. 

After some integrations by parts in $\mathbf{x}$, the final result is
\begin{align}
\label{eq:fields_bubble_heat_kernel}
     \Delta \Gamma^{(2)}_{\text{fields}} \Bigr|_{\text{reg.}}  =&  \frac{1}{128 \pi (1-2\lambda) \mu^2_{67}\sqrt{\mu_s}} \Bigg( \int d \tau d\mathbf{x}  \Big[ 8(1-\lambda)\mu_{67}^2 + \mu_5 \mu_{67} + (1-\lambda) \mu_5^2 \Big] \partial_i \bar{N}_j \partial_i \bar{N}_j \nonumber \\
     &+ (1-2\lambda) \Big[ 12(1-\lambda) \mu_{67}^2 + \mu_5 \mu_{67} + (1-\lambda) \mu_5^2 \Big] \partial_i \bar{N}_i \partial_j \bar{N}_j \Bigg) \times \int \frac{ds}{s}.
\end{align}
The irregular term is given by
\begin{align}
\label{eq:Kbubble_Irreg}
    \Delta \Gamma^{(2)}_{\text{fields}}\Bigr|_{\text{irr.}}  =&  -\frac{1}{2}\int \frac{ds}{s} \int_0^{s} ds' \int d \tau d\mathbf{x}  \left[ \partial_i \bar{N}_j \partial_i \bar{N}_j + (1-2\lambda) \partial_i \bar{N}_i \partial_j \bar{N}_j \right] \nonumber \\
    &\times \int \frac{d \omega}{2 \pi} \int \frac{d^2 q}{(2 \pi)^2} q^4 e^{-s \, 2\mu_{67} q^8} e^{i \omega (\tau - \tau'))} \Bigr|_{\tau' \rightarrow \tau} \, .
\end{align}
Upon integration over $\omega$ we now get $\delta(\tau - \tau')$ with an ill-defined limit $\tau' \rightarrow \tau$,
\begin{equation}
\label{eq:Gamma2bad}
    \Delta \Gamma^{(2)}_{\text{fields}} \Bigr|_{\text{irr.}} = -\frac{\Gamma\big(\frac{3}{4}\big)}{64 \, 2^{3/4} \pi^{2}  \mu_{67}^{3/4}} 
     \int d \tau d\mathbf{x}  \left[ \partial_i \bar{N}_j \partial_i \bar{N}_j + (1-2\lambda) \partial_i \bar{N}_i \partial_j \bar{N}_j \right]\times \int \frac{ds}{s^{3/4}} \,  \delta(\tau - \tau') \Bigr|_{\tau' \rightarrow \tau} \, .
\end{equation}
Notice that this term has a wrong scaling with $s$. The dimensions still work out right because $s$ has scaling dimension $8$ and the $\delta$-function has $-2$, but we see that this term does not contribute to the logarithmic divergence $\int\frac{ds}{s}$ of the one-loop effective action. Of course, to make this argument rigorous, we would need to regularize the $\delta$-function. However, we are going to see that this term actually cancels with a similar $\delta$-functional term in $\Delta\Gamma^{(1,1)}_{\text{fields}}$. 

\subsubsection{Effective action: Contribution $\Delta \Gamma^{(1,1)}_{\text{fields}}$}
The $\Delta \Gamma^{(1,1)}_{\text{fields}}$-term requires a bit more work. We can still use the cyclicity to shuffle the exponents,
\begin{equation}
\label{eq: FTerm}
   \Delta \Gamma^{(1,1)}_{\text{fields}}=  \int \frac{ds}{s}\int d \tau d \mathbf{x} \int \frac{d \omega }{2 \pi}\int^s_{0} ds' \int_0^{s'} ds''  \, \text{tr}\, \mathbb{D}^{(1)} e^{\Delta s\mathbb{D}^{(0)}} \mathbb{D}^{(1)}e^{-\Delta s \mathbb{D}^{(0)}}e^{-s \mathbb{D}^{(0)}}\delta(\mathbf{x}-\mathbf{x}')\Bigr|_{\mathbf{x}' \rightarrow \mathbf{x}} \, ,
\end{equation}
with
\begin{equation}
    \Delta s = s''-s' \, .
\end{equation}
Writing out the traces, we encounter the regular terms of the form
\begin{equation}
\label{eq:diagonal_exponents}
    \mathbb{X} \, e^{\Delta s \, \mathbb{d}_s[i \omega, \partial_i]} \mathbb{Y} \, e^{-\Delta s \, \mathbb{d}_s[i \omega, \partial_i]}\, ,
\end{equation}
as well as the \textit{irregular} ones with \qu{mixed} exponents,
\begin{equation}
\label{eq:mixed_exponents}
    \mathbb{X}e^{\Delta s \,\mathbb{d}_s} \, \mathbb{Y} \, e^{-\Delta s \,\mathbb{d}_{67}} \, , \quad   \mathbb{X}e^{\Delta s \,\mathbb{d}_{67}} \, \mathbb{Y} \, e^{-\Delta s \,\mathbb{d}_s} \, .
\end{equation}
Here $\mathbb{X},\mathbb{Y}$ stand for some particular matrix elements of $\mathbb{D}^{(1)}$.  There are no contributions with two irregular factors $e^{\Delta s \,\mathbb{d}_{67}}$ and $e^{-\Delta s \,\mathbb{d}_{67}}$ simply because the corresponding matrix elements of $\mathbb{D}^{(1)}$ are zero.\footnote{This is related to the fact that the interaction Lagrangian does not contain any vertices 
with one background $\bar{N}_i$ and two lapse fluctuations.} 

We compute the terms of the first type \eqref{eq:diagonal_exponents} perturbatively in derivatives of the background,
\begin{equation}
\label{eq: CommutatorExpansion}
    \mathbb{X} \, e^{\Delta s \, \mathbb{d}_s[i \omega, \partial_i]} \mathbb{Y} \, e^{-\Delta s \, \mathbb{d}_s[i \omega, \partial_i]} = \mathbb{X} \, \mathbb{Y} + \Delta s \, \mathbb{X}\left[\mathbb{d}_s,\mathbb{Y} \right] + \frac{\Delta s^2}{2} \, \mathbb{X}\left[\mathbb{d}_s, \left[\mathbb{d}_s,\mathbb{Y} \right] \right] + \dots \, .
\end{equation}
Here, the commutator is non-trivial because the matrix elements $\mathbb{Y}$ depend on $\bar{N}_i$ and derivatives thereof. We can stop the expansion at second order because every further commutator brings at least one additional derivative of the background, and we only need to keep the terms with two of them, see Eq.~\eqref{eq:eff_action_heat_kernel}. After computing all the commutators, one should not forget that $\mathbb{X}$ and commutators of $\mathbb{Y}$ are still differential operators, and so the derivatives inside $\mathbb{X}$ act both on the $\delta$-function \textit{and} on $\mathbb{Y}$. Carefully evaluating this action, we end up with an expression of the same type as $\Delta \Gamma^{(2)}_{\text{fields}}$ and we treat it accordingly. The final expression for the regular part reads,
\begin{align}
\label{eq:fields_fish_heat_kernel}
     \Delta \Gamma^{(1,1)}_{\text{fields}} \Bigr|_{\text{reg.}} =&  \frac{1}{512 \pi (1-2\lambda) \mu^2_{67}\sqrt{\mu_s}} \Bigg( \int d \tau d\mathbf{x}  \Big[ (-51+86\lambda-32 \lambda^2)\mu_{67}^2 \nonumber \\
     &+6(1-2\lambda) \mu_5 \mu_{67} - (3-2\lambda) \mu_5^2 \Big] \partial_i \bar{N}_j \partial_i \bar{N}_j \nonumber \\
     &- 2(1-2\lambda) \Big[ 8(3-4\lambda) \mu_{67}^2 + 4\mu_5 \mu_{67} + (1-2\lambda) \mu_5^2 \Big] \partial_i \bar{N}_i \partial_j \bar{N}_j \Bigg) \times \int \frac{ds}{s}.
\end{align}

To group conveniently the irregular terms (\ref{eq:mixed_exponents}), we carry over all factors involving $\mathbb{d}_{67}$ to the right. To this aim, whenever we have 
$e^{\Delta s \,\mathbb{d}_{67}}$ on the left, we write
\begin{equation}
\label{eq:mixed_trick}
  \mathbb{X} e^{\Delta s \, \mathbb{d}_{67}} \, \mathbb{Y} \, e^{-\Delta s \,\mathbb{d}_s} =\mathbb{X} \Big[ e^{\Delta s \, \mathbb{d}_{67}}\, \mathbb{Y} \,  e^{-\Delta s \,\mathbb{d}_{67}}  \Big]  e^{\Delta s \,\mathbb{d}_{67}} e^{-\Delta s \,\mathbb{d}_s}\,.
\end{equation}
We then apply the similar commutator expansion \eqref{eq: CommutatorExpansion} for the term in the square bracket, with $\mathbb{d}_{67}$ instead of $\mathbb{d}_s$.\footnote{It turns out that the terms of this type come already with two derivatives of the background, so in fact we do not need to evaluate any commutators. 
This feature can be traced back to the property that the part of the interaction Lagrangian containing the lapse perturbation $n$ \eqref{eq:Ni_n} is  proportional to the derivatives of $\bar{N}_i$.} 
Plugging the result back in \eqref{eq: FTerm}, we have
\begin{align}
\label{eq:Kfish_Irreg}
    \Delta \Gamma^{(1,1)}_{\text{fields}} \Bigr|_{\text{irr.}} =  \frac{1}{2}\int \frac{ds}{s} &\int d \tau d\mathbf{x} \int \frac{d \omega}{2 \pi}\int \frac{d^2 q}{(2 \pi)^2}q^4 \,  \mathcal{E}(s,q,\omega)\Biggr\{ \partial_i \bar{N}_j \partial_i \bar{N}_j  \left[ \frac{(2 \lambda -3 )q^4 + 4(\lambda -1)\omega^2}{4(\lambda-1)} \right] \nonumber \\
    &+ (1- 2 \lambda)  \partial_i \bar{N}_i \partial_j \bar{N}_j  \left[ \frac{(2 \lambda -1 ) \mu_{s} q^4 + 2(\lambda -1)\omega^2}{2(\lambda-1)} \right]  \Biggr\}  \, e^{i \omega (\tau - \tau'))} \Bigr|_{\tau' \rightarrow \tau} \, ,
\end{align}
with a particular combination of the exponents (cf. \eqref{eq:mixed_trick}),
\begin{align}
    \mathcal{E}(s,q,\omega) \equiv &q^4 \int_0^{s} ds'\int_0^{s'} ds''  \, \Bigr(  e^{\Delta s \, q^4(\omega^2 + \mu_s q^4)} e^{-\Delta s \, 2 \mu_{67} q^8} e^{-s \, 2\mu_{67} q^8} \nonumber \\
    &+  e^{\Delta s \, 2\mu_{67} q^8} e^{- \Delta s \, q^4(\omega^2 + \mu_s q^4)} e^{-s \,q^4 (\omega^2 + \mu_s q^4)} \Bigr) \notag\\
 =& -s \times \frac{e^{-s\, q^4(\omega^2 + \mu_s q^4)} - e^{-s \, 2\mu_{67} q^8}}{\omega^2 + (\mu_s - 2 \mu_{67}) q^4} \, .
    \label{eq:Integral s}
\end{align}
Note the appearance of a spurious pole at $\omega^2=-(\mu_s-2\mu_{67})q^4$, which at first sight may seem worrisome. However, after plugging \eqref{eq:Integral s} in \eqref{eq:Kfish_Irreg} and integrating over the momentum and frequency the spurious pole disappears and the whole expression for $\Delta \Gamma^{(1,1)}_{\text{fields}} \Bigr|_{\text{irr.}}$ splits into a sum of a properly logarithmically divergent piece and a piece with $\delta(\tau-\tau')$, 
\begin{align}
\label{eq:Fish_Irreg_HeatKernel}
    \Delta \Gamma^{(1,1)}_{\text{fields}} \Bigr|_{\text{irr.}} =& \frac{\Gamma\big(\frac{3}{4}\big)}{64 \, 2^{3/4} \pi^{2}  \mu_{67}^{3/4}} 
    \times \int \frac{ds}{s^{3/4}} \int d \tau d\mathbf{x}  \left[ \partial_i \bar{N}_j \partial_i \bar{N}_j + (1-2\lambda) \partial_i \bar{N}_i \partial_j \bar{N}_j \right]\times \delta(\tau - \tau') \Bigr|_{\tau' \rightarrow \tau} \nonumber \\
    &+ \frac{(1-2 \lambda) \sqrt{\mu_s}}{512 \pi (1-\lambda) \mu_{67}} \int d \tau d\mathbf{x}  \left[ \partial_i \bar{N}_j \partial_i \bar{N}_j + 2 \partial_i \bar{N}_i \partial_j \bar{N}_j \right] \times \int \frac{ds}{s}
\end{align}

We see that the $\delta(\tau - \tau')$-piece cancels the contribution (\ref{eq:Gamma2bad}), so that the total contribution of irregular terms into the effective action is a pure logarithmic divergence, 
\begin{equation}
\label{eq:Delta_Gamma_Leftover_HeatKernel}
    \Delta \Gamma^{(2)}_{\text{fields}} \Bigr|_{\text{irr.}} +  \Delta \Gamma^{(1,1)}_{\text{fields}} \Bigr|_{\text{irr.}} = \frac{(1-2 \lambda) \sqrt{\mu_s}}{512 \pi (1-\lambda) \mu_{67}} \int d \tau d\mathbf{x}  \left[ \partial_i \bar{N}_j \partial_i \bar{N}_j + 2 \partial_i \bar{N}_i \partial_j \bar{N}_j \right] \times \int \frac{ds}{s} \, .
\end{equation}
Note that this contribution is non-zero, so one could not simply discard the irregular terms from the start.
After identification (cf. \eqref{eq:S_To_Log})
\begin{equation}
\label{eq:s_to_log8}
    \int \frac{ds}{s} \rightarrow \log \bigg(\frac{\Lambda^8_{\text{UV}}}{k^{8}_*} \bigg)\,, 
\end{equation}
the expression (\ref{eq:Delta_Gamma_Leftover_HeatKernel})
coincides with (the $\mathbf{x}$-space representation of) the contribution to the effective action of the irregular diagrams obtained with the diagrammatic method, see Appendix~\ref{app:irreg_beta}. As discussed in the Appendix, this contribution happens to be gauge invariant, so its uniform-gauge expression coincides with the general case.

\subsection{Ghost contribution}
\label{ssec:hkghosts}
The ghost determinant is much simpler. Just like $\mathbb{D}$, the quadratic form $\mathbb{B}$ consists of three terms,
\begin{equation}
    \mathbb{B} = \mathbb{B}^{(0)} + \mathbb{B}^{(1)} + \mathbb{B}^{(2)} \, ,
\end{equation}
Explicitly from \eqref{eq:Lgh} in the \textit{uniform gauge} we have 
\bseq
\begin{align}
\label{eq: B0}
    &\mathbb{B}^{(0)}_{ij} = \delta_{ij} \big(\partial_{\tau}^2 + \mu_s \Delta^2 \big)\, , \\
\label{eq: B1}
    &\mathbb{B}^{(1)}_{ij} =  2 \, \delta_{ij} \bar{N}_k \partial_k \partial_{\tau}  -2 \big(\partial_j \bar{N}_i\big) \partial_{\tau} \, , \\
\label{eq: B2}
    &\mathbb{B}^{(2)}_{ij} = \bar{N}_k(\partial_k \partial_j \bar{N}_i) - (\partial_k \bar{N}_i) (\partial_j \bar{N}_k) -  \delta_{ij} \bar{N}_k \bar{N}_l \partial_k \partial_l - \bar{N}_k (\partial_k \bar{N}_l) \partial_l + 2  (\partial_j \bar{N}_i) \bar{N}_k \partial_k \, . 
\end{align}
\eseq
for the zeroth-, first- and second-order pieces, respectively.

Note that $\mathbb{B}^{(0)}$ is already diagonal, so we can directly proceed to the steps described in Sec.~\ref{sec:detComp}. Further, the ghost part does not have any irregular contributions, which makes the calculation straightforward. We find,
\begin{equation}
\label{eq:ghost_bubble_heat_kernel}
     \Delta \Gamma^{(2)}_{\text{ghosts}} =  \frac{1}{8 \pi \sqrt{\mu_{s}}}\partial_i \bar{N}_i \partial_j \bar{N}_j \times \int \frac{ds}{s} \, ,
\end{equation}
\begin{equation}
\label{eq:ghost_fish_heat_kernel}
    \Delta \Gamma^{(1,1)}_{\text{ghosts}}= -\frac{1}{64 \pi \sqrt{\mu_{s}}} \Big(10 \, \partial_i \bar{N}_i \partial_j \bar{N}_j + \partial_i \bar{N}_j \partial_i \bar{N}_j \Big)\times \int \frac{ds}{s} \, ,
\end{equation}
These are equal to the contribution of ghosts in the diagrammatic approach. 
Note that one should be careful here with the identification of the divergences. From \eqref{eq: B0} we see that $s$ has the scaling dimension $(-4)$, and so
\begin{equation}
    \int \frac{ds}{s} \rightarrow \log \bigg(\frac{\Lambda^4_{\text{UV}}}{k^{4}_*} \bigg)\, ,
\end{equation}
in contrast to \eqref{eq:s_to_log8}. This produces a relative factor of $2$ between the ghosts and the fields terms.

\subsection{Complete one-loop effective action}
Combining \eqref{eq:ghost_bubble_heat_kernel} and \eqref{eq:ghost_fish_heat_kernel} with the fields contribution \eqref{eq:fields_bubble_heat_kernel}, \eqref{eq:fields_fish_heat_kernel}, \eqref{eq:Delta_Gamma_Leftover_HeatKernel}, we arrive at the formula for logarithmically divergent part of the effective action:
\begin{align}
\label{eq:EffectiveActionHK}    \Delta \Gamma_{\text{log div}} =&  \frac{1}{256 \pi (1\!-\!2\lambda) \mu_{67}\sqrt{\mu_s}} \bigg[ \int d \tau d\mathbf{x}  \big( -(23-30\lambda)\mu_{67} + 2(5-6\lambda) \mu_5 + 4 (1-2\lambda) \mu_1 \big) \partial_i \bar{N}_j \partial_i \bar{N}_j \nonumber \\
     &+ 4(1-2\lambda) \big( -2(1-2\lambda) \mu_{67} - \mu_5 + 2\mu_1 \big) \partial_i \bar{N}_i \partial_j \bar{N}_j \bigg] \times \log \frac{\Lambda^4_{\text{UV}}}{k^{4}_*} \, ,
\end{align}
where we have used \eqref{eq:mus} to simplify the expression. It reproduces the result for the effective action \eqref{eq:Action Div} with the counter-terms \eqref{eq:C1}, \eqref{eq:C2} in the uniform gauge, as well as the $\beta$-functions \eqref{eq:beta_G},~\eqref{eq:beta_lambda}.

%% file: Sections/Section5.tex
\section{Conclusions}
\label{sec: Conclusion}
In this paper we made conceptual and technical progress towards formulating non-projectable Ho\v rava gravity as a \textit{quantum} theory with a Lagrangian path integral. 
We started with a phase space path integral implementing the second-class constraints present in the theory and 
transitioned to the configuration space by integrating out all the canonical momenta. We showed that after a certain change of variables one can also integrate over the auxiliary fields implementing the constraints 
and arrive at a path integral with the original action and a particular non-trivial measure. This measure is ultra-local in time, but non-local in space, and depends only on the metric and the lapse. In the fully local representation of the measure, with the auxiliary fields integrated back in, we identified a rich symmetry structure with four fermionic and four bosonic symmetries acting in the space of the lapse and auxiliary fields. Although these symmetries are linear, and therefore preserved at the quantum level, they are not enough to completely fix the structure of the sector with the auxiliary fields.  We dealt with the first-class constraints associated with the gauge symmetry of the spatial diffeomorphisms by fixing the gauge in a way similar to the one used in the projectable theory: we employed the standard Faddeev--Popov procedure jointly with the background-field method. The only difference with respect to the projectable case was the generalization of the gauge-fixing function to preserve the time reparameterization invariance. 

As a case study of the general formalism we considered the theory in $(2+1)$ dimensions on a background with flat metric, lapse set to 1 and a non-trivial space-dependent, but constant in time, shift vector. We developed a diagrammatic technique and constructed a regularization with the higher time derivatives obeying all the newly found symmetries of the action and measure. Pairing it up with the dimensional regularization in the spatial directions and focusing on the two-point function of the shift, we demonstrated cancellation of the irregular frequency divergences (IFDs) which threaten the locality of the theory. We computed the divergent part of the effective action for the shift, which turned out to be purely local. As an application of our result, we extracted the gauge-invariant $\beta$-function for the essential coupling $\lambda$ and a gauge-dependent $\beta-$function for the non-essential Newton coupling $G$. 

Finally, we repeated the computation of the effective action with an appropriate adaptation of the heat-kernel method. For simplicity, we worked with a gauge choice where the physical and gauge modes have identical dispersion relations. Due to kinetic coupling of the metric and lapse fluctuations, their Hessian is not diagonal. We developed a diagonalization procedure and then used perturbation theory in the powers of the background field to find the necessary logarithmically divergent terms. 
 The method naturally regularizes the frequency divergences via the point-splitting in time. Specifically, the IFDs arise as the factors of $\delta(\tau)$ at $\tau \rightarrow 0$, accompanied by a non-logarithmic divergence in the \qu{heat kernel time}. They cancel out in the full expression for the divergent part of the shift effective action. This is in complete agreement with the diagrammatic approach. An advantage of the heat-kernel method is that it can potentially be generalized to more complicated backgrounds than the one considered in this work, when the diagrammatic calculation becomes technically unfeasible.

At present, it is still unknown if the non-projectable Ho\v rava gravity is renormalizable or not. The key missing step in a proof of renormalizability is an argument ensuring the preservation of the path integral measure whose structure is crucial for the cancellation of IFDs. We believe, our work sheds light on the mechanism of this cancellation and sets the stage for future developments.

%% file: Appendix/AppendixA.tex
\section{Gauge fixing sector} 
\label{s:gauge fixing}

The gauge-fixing and ghost parts of the Lagrangian in a general background are given in Eq.~(\ref{eq: Quantum Lagrangian}). Specifying to the background (\ref{eq:background_shift}), (\ref{eq:background_gamma_N}) and expanding the gauge fixing term, we get,
\begin{align}
        \mathcal{L}_{\text{gf}}=  \frac{\sigma}{2 G}  \biggl[\bar{D}_t \delta \hat{N}_i (\mathcal{O}^{-1})_{ij} \bar{D}_t \delta \hat{N}_j + \frac{(\bar{D}_t \delta \hat{N}_i) ( \partial_k h_{ki} - \lambda \partial_i h)}{\sigma} 
        + \frac{( \partial_k h_{ki} - \lambda \partial_i h)\mathcal{O}_{ij}( \partial_l h_{lj} - \lambda \partial_j h)}{4 \sigma^2} \biggr] \, .
\end{align}
In order to eliminate the non-local term $\bar{D}_t \delta \hat{N}_i (\mathcal{O}^{-1})_{ij} \bar{D}_t \delta \hat{N}_j$, we integrate in an auxiliary field, which we suggestively call $\pi_i$, because it is similar to the conjugated momentum of the shift fluctuation $n_i$ \cite{Barvinsky2017AssFreed}. Doing so, we arrive at the gauge-fixing Lagrangian,\footnote{This procedure also produces a field-independent determinant, which will not be important for what follows, see \cite{Barvinsky2019} for a careful discussion.}
\begin{equation}
\label{eq:Lgf}
\begin{split}
        \mathcal{L}_{\text{gf}}=  \frac{\sigma}{2 G} & \biggl[- \frac{G^2}{ \sigma^2} \biggl( \pi_i \Delta \pi_i + \xi \pi_i \partial_i \partial_j \pi_j \biggr) + \frac{2 G}{\sigma  } \pi_i (\bar{D}_t \delta \hat{N}_i ) + \frac{(\bar{D}_t \delta \hat{N}_i ) ( \partial_j h_{i j} - \lambda \partial_i h)}{\sigma  } \\
        & + \frac{\partial_k h_{ik}- \lambda \partial_i h}{4 \sigma^2} \biggl( \Delta (\partial_j h_{ji} - \lambda \partial_i h) + \xi \partial_i \partial_j ( \partial_l h_{jl} - \lambda \partial_j h) \biggr) \biggr] \, .
\end{split}
\end{equation}
The ghost Lagrangian, in turn is given by
\begin{align}
\label{eq:Lgh}
        \mathcal{L}_{\text{gh}}=  \frac{1}{G} \biggl[ -\bar{c}_i \bar{D}_t\bar{D}_t c_i -\frac{1}{2 \sigma} \bar{c}_i \Big( \Delta^2 c_{i} + \big((1-2\lambda)(1+\xi)+\xi\big)\Delta \partial_j \partial_{i}   c_j\Big)  \biggr] + \dots \, ,
\end{align}
with
\begin{equation}
    \bar{D}_t c_i = \dot{c}_i - \bar N_{k} \partial_k  c_i +  c_k \partial_k \bar N_i \, , 
\end{equation}
and dots standing for the terms of cubic and quartic order in ghosts and fluctuations, which we will not need at one loop.

\section{Measure contribution to the quadratic Lagrangian}
\label{app:A}
On a more general background than the one considered in this paper, the field-dependent measure $\mu[N,\gamma_{ij}]$ becomes important. In this case, it is convenient to work directly with the action (\ref{eq:S_V_2}), without integrating out the fields ${\cal A}$, $\eta$, $\bar\eta$. The quadratic Lagrangian 
 $\mathcal{L}^{(2)}_{\rm flat}$ in \eqref{eq:Lqflat} then extends to\footnote{Note that the linear term in $\mathcal A$ drops out as emphasized in Sec. \ref{s:non_local_measure}}
\begin{equation}
    \mathcal{L}^{(2)}_{\rm flat} \to \mathcal{L}^{(2)}_{\rm flat} + \frac{\mu_{67}}{2G} \big( \Delta \mathcal A \Delta \mathcal A +  2\bar{\eta} \Delta^2 \eta \big) \, ,
\end{equation}
whence one reads off the propagators
\begin{equation}
\label{Aetaprops0}
    \langle \mathcal{A} \mathcal{A} \rangle =\langle  \bar{\eta} \eta \rangle = -\langle  \eta  \bar{\eta} \rangle =  \frac{i G }{ \mu_{67} q^4} \,.
\end{equation}
Note that, redefining the lapse fluctuation, 
\begin{equation}
    \tilde n = n+\mathcal{A}\,,
\end{equation}
one can remove the irregular piece form the propagator $\langle\tilde n\tilde n\rangle$, at the price of transferring it into $\langle\tilde n{\cal A}\rangle$. This field redefinition corresponds to the basis of fields $\{{\cal N},{\cal A}\}$ used in the Hamiltonian approach \cite{Bellorin2024}.

Regularization (\ref{eq: HigherDerTerm}) brings in additional quadratic contributions,
\begin{equation}
    \Delta\mathcal{L}^{(2)}_{\rm flat} = -\frac{\mu_{67}}{2G\Lambda^2} \Big(\partial_i \dot{\mathcal{A}}\partial_i \dot{\mathcal{A}} + 2\partial_i \dot{\bar{\eta}} \partial_i \dot{\eta}\Big) \, .
\end{equation}
These modify the $\langle {\cal A}{\cal A}\rangle$ and $\langle \bar\eta\eta\rangle$ propagators according to the general rule (\ref{eq:mu67sub}),
\begin{equation}
\label{Aetaprops}
    \langle \mathcal{A} \mathcal{A} \rangle =\langle  \bar{\eta} \eta \rangle = i G  \frac{1}{ \mu_{67} q^2(q^2 - \omega^2/\Lambda^2)} \,.
\end{equation}

\section{Interaction Lagrangian}
\label{app: InteractionLagrangian}
The interaction Lagrangian is given by the expansion of the quantum Lagrangian \eqref{eq: Quantum Lagrangian} around the background 
\eqref{eq:background_shift}, \eqref{eq:background_gamma_N} up to second order in $\bar{N}_i$ and the field perturbations. From there, one can read off the interaction vertices of Figs. \ref{f:3pt} and \ref{f:4pt}. 
The vertices which do not contain the lapse perturbation $n$ are the same as in the projectable Ho\v rava gravity. Since they are also the same in any number of spatial dimensions, they can be directly read off from the Appendices of Ref.~\cite{Barvinsky2019}. For completeness, we reproduce them here.\footnote{The expressions reported here may differ from those in \cite{Barvinsky2019} by an integration by parts. Additionally, here the gauge parameter $\varsigma$ of \cite{Barvinsky2019} is set to zero.} 
The vertices with a single external field $\bar N_i$ are:
\begin{subequations}
\begin{align}
 &\mathcal L_{ \bar N_i h \pi}^{(3)}  = - \dot h_{ij} \bar N
 _i \pi_j \,,\\
 &\mathcal L_{ \bar N_i n_j \pi}^{(3)}  = \frac{1}{2}(n_i \pi_j \partial_i \bar N_j- \bar N_i \pi_j \partial_i n_j) \,,\\
 &\mathcal L_{ \bar N_i \bar c c}^{(3)}  = \frac{2}{G}( -\bar c_i \bar N_j \partial_j \dot c_i+ \bar c_i \partial_j \bar N_i \dot c_j) \,,\\
 &\mathcal L_{ \bar N_i hh}^{(3)}  = \frac{1}{4G} \bar N_i \Big[ (2 \lambda -1) \dot h_{jk} \partial_i h_{jk} + 2 \lambda h_{jk} \partial_i \dot h_{jk}- \lambda h \partial_i \dot h + (2 \lambda +1) \dot h_{ij} \partial_j h \notag\\
 & \qquad\qquad\qquad\quad+ 2 \lambda h_{ij} \partial_j \dot h - 4 \dot h_{ij} \partial_k h_{jk} - 2 h_{jk} \partial_j \dot h_{ik} + h \partial_j \dot h_{ij} - 2 h_{ij} \partial_k \dot h_{ik} \Big] \,,\\
 &\mathcal L_{ \bar N_i n_j h}^{(3)}  = \frac{1}{4G} \bar N_i \biggl[ \lambda (n_j \partial_{ij}h+ h \partial_{ij}n_j)- 2 \lambda( n_j \partial_{ik}h_{jk} +h_{jk}\partial_{ik}n_j ) + \frac{4 \lambda -1}{2} \partial_i n_j \partial_j h\notag \\ 
 & \qquad\qquad\qquad\quad+\lambda( \partial_i h \partial_j n_j +  n_j\partial_{ij} h ) - \frac{1}{2} h \partial_{ij}n_j- \frac{1}{2} h \Delta n_i - \frac{1}{2} \partial_j n_i \partial_j h \notag\\
 &\qquad\qquad\qquad\quad-  n_j \partial_{jk}h_{ik} + h_{ik} \partial_{jk} n_j- \partial_jn_j \partial_k h_{ik} - 2 \lambda \partial_in_j \partial_k h_{ik} + \partial_j n_i \partial_k h_{jk}\notag\\
 &\qquad\qquad\qquad\quad +n_j \partial_{ik}h_{jk}- n_j \partial_{jk}h_{ik} + n_j \Delta h_{ij}+ h_{ij} \Delta n_j + 2 (1-\lambda) \partial_i h_{jk} \partial_k n_j \notag\\ 
 &\qquad\qquad\qquad\quad -\partial_j h_{ik} \partial_k n_j +\partial_k h_{ij} \partial_k n_j + h_{jk} \partial_{ik} n_j- 2 \lambda h_{ik} \partial_{jk}n_j+h_{jk} \partial_{jk}n_i \biggr] \,.
\end{align}
\end{subequations}
For the vertices with two $\bar N_i$ we show only the terms with derivatives acting on $\bar N_i$ since only such terms contribute to the part of the effective action we are interested in. These are:
\begin{subequations}
\begin{align}
 &\mathcal L_{ \d\bar N_i \d\bar N_j hh}^{(4)}  =  \frac{1}{2G}  \biggl[ \biggl(\frac{1}{8} h_{ij}h_{ij}- \frac{1}{16} h^2 \biggr)\biggl( (2 \lambda-1) \partial_k \bar N_l \partial_k \bar N_l - (\partial_k \bar N_k)^2 \biggr) \notag\\
 &\qquad\qquad\qquad\qquad +h_{jk} \partial_{(k} \bar N_{l)} \biggl(  h_{ij} \partial_l \bar N_i - \frac{1}{2} h \partial_l \bar N_j\biggr)+h_{ki}h_{jl} \partial_i \bar N_j \partial_{(k} \bar N_{l)} - \lambda h_{ij}h_{kl} \partial_i \bar N_j \partial_k \bar N_l\notag\\
 &\qquad\qquad\qquad\qquad-h_{jk} h_{ij} \partial_l \bar N_k \partial_i \bar N_l + 2(1-\lambda) h_{jl} h_{ij} \partial_k \bar N_k \partial_i \bar N_l- \frac{1-2\lambda}{2}h h_{il} \partial_k \bar N_k \partial_i \bar N_l\biggr] \,,\\
 &\mathcal L_{ \d\bar N_i \d\bar N_j c \bar c}^{(4)}  =  \frac{1}{G} \bar c_i \big( \partial_k \bar N_k \partial_j \bar N_i - \partial_j \bar N_k \partial_k \bar N_i\big) c_j \,.
\end{align}    
\end{subequations}

With respect to the projectable case~\cite{Barvinsky2019}, there are additional vertices containing the lapse perturbation $n$. 
Cubic interaction terms with one $\bar{N}_i$ are given by
\begin{subequations}
\label{eq:Ni_n}
\begin{align}
\label{eq: NbarLapseShiftVertex}
 &\mathcal L_{ \bar N_i n n_i}^{(3)}  =  -\frac{1}{2G} \partial_i \bar N_j \, \bigl[ n \partial_i n_j + n \partial_j n_i - 2 \lambda \, n (\partial_k n_k) \delta_{ij}  \bigr] \, ,\\
\label{eq: NbarLapseMetricVertex}
      & \mathcal L_{ \bar N_i n h}^{(3)}  = \frac{1}{2G} \partial_i \bar N_j \, \bigl[n \dot{h}_{ij} - \lambda n \dot{h} \delta_{ij}  \bigr] \,.
\end{align}
\end{subequations}
Notice that there are no cubic vertices with two $n$-legs. The quartic terms with two $\bar{N}_i$-legs following from the new part of the action are:\footnote{These are the full expressions, containing terms with and without derivatives acting on $\bar N_i$.}
\begin{subequations}
\label{eq:NiNj_n}
\begin{align}
\label{eq: NbarNbarLapseLapseVertex}
 &\mathcal L_{ \bar N_i \bar N_j nn}^{(4)}  = \frac{1}{16G} \Big[ \partial_{i} \bar N_j \partial_{i} \bar N_j + \partial_{i} \bar N_j \partial_{j} \bar N_{i} - 2\lambda \big(\partial_i \bar{N}_i\big)^2 \Big] \, n^2\,,\\
 &\mathcal L_{ \bar N_i \bar N_j nh}^{(4)}  = -\frac{\bar N_i}{2G} \Big[ n \Big( - \partial_j(h\partial_{(j}\bar N_{i)})+ \lambda \partial_{ij}(h \bar N_j)- \lambda  \partial_i h \partial_j \bar N_j+  \partial_k ( \partial_j h_{ik} \bar N_j)+ \partial_k h_{ij} \partial_k \bar N_j\notag\\
& \qquad\qquad\qquad\quad-(1-2\lambda)h_{ik} \partial_{jk}\bar N_j - 2\partial_k \left(h_{kj}\partial_{(j}\bar N_{i)} \right)-2 \partial_i h_{jk} \partial_k \bar N_j + 2\lambda \partial_{ik}(h_{jk} \bar N_j)- \Delta(h_{ij}\bar N_j)\Big) \notag\\
&
\qquad\qquad\qquad\quad+\lambda \partial_in \left( 2\partial_k(h_{jk} \bar N_j) - \partial_j (h \bar N_j) \right) + \partial_k n \Big( \partial_j h_{ik} \bar N_j - 2 \lambda h_{ik} \partial_j \bar N_j   + 2 h_{kj} \partial_{(j}\bar N_{i)} \notag\\
& \qquad\qquad\qquad\quad + \partial_i h_{jk} \bar N_j+2 h_{ij} \partial_{(j}\bar N_{k)} + \bar N_j \partial_k h_{ij} - h \delta_{kl}\partial_{(l}\bar N_{i)}\Big)\Big] \,,
\end{align}
\end{subequations}
where  $2 \, \partial_{(i}\bar N_{j)}= \partial_i \bar N_j + \partial_j \bar N_i$. 

While treating the irregular diagrams in Sec. \ref{subsec: Irreg Diag}, we have introduced the higher derivative regulator \eqref{eq: HigherDerTerm}. This term gives additional contributions to the interaction Lagrangian: 
\begin{subequations}
\label{eq:newHigherDerV}
\begin{align}
\label{eq:newHigherDerV1}
 \mathcal L_{ \bar N_i nn}^{(3), \Lambda}  =  \frac{\mu_{67}}{G \Lambda^2} \bar N_i \Big[& \partial_i \dot n \Delta n + \partial_{ij} \dot n \partial_j n + \partial_{ij}n \partial_j \dot n \Big] \,,\\
 \mathcal L_{ \bar N_i \bar N_j nn}^{(4), \Lambda}  =  \frac{\mu_{67}}{2G \Lambda^2} n \Big[& \partial_j \bar N_k \partial_j \bar N_i \partial_{ik}n - \partial_j \bar N_i \partial_k \bar N_i\partial_{jk}n+ \bar N_i \bar N_j \Delta \partial_{ij} n \notag\\
 &-\partial_in  \partial_k\big(\partial_i \bar N_j \partial_{k} \bar N_j \big) - \partial_i n \partial_j \big( \partial_k \bar N_k \partial_i \bar N_j \big)  - \partial_{ik} \bar N_j \partial_k \bar N_j \partial_i n  \notag\\
 &+\bar N_i \Big( \Delta \bar N_j \partial_{ij }n - \partial_{ijk} \bar N_k\partial_j n+ \partial_j \bar N_j \partial_i \Delta n + \partial_i \bar N_j\ \partial_j \Delta n + 2 \partial_k \bar N_j \partial_{ijk} n \Big) \Big] \,.
\end{align}    
\end{subequations}
It also introduces coupling between the background shift vector and the auxiliary fields ${\cal A}$ 
\begin{subequations}
    \label{eq:newHigherDerVAeta}
    \begin{align}
        \mathcal L_{ \bar N_i \mathcal{A}\mathcal{A}}^{(3), \Lambda}  = - \frac{\mu_{67}}{G \Lambda^2} \bar N_i \Big[& \partial_i \dot{\mathcal{A}} \Delta \mathcal{A} + \partial_{ij} \dot{\mathcal{A}}\partial_j \mathcal{A} + \partial_{ij}\mathcal{A} \partial_j \dot{\mathcal{A}} \Big] \,, \\
     \mathcal L_{ \bar N_i \bar N_j \mathcal{A}\mathcal{A}}^{(4), \Lambda}  = - \frac{\mu_{67}}{2G \Lambda^2} \mathcal{A} \Big[& \partial_j \bar N_k \partial_j \bar N_i \partial_{ik}\mathcal{A} - \partial_j \bar N_i \partial_k \bar N_i\partial_{jk}\mathcal{A}+ \bar N_i \bar N_j \Delta \partial_{ij} \mathcal{A} \notag\\
    &-\partial_i \mathcal{A}  \partial_k\big(\partial_i \bar N_j \partial_{k} \bar N_j \big) - \partial_i \mathcal{A} \partial_j \big( \partial_k \bar N_k \partial_i \bar N_j \big)  - \partial_{ik} \bar N_j \partial_k \bar N_j \partial_i \mathcal{A}  \notag\\
    &+\bar N_i \Big( \Delta \bar N_j \partial_{ij }\mathcal{A} - \partial_{ijk} \bar N_k\partial_j \mathcal{A}+ \partial_j \bar N_j \partial_i \Delta \mathcal{A} + \partial_i \bar N_j\ \partial_j \Delta \mathcal{A} + 2 \partial_k \bar N_j \partial_{ijk} \mathcal{A} \Big) \Big] \,,
    \end{align}
\end{subequations}
and $\eta$, $\bar\eta$
\begin{subequations}
\label{eq:newHigherDerVAeta1}
    \begin{align}
        \mathcal L_{ \bar N_i \bar \eta \eta}^{(3), \Lambda}  = - \frac{\mu_{67}}{ G \Lambda^2} \bar N_i \Big[& \partial_i \dot{\bar \eta} \Delta \eta + \partial_{ij} \dot{\bar \eta}\partial_j \eta + \partial_{ij}\dot{\bar \eta} \partial_j \eta \Big] \,, \\
     \mathcal L_{ \bar N_i \bar N_j \bar \eta \eta}^{(4), \Lambda}  = - \frac{\mu_{67}}{2G \Lambda^2} \bar \eta \Big[& \partial_j \bar N_k \partial_j \bar N_i \partial_{ik}\eta - \partial_j \bar N_i \partial_k \bar N_i\partial_{jk}\eta+ \bar N_i \bar N_j \Delta \partial_{ij} \eta \notag\\
    &-\partial_i \eta  \partial_k\big(\partial_i \bar N_j \partial_{k} \bar N_j \big) - \partial_i \eta \partial_j \big( \partial_k \bar N_k \partial_i \bar N_j \big)  - \partial_{ik} \bar N_j \partial_k \bar N_j \partial_i \eta  \notag\\
    &+\bar N_i \Big( \Delta \bar N_j \partial_{ij }\eta - \partial_{ijk} \bar N_k\partial_j \eta+ \partial_j \bar N_j \partial_i \Delta \eta + \partial_i \bar N_j\ \partial_j \Delta \eta + 2 \partial_k \bar N_j \partial_{ijk} \eta \Big) \Big] \,.
    \end{align}
\end{subequations}

Using these interaction terms, its is straightforward, though tedious, to obtain the corresponding vertices. A Mathematica notebook with their expressions and derivations can be found at~\cite{github}.

\section{Contribution to effective action from irregular graphs}
\label{app:irreg_beta}
Here we derive the $\Lambda$-finite contribution to the effective action from the combination of the irregular graphs shown in Figs.~\ref{f:irreg_diagrams}, \ref{f:irreg_diagrams_Aeta}. For the fish diagram with lapse and shift propagators we have,
\begin{align}
\label{eq: Fish Irreg shift}
\Gamma^3_{ij} =
      i\int \frac{ d^d q}{(2 \pi)^d} \int \frac{d \omega}{2 \pi}& \Big(\delta_{ik}  (\mathbf{q}\mathbf{Q}) + q_i Q_k - 2 \lambda \, q_k Q_i \Big) \frac{\Big(\delta_{jl}  (\mathbf{q}\mathbf{Q}) + q_j Q_l - 2 \lambda q_l Q_j \Big)}{2 \mu_{67} (\mathbf{q}+\mathbf{Q})^2\big((\mathbf{q}+\mathbf{Q})^2 + \omega^2/\Lambda^2\big)} \nonumber \\
      &\times \frac{1}{2\sigma} q^2 \Big( -\mathcal{P}_1 (\delta_{kl} - \hat{q}_k \hat{q}_l) - \frac{\rho}{1-\lambda} \mathcal{P}_2 \hat{q}_k \hat{q}_l \Big) + \dots  \equiv \hat{\Gamma}^{3}_{ij} + \dots
\end{align}
As before, dots stand for the ab initio finite in $\Lambda$ terms. Combining $\hat{\Gamma}^{3}_{ij}$ with $\hat{\Gamma}^{2}_{ij}$ from \eqref{eq: Fish Irreg metric} we get,
\begin{equation}
    \hat{\Gamma}^2_{ij} + \hat{\Gamma}^3_{ij} = i\frac{(1-2\lambda)^2\mu_s}{(1-\lambda)\mu_{67}} \int \frac{ d^d q}{(2 \pi)^d} \int \frac{d \omega}{2 \pi}  \,  \frac{\tfrac{1}{1-2\lambda} q_i q_j (\mathbf{q}\mathbf{Q})^2 - q^2 (\mathbf{q}\mathbf{Q}) \big(q_i Q_j +q_j Q_i\big) - b_d q^4 Q_i Q_j}{(\mathbf{q}+\mathbf{Q})^2\big((\mathbf{q}+\mathbf{Q})^2 + \omega^2/\Lambda^2 \big)} \, \tilde{\mathcal{P}}_s^d  \, ,
\end{equation}
with 
\begin{equation}
    b_d = 1 - (d-2)\frac{2(1-\lambda)}{1-2\lambda} \, .
\end{equation}
Notice that this combination is gauge-invariant. By rotational covariance we have
\begin{equation}
     \hat{\Gamma}^2_{ij} + \hat{\Gamma}^3_{ij} = A Q^2 \delta_{ij} + B Q_i Q_j \, .
\end{equation}
Here 
\begin{equation}
    A = i\frac{(1-2\lambda)\mu_s}{Q^4(d-1)(1-\lambda)\mu_{67}} \int \frac{ d^d q}{(2 \pi)^d} \int \frac{d \omega}{2 \pi} \frac{Q^2q^2 (\mathbf{q}\mathbf{Q})^2 - (\mathbf{q}\mathbf{Q})^4}{(\mathbf{q}+\mathbf{Q})^2\big((\mathbf{q}+\mathbf{Q})^2 + \omega^2/\Lambda^2 \big)} \tilde{\mathcal{P}}_s^d \, ,
\end{equation}
and
\begin{equation}
    B = i\frac{(1-2\lambda)\mu_s}{Q^4(d-1)(1-\lambda)\mu_{67}} \int \frac{ d^d q}{(2 \pi)^d} \int \frac{d \omega}{2 \pi} \frac{P[\mathbf{q},\mathbf{Q}]}{(\mathbf{q}+\mathbf{Q})^2\big((\mathbf{q}+\mathbf{Q})^2 + \omega^2/\Lambda^2 \big)} \tilde{\mathcal{P}}_s^d \, ,
\end{equation}
with 
\begin{equation}
    P[\mathbf{q},\mathbf{Q}] =(d-1)\Big(1-2 (d-1)\lambda + 2 (d-2) \lambda^2 \Big) q^4 Q^4 - \Big(3-4 \lambda - 2 d(1-2\lambda) \Big) q^2 Q^2(\mathbf{q}\mathbf{Q})^2 - d  (\mathbf{q}\mathbf{Q})^4
\end{equation}
Consider the first term. Integrating over the frequency and then expanding in $\Lambda^{-1}$ we get
\begin{align}
    A = i\frac{1}{Q^4} \frac{(1\!-\!2\lambda)\sqrt{\mu_s} \sqrt{\frac{1-\lambda d}{1-2\lambda}}}{2(d-1)^{3/2}(1\!-\!\lambda) \mu_{67}} \int \!\!\frac{ d^d q}{(2 \pi)^d} \Bigg[\frac{(\mathbf{q}\mathbf{Q})^2(q^2 Q^2 \!-\! (\mathbf{q}\mathbf{Q})^2)}{|\mathbf{q}+\mathbf{Q}|^4 q^2} 
    + \frac{P_1^{(4,4)}[\mathbf{q},\mathbf{Q}]}{\Lambda|\mathbf{q}+\mathbf{Q}|^5} + \frac{P_2^{(6,4)}[\mathbf{q},\mathbf{Q}]}{\Lambda^2|\mathbf{q}+\mathbf{Q}|^6} +\dots\Bigg]
\end{align}
with $P_i^{(n,m)}[\mathbf{q},\mathbf{Q}]$ polynomials of degree $n$ in $\mathbf{q}$ and $m$ in $\mathbf{Q}$. By shifting $\mathbf{q} \rightarrow \mathbf{q} - \mathbf{Q}$ one can see that only the even terms in $1/\Lambda$ contain logarithmic divergences, and these divergence are all polynomial in $Q$. To extract the logarithmic divergence $\propto \frac{1}{d-2}$ renormalizing the effective action $\sim (Q\bar{N}_i(Q))^2$ we should keep the zeroth order term in $\mathbf{Q}$, which comes only from the first expression in the square brackets. Since the integral is convergent in the IR we can expand in $\mathbf{Q}$ and then average over angles. We do the same for the coefficient $B$. The end result is\footnote{Here we exhibit the result obtained using the Schwinger parametrization procedure, where the resulting logarithmic divergence $\frac{1}{d-2}$ assumes the form of $\int \frac{ds}{s}$.}
\begin{equation}
\label{eq:DeltaGammaLeftover}
      \Delta \Gamma = \frac{(1-2 \lambda) \sqrt{\mu_s}}{256 \pi (1-\lambda) \mu_{67}} \int d \tau \int \frac{d^2Q}{(2\pi)^2}\bar N^i(Q_i) \bar N^j(-Q_i) \, \left(Q^2 \delta_{ij} + 2Q_i Q_j \right) \times \int \frac{ds}{s} \, .
\end{equation}
Note that this contribution is independent of the gauge-fixing parameters.
It corresponds to the last pieces in the counter-terms (\ref{eq:C1}), (\ref{eq:C2}), which eventually combine into the $\beta$-functions for the couplings $G$ and $\lambda$ quoted in Sec.~\ref{subsec:RG}. 

\section{Counter-terms}
\label{app:C}

Here we give the expressions, in a generic $\sigma\xi$-gauge, for the one-loop counter-terms $C_1$ and $C_2$ in \eqref{eq:Action Div}:   
\begin{align}
\label{eq:C1}
    C_1 = &\frac{1}{256 \pi (1-\lambda)(1-2\lambda)\sqrt{\mu_s}} \times \Bigg(-17  - 2 \left(\frac{1}{\sqrt{\rho}}+2 \sqrt{2}\right)\sqrt{\sigma \mu_s} \nonumber \\
    &+ \lambda \bigg(37  + 4 \left(\frac{1}{\sqrt{\rho}}+3\sqrt{2}\right) \sqrt{\sigma \mu_s}\bigg) - 2\lambda^2 \Big(11+4 \sqrt{2}\sqrt{\sigma \mu_s}\Big)\Bigg) \nonumber \\
    &+\frac{\mu_5\Big((1-2\lambda)\mu_5-2(5 - 6 \lambda)\mu_{67}\Big)}{256 \pi(1-2\lambda)\sqrt{\mu_s}\mu_{67}^2} +\frac{(1-2\lambda)\sqrt{\mu_s}}{256 \pi(1-\lambda)\mu_{67}} \\
    \label{eq:C2}
    C_2 = &\frac{1}{128 \pi (1-\lambda)\sqrt{\mu_s}} \times \Bigg(-1  -  \left(\frac{1}{\sqrt{\rho}}+2 \sqrt{2}\right)\sqrt{\sigma \mu_s} \nonumber \\
    &+ 2\lambda \bigg(2  +  \left(\frac{1}{\sqrt{\rho}}+3\sqrt{2}\right) \sqrt{\sigma \mu_s}\bigg) - 4\lambda^2 \Big(1+ \sqrt{2}\sqrt{\sigma \mu_s}\Big)\Bigg) \nonumber \\
    &+\frac{\mu_5(\mu_5-2\mu_{67})}{128 \pi\sqrt{\mu_s}\mu_{67}^2} +\frac{(1-2\lambda)\sqrt{\mu_s}}{128 \pi(1-\lambda)\mu_{67}}
\end{align}
In each of the $C$'s, the first two lines correspond to the contribution from all the diagrams \textit{without} the lapse, whereas the third line comes from the diagrams \textit{with} the lapse. 
Note that there is no gauge parameter dependence in the with-the-lapse terms.
Note also that if we extract the overall factor $1/\sqrt{\mu_s}$ and express $\mu_s$ in 
the last terms in $C_1$ and $C_2$ using Eq.~\eqref{eq:mus}, the terms explicitly proportional to $\mu_5^2$ will cancel out. In other words, both in \eqref{eq:C1} and in \eqref{eq:C2}, $\mu_5^2$ appears in the combination
\begin{align}
\label{eq:mu5cancel}
    \frac{\mu^2_5}{\sqrt{\mu_s}\mu_{67}^2}+\frac{(1-2\lambda)\sqrt{\mu_s}}{(1-\lambda) \mu_{67}}= 4 \frac{\mu_1}{\mu_{67}\sqrt{\mu_s}} \,.
\end{align}